\documentclass[twocolumn,fleqn,usenatbib]{mnras}

\usepackage[T1]{fontenc}
\usepackage{ae,aecompl}
\usepackage{multirow}
\usepackage{float}

\usepackage{graphicx}	
\usepackage{amsmath}	
\usepackage{amssymb}	
\usepackage[utf8]{inputenc}
\usepackage{cleveref} 
\usepackage{ulem}

\def\kms{\rm km\,s^{-1}}
\def\Gpcpyr{\rm Gpc^{-3}\,yr^{-1}}

\title[Binary Neutron Star Mergers and Their Host Galaxies]{Formation and Evolution of Binary Neutron Stars: Mergers and Their Host Galaxies
}
\author[Chu et al.]{
Qingbo Chu,$^{1,2}$
Shenghua Yu$^{1}$,
and Youjun Lu$^{1,2}$\thanks{E-mail: luyj@nao.cas.cn}
\\
$^{1}$National Astronomical Observatories, Chinese Academy of
Sciences, 20A Datun Road, Beijing 100101, China\\
$^{2}$School of Astronomy and Space Sciences, University of Chinese
Academy of Sciences, 19A Yuquan Road, Beijing 100049, China
}

\begin{document}
%
%

\maketitle

\begin{abstract}
In this paper, we investigate the properties of binary neutron stars (BNSs) and their mergers by combining population synthesis models for binary stellar evolution (BSE) with cosmological galaxy formation and evolution models. We obtain constraints on BSE model parameters by using the observed Galactic BNSs and local BNS merger rate density ($R_0$) inferred from Gravitational Wave (GW) observations, and consequently estimate the host galaxy distributions of BNS mergers. We find that the Galactic BNS observations imply efficient energy depletion in the common envelope (CE) phase, a bimodal kick velocity distribution, and low mass ejection during the secondary supernova explosion. However, the inferred  $R_0$ does not necessarily require an extremely high CE ejection efficiency and low kick velocities, different from the previous claims, mainly because the latest inferred $R_0$ is narrowed to a lower value ($320_{-240}^{+490}\,\Gpcpyr$). The BNS merger rate density resulting from the preferred model can be described by $R(z)\sim R_0(1+z)^{\zeta}$ at low redshift ($z\lesssim0.5$), with $R_0\sim316$-$784\,\Gpcpyr$ and $\zeta\sim1.34$-$2.03$, respectively.  Our results also show that $R_{0}$ and $\zeta$ depend on settings of BSE model parameters, and thus accurate estimates of these parameters by future GW detections will put strong constraints on BSE models. We further estimate that the fractions of BNS mergers hosted in spiral and elliptical galaxies at $z\sim0$ are $\sim81$-$84$\% and $\sim16$-$19$\%, respectively. The BNS merger rate per galaxy can be well determined by the host galaxy stellar mass, star formation rate, and metallicity, which provides a guidance in search for most probable candidates of BNS host galaxies.
\end{abstract}



\section{Introduction}
\label{sec:intro}

Gravitational wave (GW) events can be now regularly detected by the advanced Laser Interferometer Gravitational wave Observatory (LIGO) and advanced VIRGO. With O1, O2, and O3 observations of LIGO/VIRGO, more than fifty mergers of stellar compact binaries have been detected  \citep[][]{Abbott et al. 2016a, Abbott et al. 2016b, Abbott et al. 2017a, Abbott et al. 2017b, Abbott et al. 2017c, Abbott et al. 2017d, Abbott et al. 2019, GW190814, GW190425, GW190521, GW190412, GWcatalog1, GWcatalog2}.
Among them, one of the most remarkable events is GW170817, a coalescence of binary neutron stars \citep[BNSs;][]{Abbott et al. 2017c}. GW170817 is the first BNS and the only GW event up to now that was identified by electromagnetic (EM) waves \citep{Abbott et al. 2017e, Goldstein2017, Savchenko2017, Coulter2017}, which marks the beginning of a new era of GW and multimessenger astronomy. It is anticipated that many more GW170817-like objects will be detected in the near future.

EM counterpart observations of BNSs provide important information on the properties of BNS mergers and its host galaxies, which is invaluable for investigating the detailed physical processes occurred during the merger \citep[e.g.,][]{Bovard2017, Drout2017, Kasen2017, Pian2017, Siegel2017, Abbott et al. 2018a}, the formation and evolution of BNSs \citep[e.g.,][]{Abbott et al. 2017g, Chruslinska2018, Giacobbo2018, Kruckow2018, Vigna2018}, and their cosmological applications as standard sirens \citep[e.g.,][]{Baker2017, Creminelli2017, Ezquiaga2017, Langlois2018, Sakstein2017, Bahamonde2018, Boran2018, Heisenberg2019}. For example, the multi-wavelength EM observations of GW170817 not only confirm the prediction of kilonova phenomenon \citep[e.g.,][]{Arcavi2017, Chornock2017, Cowperthwaite2017, Evans2017, Nicholl2017, Smartt2017, Tanaka2017, Villar2017} but also make it possible to estimate the Hubble constant with substantial accuracy only by GW170817 \citep[][]{Abbott et al. 2017f, Guidorzi2017, Cantiello2018, Chen2018, Feeney2019}. GW170817 is hosted in the early type galaxy NGC4993, which was a surprise as BNS mergers were expected to be mostly in late type galaxies and stimulated lots of subsequent interests in studying the host galaxies of BNS mergers \citep[e.g.,][]{Mapelli2018host1, Artale2019a, Artale2019b, Ducoin2019, Toffano2019}. 

A number of models have been proposed to produce BNS GW sources in the literature. There are mainly two formation channels for BNSs, i.e., isolated evolution of binary stars \citep[hereafter BSE channel, e.g.,][]{Zwart1998, Belczynski2002, Voss2003, Belczynski2007, Bogomazov2007, Kalogera2007, Kalogera2008, Dominik2012, Dominik2013, Dominik2015, Mennekens2014, Tauris2015, Belczynski2016a, Belczynski2016b, Chruslinska2018, Giacobbo2018, Kruckow2018, Vigna2018, AndrewsMandel2019, Neijssel2019} 
and dynamical evolution in compact stellar systems \citep[hereafter dynamical channel, e.g.,][]{Rasio1992, Phinney1996, Rasio2000, Grindlay2006, Ivanova2008, Petrovich2017, Belczynski2018, Fragione2019, Ye2020}. 
For the BSE channel, several involving physical processes, such as mass transfer, common envelope (CE) evolution, and natal kicks, are still poorly understood. It is possible to constrain these physical processes by using both GW detections of BNS mergers and EM observations of the Galactic BNSs, which may help to obtain better estimates on the BNS properties and merger rate. For the dynamical channel, neutron stars are not heavy enough to sink into the centre of globular cluster (GC) and the formation of BNSs is quite inefficient via this channel. The contribution from the dynamical channel to BNS merger rate, comparing with that from the BSE channel, is probably small \citep[e.g., $\la 1\%$][]{Belczynski2018, Fragione2019, Ye2020}, if not negligible, although the estimate for BNS merger rate from this channel still has a large uncertainty. In this paper, we only consider the BSE channel for the formation of BNSs. 

Different settings on the model parameters for the above mentioned several physical processes involved in the evolution of binary stars have quite large effects on properties of the resulting BNSs \citep[e.g.,][]{Dominik2012, Chruslinska2018, Giacobbo2018, Kruckow2018}. One of the crucial uncertainties arises from the CE evolution of binaries. In the late stage of CE evolution, a binary may eject its CE if the envelope binding energy is overwhelmed by a fraction of binary orbital energy \citep[e.g.,][]{Webbink1984, Belczynski2002, Dominik2012, Giacobbo2018, Vigna2018}. This is the so-called energy conservation mechanism. Almost all the previous works employed this mechanism for the CE ejection. However, since the masses and orbital separation of a binary vary significantly before and after CE evolution, the angular momentum may also play a vital role in the CE phase \citep{Nelemans2005}. In this paper, we will consider both the energy conservation mechanism and the angular momentum evolution. Another vital uncertainty is caused by the natal kicks. Several works found that high kick velocities are required to match the pulsar proper motions (e.g., a Maxwellian distribution with a scatter of $\sigma_{\rm v}=190 \kms$ \citep{Hansen1997} or $265 \kms$ \citep{Hobbs2005}). Some other works suggest a bimodal velocity distribution composed of a first peak at low velocities and a second peak at high velocities \citep[e.g.,][]{Fryer1998, Arzoumanian2002, Verbunt2017}. In this paper, we will investigate how different natal kick distributions influence the properties of resulting BNSs.

The local BNS merger rate is constrained to be $320_{-240}^{+490}\,\Gpcpyr$ by the two BNS mergers detected up to now \citep{Abbott et al. 2017c, GW190425, O3population}, though with large uncertainties, it has already provided a benchmark for constraining the formation mechanism of BNSs. The estimated local merger rate densities from different BSE models spread in a wide range from a few tens to several hundreds $\Gpcpyr$, even as high as $10^{3}$ $\Gpcpyr$, which suggests that the accurate merger rate by future GW detections will put strong constraints on the BSE model \citep[e.g.,][]{Dominik2012, Dominik2013, Dominik2015, Chruslinska2018, Kruckow2018, Mapelli2018rate, Santoliquido2020}. In this paper, we will test our BSE models by using the local merger rate inferred from LIGO/Virgo observations, constrain the model parameters for the CE evolution and natal kicks, and envisage further possible constraints by future BNS GW observations. 

Besides the formation mechanism and merger rate of BNSs, it is also significant to understand the properties of their host galaxies. On the one hand, a thorough understanding of the host galaxy properties of BNS mergers will be helpful for searching their EM counterparts by picking out the most probable host candidates, since GW BNS events can be localized at the best to within a few square degrees in future \citep{Abbott et al. 2018b}. Previous studies show that the stellar mass of BNS host galaxy is an excellent tracer and most BNS mergers locate in massive galaxies (with stellar mass $> 10^{9}M_\odot$) \citep[e.g.,][]{Mapelli2018host1, Artale2019a, Artale2019b, Toffano2019}. In addition, there may be a positive correlation between the star formation rate (SFR) and the merger rate per galaxy. On the other hand, the observationally determined properties of BNS host galaxies may put strong constraints on the formation channels and physical processed involved in the formation of BNSs, and may be used to calibrate the population synthesis models. For instance, the early-type host galaxy of GW170817 indicates a long delay time of the BNS merger from its formation \citep[][]{Blanchard2017, Hjorth2017, Im2017, Levan2017}. In this paper, we will statistically study the properties of BNS host galaxies in more details by implemented our BSE model results into several cosmological galaxy formation and evolution models/simulations, including the Millennium-II model \citep[hereafter Millennium-II;][]{Guo2011}, Illustris-TNG simulation \citep[hereafter Illustris-TNG;][]{Pillepich2018}, and EAGLE simulation \citep[hereafter EAGLE;][]{Schaye2015}.

In this paper, we systematically investigate the properties of BNS systems in a large parameter space, and study their host galaxies by adopting several cosmological galaxy formation and evolution models/simulation. Besides a few crucial parameters in binary stellar evolution, such as the metallicity, initial mass functions, and natal kicks, we also employ two phenomenological recipes to study the binary CE evolution by considering either the angular momentum evolution for CE ejection or the energy conservation mechanism. Utilizing these models, we calculate the BNSs formed in each galaxy from high redshift to local Universe, follow their evolution due to GW radiation, and obtain the BNS mergers in each galaxy at each given redshift and thus the statistical distributions of the host galaxy properties of BNS mergers. The paper is organized as follows. In Section~\ref{sec:method}, we describe the population synthesis models and introduce the several cosmological models/simulations for galaxy formation and evolution that we adopted. In Section~\ref{sec:results}, we show the acquisition of the simulated BNS samples, the comparison between the observed Galactic BNS systems and mock BNSs in Milky-Way-like galaxies at redshift $z\sim 0$, the BNS merger rates in the local Universe and its evolution with redshift, and the BNS host galaxy properties and its cosmic evolution. Discussions and conclusions are given in Sections~\ref{sec:discussion} and \ref{sec:conclusion}, respectively.

\section{Methods}
\label{sec:method}

To estimate the BNS merger rate evolution and properties of BNS (merger) host galaxies, it is necessary to convolve the BSE and population synthesis model for the formation of BNSs with a cosmological galaxy formation and evolution model. In this way, the BSE and population synthesis model provides the correspondence between star formation and BNS formation, while the cosmological galaxy formation model provides information of (binary) star formation in each galaxy at any given time and the hierarchical assembly history of each galaxy. Combining them together, one can obtain detailed information on how BNSs and BNS mergers distributed in galaxies at any given cosmic time.

\subsection{Cosmological galaxy formation and evolution}
\label{subsec:galform}

Currently, there are a number of cosmological hydrodynamical simulations for galaxy formation and evolution \citep{Vogelsberger2020}, including the most popular ones, EAGLE simulation \citep{Schaye2015}, Illustris simulation and its updated version Illustris-TNG \citep{Pillepich2018}. EAGLE was run with {\bf GADGET-3} code \citep{Springel2005}, and Illustris-TNG was run with the moving-mesh code {\bf Arepo} \citep{Springel2010}. Besides the cosmological hydrodynamical simulations, the updated Millennium-II galaxy formation and evolution model/simulation \citep{Guo2011} also provides mock galaxies formed throughout the universe by implementing semianalytical models into cosmological N-body simulations. Note that we adopt these three different galaxy formation and evolution models to show whether the estimates of BNS merger rate and properties of host galaxy are robust and how it may be dependent on different models. The (public available) databases used in this paper include the EAGLE {\bf RefL0100N1504} (http://icc.dur.ac.uk/Eagle/), mini-Millennium-II (https://wwwmpa.mpa-garching.mpg.de/ galform/millennium-II/index.html), and Illustris-TNG {\bf TNG100-1} (https://www.tng-project.org/), respectively.

In these simulations/models, the sub-grid physics (including star formation, stellar evolution and chemical enrichment, cooling and heating, stellar feedback, supermassive black hole growth and AGN feedback, etc.) are considered, though with differences in detailed treatments, and the formation and evolution of galaxies with high resolution are obtained across the cosmic time. Therefore, one can have not only a complete catalog of mock galaxies but also the detailed assembly and star formation history of each mock galaxy from these simulations. 
For EAGLE, there are only $29$ snapshots and time resolution between the available snapshots from high to low redshift increases from $\sim0.1\rm Gyr$ to $\sim1\rm Gyr$. For Millennium-II and Illustris-TNG, there are $68$ and $100$ snapshots, respectively, and the time resolution from high to low redshift increases from $\sim0.02\rm Gyr$ to $\sim0.3\rm Gyr$, from $\sim0.05\rm Gyr$ to $\sim0.2\rm Gyr$, respectively. These time resolutions are sufficient for the study of BNS merger rate evolution.
The statistical properties of mock galaxies resulting from these three simulations (including stellar mass function, mass-metallicity relation, etc.) are distinguishable from each other because of the differences in treatments of sub-grid physics. Note here, we neglect the slight differences in the choice of the cosmological parameters by Millennium-II, Illustris-TNG, and EAGLE, and adopt the $\Lambda$CDM cosmology with $h_{0}=0.677$, $\Omega_{\rm M}=0.309$, and $\Omega_{\Lambda}=0.691$ \citep{Planck2016} throughout the paper.

\begin{figure*}
\begin{center}
\includegraphics[width=16cm]{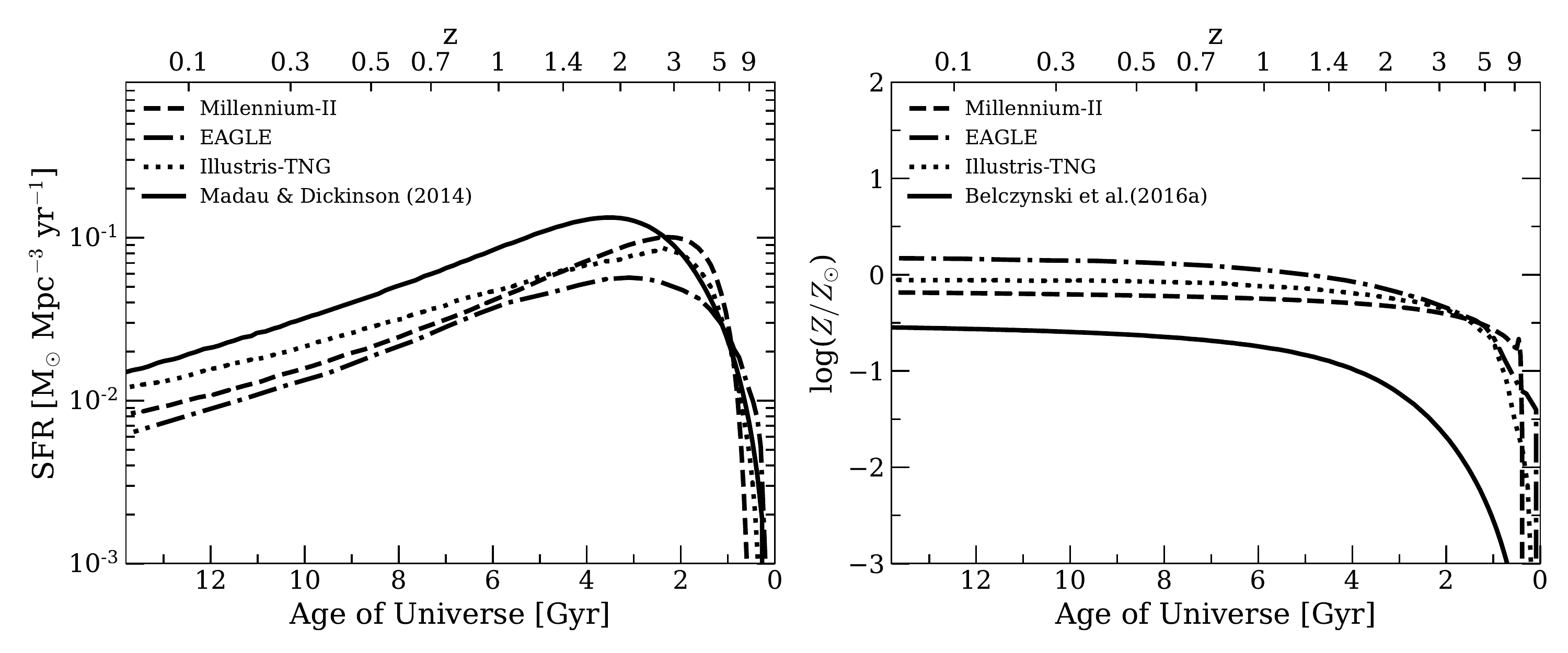}
\caption{Global SFR density evolution (left panel) and metallicity evolution (right panel) from different cosmological galaxy formation and evolution models/simulations and observations. Dotted, dot-dashed, and dashed lines represent the SFR history and metallicity evolution obtained from Illustris-TNG, EAGLE, and Millennium-II, respectively. Solid lines represent the extinction-corrected observational SFR evolution given by \citet[][left panel]{Madau2014} and mean metallicity evolution in \citet[][right panel]{Belczynski2016a}, respectively.
}
\label{fig:f1}
\end{center}
\end{figure*}

Figure\,\ref{fig:f1} shows the global SFR and metallicity evolution obtained from different cosmological galaxy formation models/simulations. For comparison, we also show the extinction-corrected observational specific SFR obtained in \citet{Madau2014} and the mean metallicity evolution given by \citet{Belczynski2016a}, respectively, in this Figure. Apparently, the total observational determined star formation over the full range of redshift is higher than those from cosmological simulations/models (i.e., $\sim 1.59$, $2.23$, and $1.55$ times of that from Millennium-II,  EAGLE, and Illustris-TNG, respectively). Among these simulations/models, Illustris-TNG results in the highest SFR density at low redshift, while at higher redshift ($z\gtrsim1.5$), Millennium-II leads to relatively higher star forming activities. It is notable that the evolution of SFR resulting from Illustris-TNG is flatter than those from the other two. The evolution curve of SFR given by \citet{Madau2014} peaks at a lower redshift while those from simulations peak at higher redshift.   However, one should also note that some recent observations do suggest that the cosmic SFR at redshift beyond the SFR peak may be higher than that given by \citet{Madau2014} \citep[e.g.,][]{Gruppioni2020}. 
These differences in SFR will lead to distinctions in the estimated local BNS merger rates and the shapes of the merger rate evolution curves. 

The metallicity evolutions from three cosmological galaxy formation and evolution simulations/models are comparable with each other, but these simulations/models generally result in higher metallicity on average than that in \citet{Belczynski2016a} at any redshift, which may also have some effect on the BNS merger rate and its evolution. Alternatively, one may also adopt the cosmic chemical evolution of galaxies at those in \citet{Chruslinska2019b}, \citet{Maiolino2019}, and \citet{Boco2021}, etc.
We note here that because of different subhalo databases, the metallicity evolution shown in Figure~\ref{fig:f1} obtained from Illustris and EAGLE represents the mean metallicity of star-forming gas particles, while that obtained from Millennium-II represents the mean metallicity of star particles. From Millennium-II, we can only extract both the mean metallicities of star particles. We find that the mean metallicity of star-forming gas particles is slightly lower ($\sim0.25$ dex) than that of star particles according to Illustris-TNG. Therefore, the adoption of mean metallicity of star particles as a surrogate of star-forming gas particle for Millennium-II may only have small effect on our results.

The probability (or rate) for BNS formation at a given time in each mock galaxy can be obtained by combining the resulting BNSs from the BSE and population synthesis models as templates and the star formation and metallicity evolution history of this galaxy. For a given galaxy at a given snapshot, we estimate the total new-born stars by multiplying the SFR with the snapshot time interval. Then we can calculate the number of new-born BNS systems within the total new-born stars according to the specific BSE model at a given metallicity via appropriate mass scaling. According to this probability, we randomly assign newly formed BNSs to individual galaxies across the cosmic time. After the formation of a BNS, it evolves then due to GW radiation. Thus we can get mock catalogs for BNS mergers and correspondingly GW events as a function of the cosmic time $t$. Each BNS merger event is characterized by the masses of its two components, i.e., $m_{\rm NS,1}$ and $m_{\rm NS,2}$, BNS formation time $t(z)$\footnote{The difference between BNS formation time and BNS progenitor formation time is about $10$\,Myr, and simply ignore this difference.}, BNS merger time $t(z)+t_{\rm d}$, properties (e.g., stellar mass, metallicity, morphology) of its host and progenitor galaxies.

We compute the spatial density of BNS merger rate and investigate the properties of their host galaxies by means of combining catalogues of BNSs from population-synthesis simulations with the outputs of the cosmological galaxy formation and evolution models/simulations.

Note that the typical mass of the "gas/star particles" used in the cosmological simulations is few times $10^{6}M_{\odot}$, and there is a limitation for the study of dwarf host galaxies with mass smaller than a few $10^{6}M_{\odot}$ due the mass resolution of the cosmological simulations. We find that the dwarf galaxies with mass smaller than a few $10^{6}M_{\odot}$ make little contribution (less than $1\%-3\%$) to the BNS merger rate if extrapolating the distribution function from $10^{7}M_{\odot}$ to smaller dwarf galaxies, as the number of BNS mergers are positively correlated with galaxy stellar mass.

\subsection{BSE and Population synthesis models}
\label{subsec:popmod}

\begin{table}
\begin{center}
\caption{Parameter settings for binary stellar evolution models}
\label{tab:t1}
\setlength{\tabcolsep}{3.2mm}{
\begin{tabular}{c|c|c|c} \hline
Model &  $\alpha\lambda$ or $\gamma$ & $\sigma_{\rm k}\ [\rm km\ \rm s^{-1}]$ \\
\hline
$\alpha0.1\textrm{kl}$  &  $\alpha\lambda=0.05$  & 30  \\
$\alpha0.1\textrm{kh}$ &  $\alpha\lambda=0.05$  & 190  \\
$\alpha0.1\textrm{kb}$ &  $\alpha\lambda=0.05$  & 190/30 \\
$\alpha1.0\textrm{kl}$  &  $\alpha\lambda=0.5$  & 30  \\
$\alpha1.0\textrm{kh}$ &  $\alpha\lambda=0.5$  & 190 \\
$\alpha1.0\textrm{kb}$ &  $\alpha\lambda=0.5$  & 190/30\\
$\alpha10.\textrm{kl}$  &  $\alpha\lambda=5.0$  & 30 \\
$\alpha10.\textrm{kh}$ &  $\alpha\lambda=5.0$  & 190\\
$\alpha10.\textrm{kb}$ &  $\alpha\lambda=5.0$  & 190/30\\
$\gamma1.1\textrm{kl}$   & $\gamma=1.1$  & 30\\
$\gamma1.1\textrm{kh}$  & $\gamma=1.1$  & 190\\
$\gamma1.1\textrm{kb}$  & $\gamma=1.1$  & 190/30\\
$\gamma1.3\textrm{kl}$   & $\gamma=1.3$  & 30\\
$\gamma1.3\textrm{kh}$  & $\gamma=1.3$  & 190\\
$\gamma1.3\textrm{kb}$  & $\gamma=1.3$  & 190/30\\
$\gamma1.5\textrm{kl}$   & $\gamma=1.5$  & 30\\
$\gamma1.5\textrm{kh}$  & $\gamma=1.5$  & 190\\
$\gamma1.5\textrm{kb}$  & $\gamma=1.5$  & 190/30\\
$\gamma1.7\textrm{kl}$   & $\gamma=1.7$  & 30\\
$\gamma1.7\textrm{kh}$  & $\gamma=1.7$  & 190\\
$\gamma1.7\textrm{kb}$  & $\gamma=1.7$  & 190/30\\
\hline
\end{tabular}}
\end{center}
\begin{flushleft}
\footnotesize{Column 1: model name; column 2: values of $\alpha \lambda$ or $\gamma$ in the CE evolution phase; column 3: the root-mean-square (rms) of the assumed Maxwellian distribution for the natal kick velocities. For each model listed in this Table, we perform calculations for $3$-set of sub-models by considering mass ejection with $\beta=0.9$, $0.8$, and $0.6$, respectively. The definition of $\beta$ is given by Equation~(\ref{eq:beta}) in Section~\ref{subsec:popmod}, with a small $\beta$ indicating a larger fraction of mass ejected by the SN explosion.}
\end{flushleft}
\end{table}

\begin{table*}
\begin{center}
\caption{Properties of the detected Galactic BNSs.}
\label{tab:t2}
\setlength{\tabcolsep}{3.0mm}{
\begin{tabular}{c|c|c|c|c|c|c|c|l} \hline
\multirow{2}{*}{System} &  $P_{\rm orb}$  & \multirow{2}{*}{$e$} & $M_{\rm psr}$ & $M_{\rm comp}$ & $M_{\rm total}$ & \multirow{2}{*}{mass ratio} & $\tau_{\rm GW}$ & \multirow{2}{*}{Reference} \\
                        &   [day]        &                    & [$M_\odot$] & [$M_\odot$]  & [$M_\odot$] & &  [$10^{10}$\,yr] & \\
\hline
\multicolumn{9}{c}{Milky Way field}\\ \hline
J0737$-$3039          &  0.102  & 0.088 & 1.338 & 1.249 & 2.587 & 0.93 & 0.0086 & \citet{Tauris2017} \\
B1534+12              &  0.421  & 0.274 & 1.333 & 1.346 & 2.678 & 0.99 & 0.27 & \citet{Tauris2017} \\
J1756$-$2251          &  0.320  & 0.181 & 1.341 & 1.230 & 2.570 & 0.92 & 0.17 & \citet{Tauris2017} \\
J1906+0746$^{\star}$  &  0.166  & 0.085 & 1.291 & 1.322 & 2.613 & 0.98 & 0.031 & \citet{Tauris2017} \\
J1913+1102            &  0.206  & 0.090 & 1.62  & 1.27  & 2.88  & 0.78 & 0.047 & \citet{Ferdman2020} \\
J1946+2052            &  0.078  & 0.064 & <1.31 & >1.18 & 2.50  & - & $0.0047-0.0049^{\vartriangle}$ &  \citet{Stovall2018}\\
J0453+1559            &  4.072  & 0.113 & 1.559 & 1.174 & 2.734 & 0.75 & 150 & \citet{Tauris2017} \\
J1411+2551            &  2.616  & 0.170 & <1.62 & >0.92 & 2.538 & - & 49-51$^{\vartriangle}$ & \citet{Martinez2017} \\
J1518+4904            &  8.634  & 0.249 & 1.41  & 1.31  & 2.718 & 0.93 & 920 & \citet{Tauris2017} \\
J1753$-$2240          & 13.638  & 0.304 & - & -  & - &  -  & - & \citet{Tauris2017} \\
J1755$-$2550$^{\star}$& 9.696   & 0.089 & - & >0.40 & - &  -  & - & \citet{Tauris2017} \\
J1811$-$1736          & 18.779  & 0.828 & <1.64 & >0.93 & 2.57 & - & 181-187$^{\vartriangle}$ & \citet{Tauris2017} \\
J1829+2456            & 1.176   & 0.139 & <1.38 & >1.22 & 2.59 & - & 5.9-6.1$^{\vartriangle}$ & \citet{Tauris2017}\\
J1930$-$1852          & 45.060  & 0.399 & <1.32 & >1.30 & 2.59 & - & 55400-57200$^{\vartriangle}$ & \citet{Tauris2017}\\
J0509+3801            & 0.380   & 0.586 & 1.34  & 1.46  & 2.805 & 0.92 & 0.058 &  \citet{Lynch2018} \\
J1757$-$1854          & 0.184   & 0.606 & 1.338 & 1.395 & 2.733 & 0.96 & 0.0076 & \citet{Cameron2018} \\
B1913+16              & 0.323   & 0.617 & 1.440 & 1.389 & 2.828 & 0.97 & 0.030 & \citet{Tauris2017} \\ \hline
\multicolumn{9}{c}{Milky Way Globular Cluster}    \\ \hline
B2127+11C             & 0.335  & 0.681 & 1.358 & 1.354 & 2.713 & 0.997 & 0.022 & \citet{Tauris2017} \\
J1807$-$2500$^{\star}$& 9.957  & 0.747 & 1.366 & 1.206 & 2.572 & 0.88 & 110 & \citet{Tauris2017} \\
J0514$-$4002$^{\star}$& 18.8   & 0.89  & 1.25  & 1.22  & 2.473 & 0.98 & 47 & \citet{Ridolfi2019} \\ \hline
\end{tabular}}
\end{center}
\begin{flushleft}
\footnotesize{Column 1: BNS systems; column 2: orbital period in unit of day; column 3: eccentricity; column 4: pulsar mass in unit of solar mass $M_\odot$; column 5: companion mass in unit of $M_\odot$; column 6: total mass in unit of $M_\odot$; column 7: mass ratio; column 8: merger time in unit of the Hubble time $10^{10}$\,yr. Notes: Both J0737-3039 components are radio pulsars. The two systems with a mark $^{\star}$ need further confirmation, which could be alternatively neutron star + white dwarf binaries. The BNS systems with mark $^{\vartriangle}$ only have poor mass constraints, thus their merger time are calculated by assuming mass ratio in the range from $0.7-1$.}
\end{flushleft}
\end{table*}

In this section, we briefly review the initial, boundary conditions, and important physical processes in binary stellar evolution and for population synthesis to obtain the BNS samples. Population synthesis, including individual stellar-evolution tracks, was carried out using a modified version of {\bf BSE} code described by \citet{Yu2010, Yu2011, Yu2015} and \citet{Hurley2000, Hurley2002} in an initial study of the present Galactic double degenerates and neutron star populations. In this paper, we vary some of the main parameters to systematically study the birth- and merge-rates of binary neutron star (BNS) population in a galaxy, which are summarized as follows.

We set the initial mass function (IMF) for the primary components of initial binaries as the frequently used power-law form 
\begin{equation}
\mathfrak{F}(m_{1})\propto \left\{
\begin{array}{cl}
    m_{1}^{-0.3},& \ 0.01\leq m_{1}/M_\odot<0.08,  \\
    m_{1}^{-1.3},& \ 0.08\leq m_{1}/M_\odot<0.50,  \\
    m_{1}^{-2.3},& \ 0.50\leq m_{1}/M_\odot<1,  \\
    m_{1}^{-\alpha_{3}},& \ 1\leq m_{1}/M_\odot<100. 
\end{array}
\right.
\end{equation}
with $\alpha_{3}=2.5$. This $\alpha_3$ value is set according to \citet{Kroupa2002}, in which the slope at high mass part is estimated to be $\alpha_3=_{+2.3\pm0.3}^{+2.7\pm0.3}$ for $m_{1} \geq 1M_\odot$. 

We assume a uniform distribution for mass ratio $q=m_2/m_1$ ($m_2\leq m_1$) \citep[see also][]{Eggleton1989, Mazeh1992, Goldberg1994}, i.e., 
\begin{equation}
\mathfrak{F}(q)= \rm constant \ \ {\rm for} \  q\in (0,1],
\end{equation}	
and we set $m_2 >3M_\odot$.

We assume the semi-major axis ($a$) distribution of initial  binaries is constant in logarithm for wide binaries and falls off smoothly at small $a$ as that in \citet{Han1998} and \citet{Han2003}
\begin{equation}
\frac{{\rm d}a}{{\rm d}n}=\left\{
\begin{array}{c}
\alpha_a\frac{a}{a_{0}})^{k},a\leqslant
a_{0},\\
\alpha_a,a_{0}<a<a_{1}.
\end{array}
\right.
\label{eq_a}
\end{equation}	
where $\alpha_q\simeq0.060$, $a_{0}=1.15R_\odot$, $a_{1}=1.86\times 10^{7}R_\odot$ and $k=0.94$. 

The grid of metallicity is set to be $Z=10^{-2.3} Z_\odot$, $10^{-1.8} Z_\odot$, $10^{-1.3} Z_\odot$, $10^{-0.8} Z_\odot$, $10^{-0.4} Z_\odot$, and $1 Z_\odot$, to cope with metallicity bins of zero-age main sequence (ZAMS) binaries $Z/Z_\odot \in (0,10^{-2.0}]$, ($10^{-2.0}, 10^{-1.5}$), [$10^{-1.5}, 10^{-1.0}$), [$10^{-1.0}, 10^{-0.6}$), [$10^{-0.6}, 10^{-0.2}$), and $[10^{-0.2}, 5)$, separately. Here, we take the solar metallicity as $Z_\odot=0.02$.

\begin{itemize}
\item {\bf Common Envelope:} for the formation of BNSs (or binary black holes), one of the most uncertain physical processes involved in is the common envelope (CE) evolution. In this paper, we adopt two different formalisms to deal with the CE ejection process. The first one is denoted as the $\alpha$-formalism, in which the CE ejection of a binary star requires that the envelope binding energy, including gravitational-binding and recombination energies, must represent a significant fraction of the orbital energy \citep{Webbink1984} as described by
\begin{equation}
\label{eq:alpha-f}
\frac{G(m_1-m_{1\rm c})m_1}{\lambda r_{\rm L}} = \alpha_{\rm CE} \left( \frac{Gm_{1\rm c}m_2}{2a_{\rm f}} - \frac{Gm_1 m_2}{2a_{\rm i}}\right),
\end{equation}
where the CE ejection efficiency $\alpha_{\rm CE}$ represents the fraction of orbital energy that is used to eject CE, $r_{\rm L}$ is the Roche lobe radius, $\lambda$ is a structure parameter depending on the evolutionary state of the donor, $a_{\rm i}$ and $a_{\rm f}$ are the semimajor axis of the binary before and after the CE stage, $G$ is the gravitational constant. In this paper, we consider $\alpha_{\rm CE}=0.1$, $1.0$ and $10.0$, $\lambda=0.5$ to investigate cases with high and low CE ejection efficiency respectively. 
Note that here $\lambda$ is taken to be a constant rather than the most popular one, $\lambda_{\rm Nanjing}$, from \citet{Xu2010}, which depends on the evolutionary stage of the donor. For massive stars, the envelope ejection efficiency decreases considerably and $\lambda$ changes to a small value, i.e., $\lambda<0.2$ \citep{Dewi2000, Dewi2001, Podsiadlowski2003, Xu2010, Dominik2012}. We here only consider the evolution of BNSs and hence take a slightly higher value of $\lambda$, i.e., $\lambda=0.5$, because the binding energy of lower mass stars would be smaller.

The second one is denoted as the $\gamma-$formalism \citep[e.g.,][]{Nelemans2005}, in which the angular momentum lost by a binary system undergoing non-conservative mass transfer is assumed to be parameterized by the decrease of primary mass times a factor $\gamma$, i.e.,
\begin{equation}
\frac{J_{\rm i}-J_{\rm f}}{J_{\rm i}} = \gamma \frac{m_1-m_{1\rm c}}{m_1+m_2},
\end{equation}
where $J_{\rm i}$ and $J_{\rm f}$ are the angular momentum of the binary before and after the CE stage, $m_1$, $m_{1\rm c}$, and $m_2$ are the masses of binary component in consideration before and after the CE stage, and the masses of the companion, respectively. We consider $\gamma=1.1$, $1.3$, $1.5$, and $1.7$, respectively, to investigate the influence of angular momentum loss on the rates.

In most previous works \citep[e.g.,][]{Dominik2012, Chruslinska2018, Giacobbo2018, Vigna2018}, the $\alpha$-formalism is adopted to cope with the CE evolution. Especially, \citet{Mapelli2018rate} found that only if a large $\alpha$ and low natal kicks are assumed, the estimated local merger rate can match that inferred from the LIGO/Virgo O1/O2 observations well. Different from previous studies, we find that the gamma prescription for CE can also well reproduce the Galactic BNSs and provide a suitable local merger rate, which means the angular momentum evolution of a binary may play an important role in the CE evolution phase, and makes the gamma prescription an alternative way for the CE ejection models.

\item {\bf Kick velocity:} During the formation process of a neutron star, it may receive a natal kick. The kick velocity can be hundreds of km/s and such a kick may cause the break-up or significant orbital re-configuring of a binary system. In the past two decades, many studies have been done to investigate the magnitude and distribution of this kick velocity both theoretically and observationally, \citep[e.g.,][]{Hansen1997, Fryer1998, Arzoumanian2002, Hobbs2005, Wang2006, 2006ApJ...643..332F, Wongwathanarat2010, Beniamini2016, Verbunt2017}, though it is still not well understood, yet. In this paper, we simply assume that the kick velocities have a Maxwellian distribution following the best estimate of \citet{Hansen1997} as
\begin{equation}
\label{eq:kickv}
\frac{{d}N}{N{d}v_{\rm k}}=\left( \frac{2}{\pi} \right)^{1/2}\frac{v_{\rm k}^{2}}{\sigma_{\rm k}^{3}}
\exp^{-v_{\rm k}^{2}/2\sigma_{\rm k}^{2}},
\end{equation}
where $v_{\rm k}$ is the kick velocity and $\sigma_{\rm k}$ is its dispersion, ${d}N/N$ is the normalized number in a kick velocity bin ${d}v_{\rm k}$. We take standard value of $\sigma_{\rm k}=190\,\kms$, so that the probable kick velocity $v_{\rm kp}\approx 268\,\kms$.
Several works suggest a bimodal velocity distribution composed of a first peak at low velocities and a second peak at high velocities \citep[e.g.,][]{Fryer1998, Arzoumanian2002, Verbunt2017}. It has been pointed out that the low natal kicks can well explain the observed Galactic BNS systems \citep{AndrewsMandel2019} and the local BNS merger rate \citep{Mapelli2018rate, Baibhav2019}.
We also alternatively assume different models for the kick velocity distribution, i.e., it follows Equation~(\ref{eq:kickv}) but with $\sigma_{\rm k}=30\,\kms$. In addition, we take the bimodal kick velocity distribution into consideration by assuming that neutron stars randomly receive a high kick or a low kick at their birth following the Maxwellian distribution (Eq.~\ref{eq:kickv}) with $\sigma_{\rm k}=190\,\kms$ and $\sigma_{\rm k}=30\,\kms$, respectively.

As for other parameters in our population synthesis, a Monte Carlo procedure is adopted to generate the individual kick velocities for neutron stars. Other parameters associated with the kick velocity (e.g., the directions) are assumed to follow a uniform distribution. Our method to model the kick is the same as that in \citet{Brandt1995}.

\item {\bf Formation route for BNSs:}
We follow the evolution of each main sequence binary in our simulations, and consider the formation of neutron stars in the following three cases: i) if a star has core mass of $m_{\rm c}\lesssim 2.25 M_\odot$ at shell helium ignition, it evolves through double-shell thermal-pulses up to the asymptotic giant branch. The star may become a neutron star if its core mass grows and eventually exceeds $2.25 M_\odot$; ii) if the core mass of the star on the thermal-pulsing asymptotic giant branch does not exceed $2.25 M_\odot$ but it is heavy enough ($m_{\rm c}\gtrsim1.6 M_\odot$) to become an electron-degenerate oxygen-neon white dwarf which may become a neutron star via accretion-induced collapse; iii) if a star has a core mass of $m_{\rm c}\gtrsim 2.25 M_\odot$ at the start of the early asymptotic giant (or red giant) branch, it will  become a neutron star without ascending the thermal-pulsing asymptotic giant branch. If the core mass of a star at the time of supernova explosion is sufficiently high ($\gtrsim7 M_\odot$), it will most likely become a black hole unless significant mass loss takes place. 
When a neutron star is formed in one of the cases given above, we assume that its gravitational mass is given by 
\begin{equation}
m_{\rm NS}=0.89+0.23m_{\rm c,SN}
\end{equation}
where $m_{\rm c,SN}$ is the mass of the CO-core at the time of supernova explosion, corresponding to the minimum and maximum possible mass of a neutron star as $\sim1.26M_\odot$ and $\sim2.5M_\odot$ \citep[stiff equation of state, e.g., MS1 or MS1b from][]{Muller1996}, respectively.

For the formation of BNS, possible evolution stages may include different mass transfer processes, such as CE, stable RLOF, and exposed core phase. We therefore diagnose the evolution channels of BNS according to whether they experience these evolution phases or not \citep[see][for details]{Yu2015}. The evolution channels diagnosed in our simulations are listed below. 

\begin{enumerate}
\item CE ejection + CE ejection;
\item  Stable RLOF + CE ejection;
\item  CE ejection + stable RLOF;
\item  Stable RLOF + stable RLOF;
\item  Exposed core + CE ejection;
\item  Solo CE ejection;
\item  Solo RLOF.
\end{enumerate} 

In the above items, RLOF means Roche lobe overflow. 

\item {\bf Mass ejection:}
Right before the secondary SN, the binary, comprised of a NS with a naked helium star companion, may undergo a phase of mass transfer, the so-called Case BB mass transfer phase \citep[][and references therein]{Tauris2015}, which is currently not well understood but significantly influence the properties of resulting BNSs. Here, we follow the method in \citet{AndrewsZezas2019}, focused on the second SNe forming the BNS, and set
\begin{equation}
\beta=\frac{m_{\rm NS,1}+m_{\rm NS,2}}{m_{\rm NS,1}+m_{\rm He}},
\label{eq:beta}
\end{equation}
the ratio of the post-SN to pre-SN total system masses, as a free parameter to systematically investigate the effect of mass ejection on BNS formation. From the modified BSE code, we obtain the masses of NSs ($m_{\rm NS,1}$, $m_{\rm NS,2}$) and the masses of pre-SN system ($m_{\rm NS,1}$, $m_{\rm He,0}$). For different values of $\beta$, we calculate $m_{\rm He}$ and compare it with $m_{\rm He,0}$. If $m_{\rm He}<m_{\rm He,0}$, the companion is stripped from $m_{\rm He,0}$ to $m_{\rm He}$ and we recalculate the secondary SN process. Otherwise, the results remain unchanged. To complement the population synthesis methods, we use this recipe, similar to that in \citet{AndrewsZezas2019}, to calculate three sub-models with $\beta=0.9$, $0.8$, and $0.6$, corresponding to heavily-stripped, stripped, and slightly-stripped assumptions, respectively. 

In our simulations, we calculate $10^{7}$ binary systems with a total mass of $\sim 1.8\times 10^8 M_\odot$ for each case. 
We summarize all of our models in Table~\ref{tab:t1}.

\end{itemize}

\subsection{Observed Galactic BNSs}
\label{subsec:observe_BNS}

With the effort in pulsar searches and observations in the past several decades, $20$ BNSs have been found in the Milky Way (MW). Among them, $17$ BNSs are in the MW field and the rest $3$ are in MW globular clusters \citep[see][]{Tauris2017, Stovall2018, Martinez2017, Lynch2018, Cameron2018, Ridolfi2019}. Table~\ref{tab:t2} lists the properties of these Galactic BNSs, including their orbital period ($P_{\rm orb}$), eccentricity ($e$), primary pulsar and its companion masses $M_{\rm psr}$ and $M_{\rm comp}$, respectively.

In the present paper, we assume that all the observed $17$ BNSs in the MW field were formed through the BSE channel. Other $3$ BNSs found in globular clusters (Table~\ref{tab:t2}) are suggested to be formed via the dynamical evolution channel, i.e., the second exchange encounters \citep{Prince1991, Lynch2012, Verbunt2014, Tauris2017, Ridolfi2019}, which are excluded in the matching of our model results with the observed MW BNSs and thus constraining the BSE models for BNSs in Section~\ref{subsec:GalBNSs}. We also note that the short-period, high eccentricity Galactic BNSs (J0509$+$3801, J1757$-$1854, and B1913$+$16) may be better explained by the dynamical origin, however, the BSE channel for the origin of these three BNSs is not ruled out, yet \citep{AndrewsMandel2019}. We will further discuss the effects on our results if assuming that these three short-period, high eccentricity BNSs were formed via the dynamical channel in Appendix~\ref{sec:appA}.

\section{Results}
\label{sec:results}

\subsection{The properties of simulated BNSs}
\label{subsec:results_1}

\begin{figure*}
\begin{center}
\includegraphics[width=16cm]{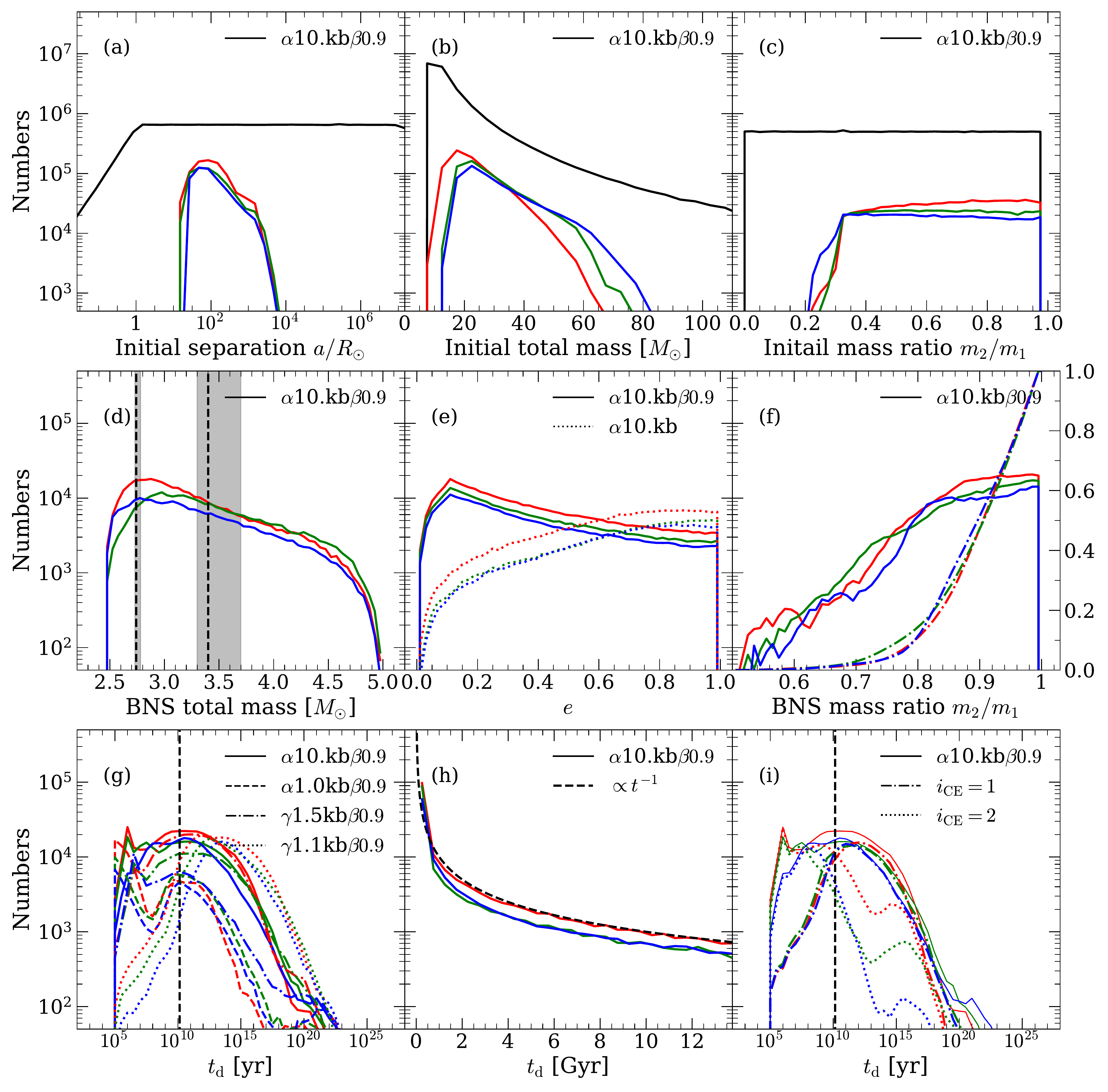}
\caption{Initial condition and acquisition of population synthesis models, mostly the results of the model $\alpha10.{\rm kb}\beta0.9$. In each panel, the red, green, and blue lines represent the cases with progenitors metallicity of  $Z<10^{-1.5}Z_\odot$, $10^{-1.5}Z_\odot \leq Z \leq 10^{-0.5}Z_\odot$, and $Z>10^{-0.5}Z_\odot$, respectively. Panels (a)-(c) are for the initial semimajor axis, total mass, and mass ratio distribution of the binaries, with the black solid lines indicating the distributions of all binary progenitors. Panel (d) shows the total mass distribution of BNSs resulting from the model $\alpha10.{\rm kl}$. In this panel, the black dashed lines and grey regions indicating the total mass ranges of GW170817 \citep[$2.74^{+0.04}_{-0.01}M_\odot$;][]{Abbott et al. 2017c} and GW190425 \citep[$3.4^{+0.3}_{-0.1}M_\odot$;][]{GW190425}. Panel (e) shows the eccentricity distribution of BNSs resulting from the model $\alpha10.{\rm kb}\beta0.9$ (solid lines) and $\alpha10.{\rm kb}$ (dotted lines), respectively. Panel (f) shows the mass ratio distribution of BNSs resulting from the model $\alpha10.{\rm kb\beta0.9}$, where the dash-dot lines are for the cumulative distributions of mass ratio. Panel (g) shows the delay time distributions of BNSs from the model $\alpha10.{\rm kb\beta0.9}$ model (solid lines), $\alpha1.0{\rm kb\beta0.9}$ (dashed lines), $\gamma1.5{\rm kb\beta0.9}$ (dash-dot lines), and $\gamma1.1{\rm kb\beta0.9}$ (dotted lines), respectively, where the black dashed line indicating the Hubble time. Panel (h) shows the delay time distributions of BNSs merging within a Hubble time from the model $\alpha10.{\rm kb\beta0.9}$, where the black solid line indicating $p(t_{\rm d})\propto t_{\rm d}^{-1}$. Panel (i) shows the delay time distributions of BNSs from the model $\alpha10.\rm kb\beta0.9$ model via different formation channels, where the dash-dot lines and dotted lines represent the count number of BNSs formed via the channels with one CE ejection $i_{\rm CE}=1$ and double CE ejection $i_{\rm CE}=2$, respectively, and the black dashed line indicates the Hubble time. 
}
\label{fig:f2}
\end{center}
\end{figure*}

A number of recent works have already investigated the influence of crucial parameters of $\alpha$-formalism in binary stellar evolution on the distributions of the resulting BNS parameters \citep{Dominik2012,Giacobbo2018,Vigna2018}. Here, we take the model $\alpha10.\rm kb\beta0.9$ as an example to show the distributions of initial semimajor axis (panel (a) of Fig.~\ref{fig:f2}), total mass (panel (b)), mass ratio (panel (c)) of the main sequence-main sequence binaries and the BNS progenitors, respectively, the distributions of BNS total mass (panel (d)), eccentricity (panel (e)), and mass ratio (panel (f)), respectively, and the delay time distributions of BNS mergers (panels (g)-(i)).
As seen from Figure~\ref{fig:f2}, only binaries that have appropriate semimajor axes, around $10\sim10000\ R_\odot$ can form BNSs efficiently, especially when $a\sim100\ R_\odot$ (panel (a)). For low metallicity progenitors, $Z<10^{-1.5}Z_\odot$, the BNSs are mostly formed from binary stars with a total mass between $15\ M_\odot$ and $20\ M_\odot$; for higher metallicity progenitors, $Z\geq10^{-1.5}Z_\odot$, the BNSs are mostly formed from binaries with a higher total mass between $20\ M_\odot$ and $25\ M_\odot$ because the severe mass loss via stellar winds hinders low mass binaries to form BNSs (panel (b)). In our simulations, the efficiency of BNS formation is not sensitive to the initial mass ratio $q$ when $q\gtrsim 0.2$, but it is apparent that for too low mass ratio progenitors, the secondary star is always too light to form a neutron star (panel (c)).

In our simulations, the distributions of BNS total mass range from $\sim2.5M_\odot$ to $\sim5.0M_\odot$, and peak at $\sim2.75M_\odot$ (a mass similar that of GW170817 $2.74^{+0.04}_{-0.01}M_\odot$; \citealt{Abbott et al. 2017c}). The fraction of BNSs with total mass larger than that of GW190425 \citep[$3.4^{+0.3}_{-0.1}M_\odot$;][]{GW190425} is about $30\%$. In most instances, the primary neutron star mass is slightly higher than the secondary one and the mass ratio $\sim 1$. In addition, we show the cumulative distributions of mass ratio of BNSs (the dashed-dot lines in panel (f)). The fraction of a large mass ratio, $q>0.9$, is $55.2\%$, $54.9\%$ and $50.7\%$ for $Z<10^{-1.5}Z_\odot$, $10^{-1.5}Z_\odot \leq Z \leq 10^{-0.5}Z_\odot$, and $Z>10^{-0.5}Z_\odot$, respectively. 

Different from previous works, we also investigate the influence of the mass ejection during the secondary SNe on the distributions of BNS orbital parameters. We find that the mass ejection highly influences the eccentricities of resulting BNSs, which tend to increase with increasing ejection mass (panel (e)). To comparing with observational distributions of BNS semimajor axis and eccentricity, one needs to further consider the orbital decay of BNSs by their GW radiation directly.

In panel (g) of Figure~\ref{fig:f2}, we show the delay time distribution for BNSs resulting from the models $\alpha10.{\rm kb}\beta0.9$ and $\alpha1.0{\rm kb}\beta0.9$. We find that in the condition of a small value of $\alpha\lambda$ (e.g., $0.5$), it is inefficient for binaries to eject the CE and the semimajor axes of binaries shrink considerably during the CE phase, resulting in BNSs with small semimajor axes and short timescales of merger. By contrast, if we adopt a large value of $\alpha\lambda$ (e.g., $\alpha\lambda=5$), BNSs are formed with larger semimajor axes and prefer to merge with longer time delay.

For the model $\alpha10.{\rm kb}\beta0.9$, we find that the majority of BNSs, $\sim 53\%-56\%$, can hardly merge within a Hubble timescale. Panel (h) shows the distribution of delay time for those BNSs that can merge within a Hubble time, which can be approximate as a power-law $p(t_{\rm d})\propto t^{\Gamma}$ with $\Gamma\sim -1$, consistent with previous studies \citep{Belczynski2016a, Lamberts2016, Giacobbo2018}.

It is worthy to note that the delay time distributions for BNSs have multiple peaks (panel (g) of Fig.~\ref{fig:f2}), corresponding to different formation channels. The first peak at a small delay time is composed of BNSs which go through CE phase twice. A small part of these BNSs ($\sim7\%$) are formed from close binaries via the formation channel (i) (i.e., CE ejection $+$ CE ejection), with the first CE phase occurring at an Early Asymptotic Giant Branch (EAGB) star $+$ a MS star stage and the second CE phase occurring at a NS $+$ a Core Helium Burning (CHeB) star stage. The majority of these BNSs are formed via the formation channel (ii) (i.e., stable RLOF $+$ CE ejection), with the first CE phase occurring at a NS $+$ a CHeB star stage and the second CE phase occurring at a NS $+$ a Naked Helium Star Hertzsprung Gap (HeHG) stage in a short interval. Hence, the orbits shrink much more significantly than that with a single CE ejection formation channel and the resulting BNSs can merge on a relatively shorter timescale (panel (i)).

We also show the delay time distributions of BNSs inferring from $\gamma$-formalism (i.e., models $\gamma1.1\rm kb \beta0.9$ and $\gamma1.5\rm kb \beta0.9$) in panel (g) of Figure~\ref{fig:f2}. We find that in the simulations with a small value of $\gamma$ (e.g., $1.1$), the angular momentum loss is relatively lower than that in other simulations in the CE phase and the semimajor axes of binary systems become larger. Therefore, the merging timescales of BNSs are relatively long in these simulations. For cases with larger $\gamma$ (e.g., $1.5$), the orbital angular momentum loss becomes severer and the binary systems get closer, resulting in the decrease of orbital separation and merging timescale. 

From panel (g) of Figure~\ref{fig:f2}, we see that $\alpha$ and $\gamma$ formalism can form BNSs with similar orbits. Hence, it is not immediately clear if one could distinguish the two prescription. However, the principle to model both CE+CE channel with the gamma prescription and RLOF+CE with non-conservative stable mass transfer is based on the angular momentum loss assumptions. The equation for the CE ejection with the gamma prescription is slightly different from the equation for the stable mass transfer (Eq.(13) and (26) in \cite{Yu2010}), which may illuminate the formation channel.

By performing $84$ different simulations in total, we find that the assumptions for the CE phase can have major impacts on the semimajor axis distribution of resulting BNSs, while the mass ejection can have major impacts on the eccentricity distributions. The magnitude of kick velocity mainly influences the formation efficiency of BNSs and the distribution of eccentricity, as binary systems can easily be broken or evolve to higher elliptical orbits with a large kick velocity, which has been frequently discussed in literature \citep{Dominik2012, Chruslinska2018, Vigna2018, Giacobbo2018, Jiang2020, Santoliquido2020}.
Metallicity has some influence on the properties of BNSs population when considering the formation and evolution of their progenitors, but not significant comparing with the above several parameters.

\subsection{Mock Galactic BNSs}
\label{subsec:GalBNSs}

Milky Way is a disc-dominated galaxy with disk mass $\sim 5.2 \times 10^{10} M_\odot$ and bugle mass $\sim 0.9 \times 10^{10} M_\odot$ \citep{Licquia2015}. Since there are still some uncertainties in our understanding of the MW star formation and metallicity enrichment history, we alternatively adopt the detailed star formation and metal enrichment history of MW-like galaxies resulting from cosmological galaxy formation simulations as a surrogate to study the formation BNSs in our MW. Here, we define a galaxy resulting from a cosmological simulation (i.e.,  Millennium-II, Illustris-TNG, or EAGLE) as a MW-like galaxy if (i) its total stellar mass is in the range of $6.1 \pm 1.5 \times 10^{10} M_\odot$, and (ii) the bulge mass is in the range of $0.9 \pm 0.5 \times 10^{10} M_\odot$. Adopting Millennium-II results\footnote{Millennium-II directly gives the bulge information for each mock galaxies, which is convenient for our study on the morphological types of host galaxies of BNS mergers.}, we find $647$ galaxies satisfying the above criteria. According to their star formation and metal enrichment histories, we randomly assign newly formed BNSs in these MW-like galaxies across the cosmic time and follow their orbital evolution. We then obtain the merger rates, and the properties of survived/merging BNSs at any given cosmic time/redshift $z$ (including the present day $z=0$).

Since there is at least one pulsar component in all observed Galactic BNSs, it is better to compare the simulated pulsar-NS and pulsar-pulsar binaries rather than all BNSs with them. To do so, we need to identify pulsar-NS and pulsar-pulsar binaries output from all models. It is believed that pulsars are young neutron stars though the actual period/age for pulsars to exist remains an unsolved problem \citep[e.g.,][]{Chen1993, Arons2000, Zhang2003, Zhou2017}. Here we assume that the lifetime of a pulsar is about $10^8$\,yr and all neutron stars are pulsars in a period of $10^8$\,yr after their formation \citep{Belczynski2018b}.

We follow the method adopted in \citet{Vigna2018} to obtain the most preferable one among the models listed in Table~\ref{tab:t1} by matching the model pulsar-NS/pulsar-pulsar and all NS-NS binaries to the observed Galactic BNSs on the $e-P_{\rm orb}$ (or $e-a$) plane.
We denote the model $\alpha10.\textrm{kb}$ as the reference model and define a Bayes factor ($K_i$) below to measure the relative goodness of the model that matches the Galactic BNS observations, i.e.,
\begin{equation}
\log K_{i}=\log L_{i}-\log L_{\alpha10.\textrm{kb}},
\end{equation}
where $L_i$ is the likelihood of the model ${\rm M}_i$, which is given in Appendix~\ref{sec:appA}. We ignore the selection effect in the observed $e-P_{\rm orb}$ distribution, which may be considered in future with better understanding of the selection.

\begin{table*}
\begin{center}
\caption{Logarithmic values of the relative Bayes factor, $\log K_{i}$, obtained by matching the detected Galactic BNSs with those resulting from each model if considering all survived BNSs or those survived BNSs with at least one pulsar component.}
\label{tab:t3}
\setlength{\tabcolsep}{1.8mm}{
\begin{tabular}{l|c|c|c|c|c|c|c|c|c}
\hline
\multirow{2}{*}{Model Name} & \multicolumn{4}{c}{$\log K_i$ for survived pulsar-BNSs}  & & \multicolumn{4}{c}{$\log K_i$ for all survived BNSs} \\ \cline{2-5} \cline{7-10}
&    & $\beta=0.6$  & $\beta=0.8$  & $\beta=0.9$ & &     & $\beta=0.6$  & $\beta=0.8$  & $\beta=0.9$ \\
\hline
$\alpha0.1{\textrm{kl}}$ & -77.59 & -72.99 & -46.91 & -33.23 &  & -80.42 & -48.03 & -30.04 & -39.80 \\
$\alpha0.1{\textrm{kh}}$ & -77.83 & -82.53 & -56.23 & -75.25 &  & -60.76 & -65.22 & -65.77 & -63.60 \\
$\alpha0.1{\textrm{kb}}$ & -63.13 & -50.46 & -25.32 & -16.84 &  & -73.02 & -51.90 & -34.55 & -33.13 \\
$\alpha1.0{\textrm{kl}}$ & -58.73 & -29.39 & -14.64 & -10.92 &  & -67.42 & -30.63 & -26.51 & -31.51 \\
$\alpha1.0{\textrm{kh}}$ & -58.72 & -54.09 & -75.89 & -59.00 &  & -62.71 & -61.43 & -64.30 & -69.16 \\
$\alpha1.0{\textrm{kb}}$ & -23.48 & -11.59 & 0.17   & 5.66   &  & -41.84 & -23.31 & -27.21 & -25.82 \\
$\alpha10.{\textrm{kl}}$ & -23.70 & -22.29 & -0.90  & -7.54  &  & -14.80 & -13.28 & -19.14 & -25.60 \\
$\alpha10.{\textrm{kh}}$ & -21.36 & -11.31 & -4.94  & -7.47  &  & -27.84 & -40.31 & -27.21 & -32.41 \\
$\alpha10.{\textrm{kb}}$ & 0      & -0.50  & 8.11   & 9.74   &  & -20.20 & -22.56 & -24.33 & -8.60  \\
$\gamma1.1{\textrm{kl}}$ & -57.91 & -58.79 & -51.15 & -50.82 &  & -57.99 & -52.72 & -51.08 & -50.37 \\
$\gamma1.1{\textrm{kh}}$ & -47.54 & -58.61 & -47.54 & -51.62 &  & -30.48 & -35.38 & -17.60 & -33.57 \\
$\gamma1.1{\textrm{kb}}$ & -57.39 & -53.55 & -50.16 & -50.06 &  & -47.35 & -47.03 & -39.38 & -44.16 \\
$\gamma1.3{\textrm{kl}}$ & -25.96 & -25.41 & -13.42 & -11.20 &  & -35.51 & -35.44 & -25.53 & -27.19 \\
$\gamma1.3{\textrm{kh}}$ & -3.86  & -3.15  & -4.69  & -6.84  &  & -21.35 & -10.49 & -11.55 & -27.63 \\
$\gamma1.3{\textrm{kb}}$ & -7.68  & -8.81  & 4.71   & 6.65   &  & -19.49 & -23.50 & -20.16 & -15.18 \\
$\gamma1.5{\textrm{kl}}$ & -25.55 & -22.61 & -0.89  & -7.66  &  & -32.65 & -20.14 & -26.57 & -32.14 \\
$\gamma1.5{\textrm{kh}}$ & -36.29 & -36.81 & -39.94 & -17.69 &  & -33.11 & -38.38 & -27.90 & -33.18 \\
$\gamma1.5{\textrm{kb}}$ & -20.80 & -4.66  & 6.57   & 8.80   &  & -36.85 & -17.09 & -19.65 & -25.29 \\
$\gamma1.7{\textrm{kl}}$ & -25.78 & -33.98 & -15.18 & -13.21 &  & -29.96 & -29.27 & -21.57 & -31.63 \\
$\gamma1.7{\textrm{kh}}$ & -22.83 & -3.25  & -17.22 & -4.03  &  & -28.93 & -11.74 & -40.35 & -32.57 \\
$\gamma1.7{\textrm{kb}}$ & -6.46  & -1.66  & 3.53   & 5.65   &  & -19.11 & -17.63 & -20.62 & -20.48 \\
\hline
\end{tabular}}
\end{center}
\begin{flushleft}
\footnotesize{Column 1: Name for main BSE models; columns 2, 3, 4, and 5 (or 6, 7, 8, and 9) list $\log K_{i}$ values obtained from sub-models. A positive (or negative) $\log K_{i}$ means the given model is preferred to (or disfavored comparing with) the model $\alpha10.\rm kb$. Here, $\log L_{\alpha10.\rm kb}=-67.39$. 
}
\end{flushleft}
\end{table*}

\begin{figure*}%
\begin{center}
\includegraphics[width=16cm]{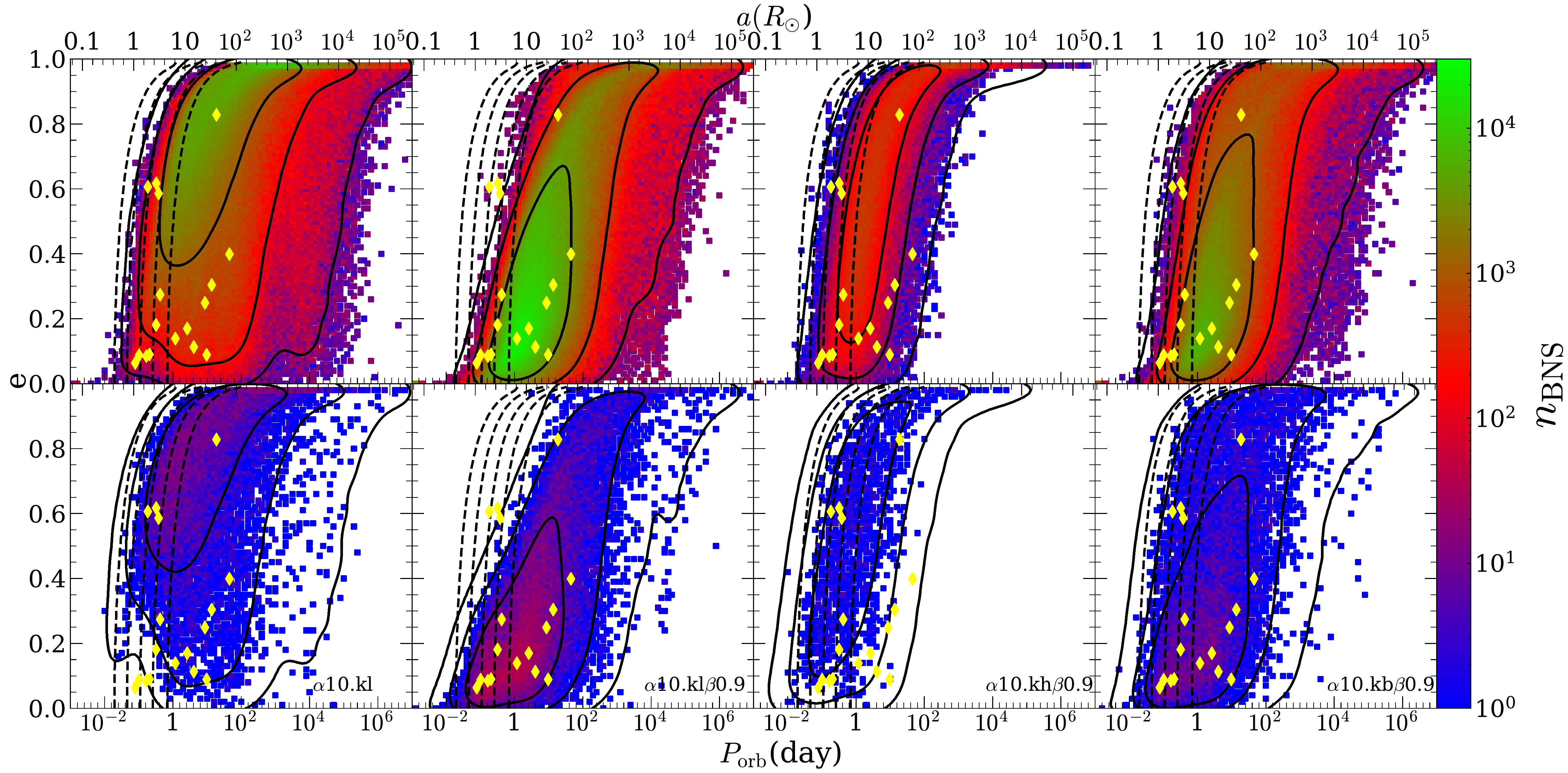}
\caption{Distribution of simulated BNSs on the eccentricity-orbital period ($e-P_{\rm orb}$) plane or equivalently the eccentricity-semi-major axis ($e-a$) plane for the model $\alpha10.\textrm{kl}$, $\alpha10.\textrm{kl}\beta0.9$, $\alpha10.\textrm{kh}\beta0.9$ and $\alpha10.\textrm{kb}\beta0.9$, respectively. Top panels: all survived BNSs at $z=0$ in a MW-like galaxy. Bottom panels: survived pulsar-NS or pulsar-pulsar binaries at $z=0$ in a MW-like galaxy. In each panel, the yellow diamonds mark the observed Galactic BNSs. Other color symbols show the number of BNSs in each $e-P_{\rm orb}$ or $e-a$ pixel (pixel size: $\Delta\log P_{\rm orb}=0.1$, $\Delta e=0.01$), with the correspondence between BNS number and color indicating by the color bar on the right hand side (rhs). The black solid lines indicate the $68.3\%$, $95.5\%$ and $99.7\%$ confidence intervals from inside to outside, respectively. The black dash lines represent the merger time of $t_{\rm d}=1$, $10$, $100$, $1000$\,Myr, and $t_{\rm Hubble}$, from left to right, respectively. The lines of constant $t_{\rm d}$ and the orbital semi-major axis scale on top assume a mass of 1.35 $M_\odot$ for both NSs. 
}
\label{fig:f3}
\end{center}
\end{figure*}
\begin{figure*}%
\begin{center}
\includegraphics[width=16cm]{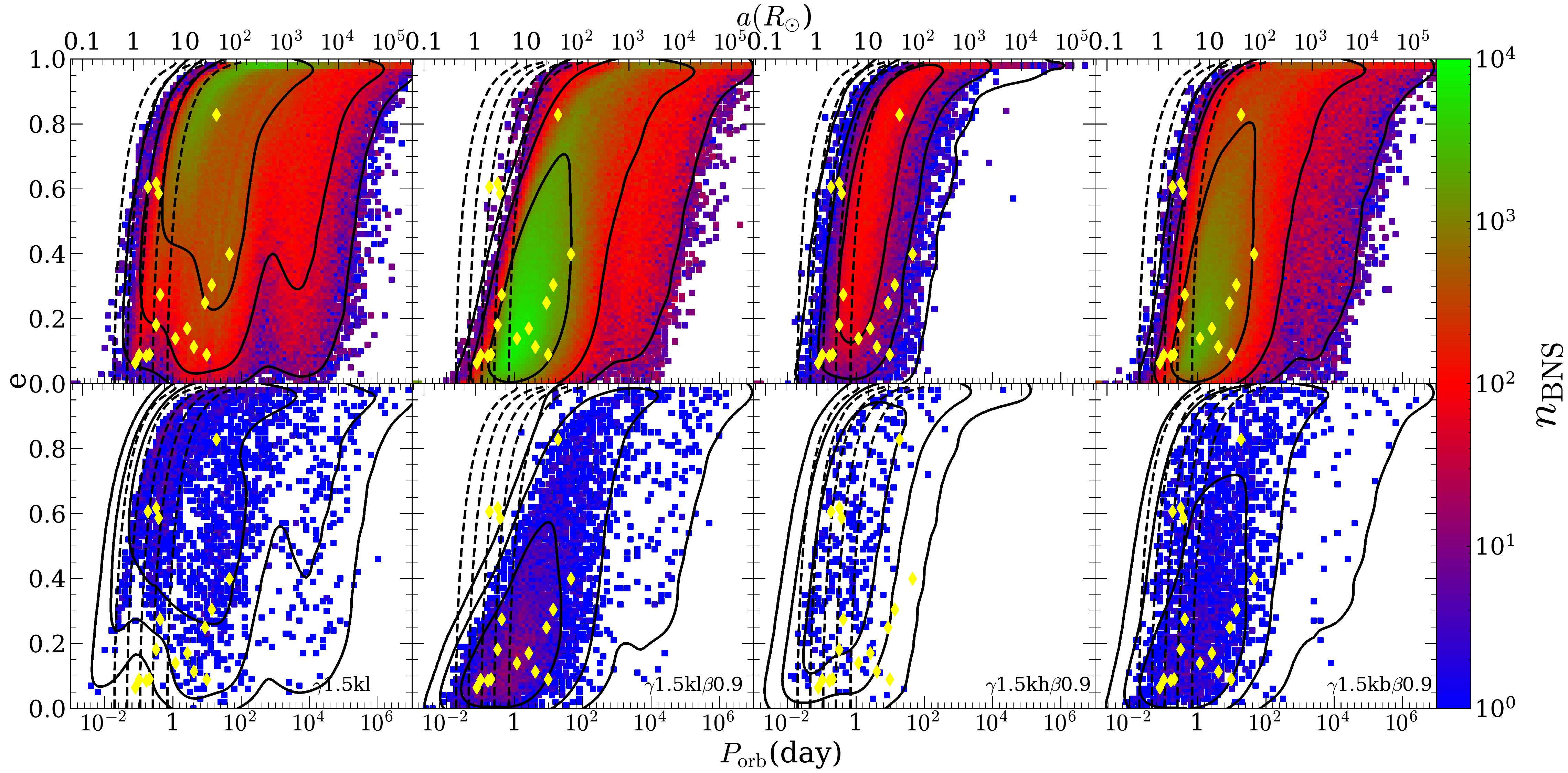}
\caption{Legend/ similar to that of Fig.~\ref{fig:f3}. But in this figure we show the results obtained from the models $\gamma1.5\textrm{kl}$, $\gamma1.5\textrm{kl}\beta0.9$, $\gamma1.5\textrm{kh}\beta0.9$ and $\gamma1.5\textrm{kb}\beta0.9$, respectively. 
}
\label{fig:f4}
\end{center}
\end{figure*}

Table~\ref{tab:t3} lists values of the logarithm Bayes factor, $\log K_{i}$, obtained by using all the survived BNSs or only those survived BNSs with at least one pulsar component at $z=0$ generated from each model. As seen from this table, the resulting distributions of pulsar-NS/pulsar-pulsar binaries can match the distribution of Galactic BNSs better than that of all NS-NS binaries. The model $\alpha10.{\textrm{kb}}\beta0.9$ results in the largest $\log K_i$, which suggests it is the most preferable one among all the models investigated in the present paper. For high mass ejection models (i.e., $\alpha10.{\textrm{kl}}$), the eccentricity distribution of predicted BNSs is apparently higher than that of the observations. For low mass ejection models, (i) if we adopt the low kick assumption, most Galactic BNSs can be reproduced but those with short orbital period and high eccentricity, especially J0509$+$3801, J1757$-$1854, and B1913$+$16, are hard to be reproduced; (ii) if we adopt the high kick assumption, the short-period Galactic BNSs can be well reproduced but long-period ones are hard to be reproduced. Hence, the model $\alpha10.{\textrm{kb}}\beta0.9$ with bi-modal kick velocity distribution is preferred compared with other models when matching all the Galactic BNSs (Fig.~\ref{fig:f3}), as the model $\gamma1.5{\textrm{kb}}\beta0.9$ is under the $\gamma$-formalism (Fig.~\ref{fig:f4}).

\subsection{BNS merger rate in galaxies and the local Universe}
\label{subsec:Galmrgrate}

\begin{table*}
\begin{center}
\caption{BNS merger rate densities in unit of ${\rm Gpc}^{-3}\,{\rm yr}^{-1}$ in the local Universe, ${\rm Myr}^{-1}$ in the MW-like galaxy at redshift $z=0$ and the fitting parameters of the evolution of the merger rate density as a function of redshift resulting from the preferred models.}
\label{tab:t4}
\setlength{\tabcolsep}{1mm}{
\begin{tabular}{c|c|c|c|c|c|c|c|c|c|c|c|c|c|c|c|c|c}
\hline
\multirow{2}{*}{Model Name}& \multicolumn{4}{c}{Millennium-II}  & \multicolumn{4}{c}{EAGLE}  & \multicolumn{4}{c}{Illustris-TNG}  & \multicolumn{4}{c}{Madau \& Dickinson (2014)}  & $R_{\rm MW}$\\ \cline{2-17}
&$R_{\rm Mil,0}$&$\zeta$&$z_{\star}$&$\xi$  &$R_{\rm EAG,0}$&$\zeta$&$z_{\star}$&$\xi$  &$R_{\rm Ill,0}$&$\zeta$&$z_{\star}$&$\xi$ & $R_{\rm MD,0}$&$\zeta$&$z_{\star}$&$\xi$ &$[{\rm Myr}^{-1}]$\\
\hline
SFR         &-&2.22&3.22&6.25      &-&2.78&1.79&4.47      &-&1.94&2.87&5.24      &-&2.7&1.9&5.6& -  \\ \hline
$\alpha1.0{\textrm{kb}}\beta0.9$ & 237   & 1.95 & 4.49 & 6.39 & 177   & 2.59  & 2.72 & 3.96  & 311   & 1.69  & 4.23 & 5.05  & 491    & 2.51  & 3.05 & 4.94  & 44.4  \\
$\alpha10.{\textrm{kh}}\beta0.9$ & 135   & 1.84 & 4.30 & 6.17 & 99.8  & 2.31  & 2.80 & 3.90  & 167   & 1.56  & 4.14 & 5.08  & 272    & 2.32  & 3.01 & 5.24  & 25.2  \\
$\alpha10.{\textrm{kb}}\beta0.9$ & 421   & 1.58 & 4.28 & 5.98 & 316   & 2.03  & 2.75 & 3.72  & 499   & 1.34  & 4.08 & 4.93  & 784    & 1.98  & 3.06 & 5.10  & 80.5  \\
$\gamma1.3{\textrm{kl}}\beta0.9$ & 447   & 1.07 & 3.46 & 5.21 & 332   & 1.33  & 2.45 & 3.23  & 480   & 0.85  & 3.27 & 4.03  & 784    & 1.57  & 2.74 & 4.33  & 86.2  \\
$\gamma1.3{\textrm{kb}}\beta0.9$ & 261   & 1.28 & 3.99 & 5.52 & 194   & 1.58  & 2.77 & 3.44  & 291   & 1.04  & 3.87 & 4.52  & 477    & 1.76  & 2.90 & 4.67  & 50.1  \\
$\gamma1.5{\textrm{kb}}\beta0.9$ & 199   & 1.66 & 4.23 & 6.11 & 137   & 2.08  & 2.98 & 4.18  & 228   & 1.43  & 4.15 & 5.44  & 429    & 1.98  & 2.79 & 4.75  & 35.5  \\
$\gamma1.7{\textrm{kh}}\beta0.9$ & 123   & 1.74 & 4.45 & 6.18 & 81.3  & 1.91  & 3.51 & 3.82  & 138   & 1.45  & 4.45 & 5.17  & 229    & 2.35  & 2.95 & 5.42  & 21.3  \\
$\gamma1.7{\textrm{kb}}\beta0.9$ & 369   & 1.33 & 4.90 & 6.44 & 226   & 1.30  & 4.81 & 4.14  & 368   & 1.08  & 5.13 & 5.57  & 683    & 2.10  & 2.90 & 5.18  & 62.9  \\ \hline
\end{tabular}}
\end{center}
\begin{flushleft}
\footnotesize{Column 1: model name; columns 2-5, 6-9, 10-13, and 14-17: the local merger rate density and fitting parameters resulting from Millennium-II, EAGLE, Illustis-TNG and observational extinction-corrected SFR in \citet{Madau2014} and the metallicity redshift evolution in \citet{Belczynski2016a}, respectively; column 18: the Galactic BNS merger rate in unit of $\rm Myr^{-1}$. 
}
\end{flushleft}
\end{table*}

\begin{figure*}
\begin{center}
\includegraphics[width=16cm]{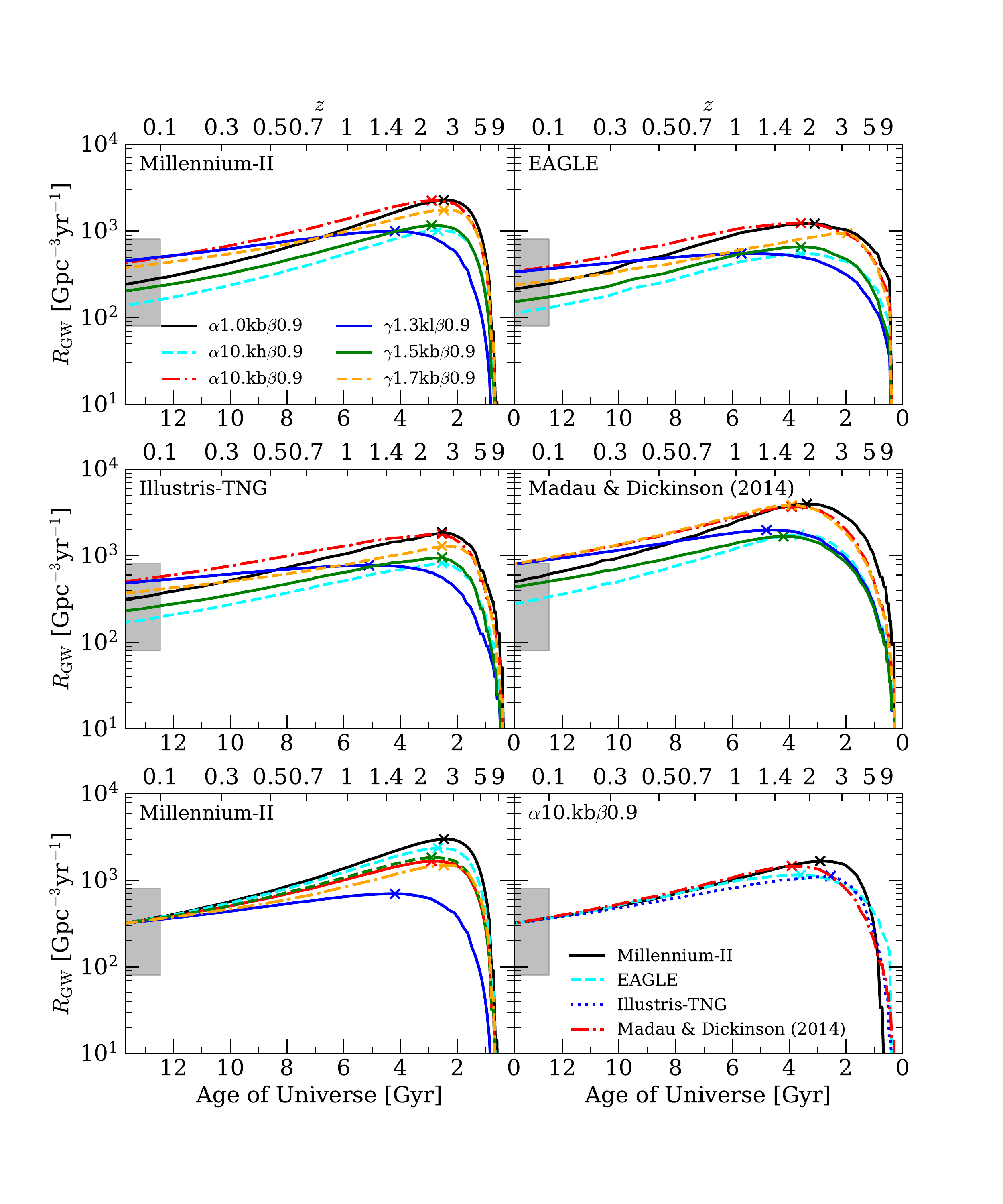}
\caption{BNS merger rate comoving densities as a function of the cosmic time (bottom $x$-axis) or redshift (top $x$-axis) for the preferred BSE models (see Table~\ref{tab:t4}). In each panel, the grey region indicate the range of the local BNS merger rate inferred from detections, $80-810 {\rm Gpc}^{-3}\,{\rm yr}^{-1}$ \citep{O3population}. Top left, top right, and middle left panels show the results obtained by using Millennium-II, EAGLE, and Illustris-TNG, respectively, while middle right panel shows the results obtained by using the observational extinction-corrected SFR in \citet{Madau2014} and the mean metallicity redshift evolution in \citet{Belczynski2016a}. Bottom left panel shows the difference in the evolution of merger rate densities of different models, which are re-normalized to the observational determined local merger rate density, $320 {\rm Gpc}^{-3}\,{\rm yr}^{-1}$ at $z=0$ \citep{O3population}, in Millennium-II. Bottom right panel shows difference in the evolution of merger rate densities of model $\alpha10.{\textrm{kb}}\beta0.9$ caused by different cosmological simulations, which are also scaled by detections. The cross symbols denote the peaks.
}
\label{fig:f5}
\end{center}
\end{figure*}

The merger rate density is an important global parameter for studying BNSs. Not only does it help to guide the design of future GW detectors and predict the detection rate/number of BNSs and its EM counterparts, but also to test the formation models of BNSs. In this paper, we calculate the BNS merger rate in the local Universe for different BSE models combining with Millennium-II, EAGLE, and Illustris-TNG, respectively. We also adopt the extinction-corrected specific star formation rate given by \citet{Madau2014} and the mean metallicity redshift evolution in \citet{Belczynski2016a} to estimate the merger rate density for comparison.

In our calculations, we consider that half of the stars are in binary systems ($f_{\rm bin}=0.5$) and the galaxy merger has no effect on the BNS systems. The BNS merger rate can be calculated as
\begin{equation}
R_{\rm GW}\simeq N_{\rm mrg} \left(\frac{l_{\rm box}}{\rm Gpc}\right)^{-3} \left(\frac{\Delta t}{\rm yr}\right)^{-1} \Gpcpyr.
\end{equation}
Here $N_{\rm mrg}$ is the number of BNS mergers per time bin $\Delta t$ in the entire volume $l_{\rm box}^{3}$ of cosmological simulations (Millennium-II: $l_{\rm box}=137 \textrm{Mpc}$; EAGLE: $l_{\rm box}=100 \textrm{Mpc}$; Illustris-TNG: $l_{\rm box}=110.7 \textrm{Mpc}$). We obtain the BNS merger rates and their evolution with redshift resulting from three different cosmological galaxy formation and evolution simulations/models and the observations \citep{Madau2014, Belczynski2016a}.

To quantitatively compare the evolution of BNS merger rate density ($R(z)$), resulting from the combination of different BSE models with different cosmic SFR as histories, we further perform a fit to $R(z)$ resulting from each model as
\begin{eqnarray}
R(z) & = & R_*\frac{2[(1+z)/(1+z_*)]^{\zeta}}{1+[(1+z)/(1+z_{\star})]^{\xi}}  \nonumber \\
& = & R_0\left[1+\frac{1}{(1+z_*)^{\xi}}\right] \frac{[(1+z)]^\zeta}{1+[(1+z)/(1+z_{\star})]^{\xi}}. \nonumber \\
\label{eq:eq8}
\end{eqnarray}
Here $R_*$ and $R_0$ are the BNS merger rate density at $z_*$ and the local BNS merger rate density ($z=0$), respectively; $z_{\star}$, $\zeta$, and $\xi$ are the shape parameters. We list the estimated local BNS merger rates and their best-fit values of several preferred models in Table~\ref{tab:t4}, which are compatible with O3 observational constraint on the local BNS merger rate, $80\,\Gpcpyr$-$810\,\Gpcpyr$ \citep{O3population}. We also present our results for other models in Appendix~\ref{sec:appB}. 

As seen from Tables~\ref{tab:t4} and \ref{tab:t9}-\ref{tab:t12} (in the Appendix), the resulting $R_0$ are $\sim0.4$-$782$, $0.4$-$588$, $0.5$-$901$, and $1.0$-$1404\,\Gpcpyr$ for those models using Millennium-II, EAGLE, Illustris-TNG, and the extinction-corrected SFR in \citet{Madau2014}, respectively. The value of $R_0$ highly depends on the assumptions of CE evolution, distribution of natal kick velocities, and mass ejection. 
For those BSE models that assume a small $\alpha$ (e.g., $0.1$) or $\gamma$ (e.g., $1.1$), the formation or merger of BNSs are suppressed, which both lead to relatively low $R_0$. For those BSE models that assume large natal kick velocities or mass ejection, it is more efficient in leading to elongation or breakup of binary systems, and thus result in relatively lower $R_0$ when considering the same CE model. Obviously, the value of $R_0$ is also directly affected by the assumed star formation history within the population synthesis models considered in this paper. For instance, the observational determined SFR has the highest value, and it is about a factor of $\sim 1.59$, $2.23$, and $1.55$ times of that given by Millennium-II, EAGLE, and Illustris-TNG, which is directly reflected in the estimates of $R_0$. Illustris-TNG produces a higher SFR than Millennium-II does at low redshift and the values of $R_0$ obtained from Illustris-TNG are higher than those from Millennium-II, although the total stars formed in these two simulations are almost equal. 

Different from the previous claims \citep{Mapelli2018rate, Jiang2020}, we find that the inferred $R_0$ from the detections does not necessarily require an extremely high CE ejection efficiency and low kick velocities, mainly because the latest inferred $R_0$ is narrowed to a lower value $80-810\,\Gpcpyr$ though still with large uncertainties \citep{O3population}. Future GW detections will provide more accurate estimation of local merger rate and put strong constraints on the BSE model. Note also that the uncertainties arising from the initial distributions of binaries (e.g., IMF) and the estimates of cosmic SFR histories \citep[e.g.,][]{Chruslinska2019,Neijssel2019,Broekgaarden2021} are not negligible. Further information from future GW detection on the merger rate evolution may put much stronger constraints on BSE models.

Figure~\ref{fig:f5} shows the BNS merger rate densities as a function of redshift ($R(z)$) or cosmic time resulting from several preferred models (see Tab.~\ref{tab:t4}). Top-left, top-right, and middle-left panels show the results obtained by using Millennium-II, EAGLE, and Illustris-TNG, respectively, while middle-right panel shows the results obtained by using the observational extinction-corrected SFR in \citet{Madau2014} and the mean metallicity redshift evolution in \citet{Belczynski2016a}.
The evolution of BNS merger rate densities is mainly determined by the convolution between the cosmic SFR history and distribution of delay time.
As seen from the figure, for those BSE models that produce BNS mergers with relative large delay time (e.g., $\gamma1.3\rm kl\beta0.9$), the resulting $R(z)$ curve are relative flat (i.e., $\zeta=1.07$ if using Millennium-II) and peak at lower redshift (e.g., $z\sim 1.6$ if using Millennium-II). In contrast, for BSE models that produce BNS mergers with small delay time (e.g., $\alpha1.0\rm kb\beta0.9$), the resulting BNS merger rate density is almost proportional to the SFR distribution (i.e., $\zeta=1.95$ if using Millennium-II), and peak at a redshift similar to that of the SFR history (e.g., $z\sim2.8$ if using Millennium-II).

Bottom-right panel of Figure~\ref{fig:f5} shows the shape differences between $R(z)$ curves resulting from the combination of BSE model $\alpha10.\rm kb\beta0.9$ with different galaxy formation and evolution models/simulations or the observational SFR history. The different choices of the cosmic SFR histories do have some effect on the shape of resulting $R(z)$ curve, as expected, but it is not as significant as that caused by the choice of different BSE models (see bottom-left panel). Adopting Illustris-TNG, the resulting $R(z)$ curve is relatively flat, $\zeta=1.34$, and peaks at a higher redshift $z\sim2.6$. When considering the cosmic SFR history given by \citet{Madau2014}, the $R(z)$ curve becomes steeper, $\zeta=1.98$, and peaks at a lower redshift $z\sim1.7$.
Future ground-based GW detectors, such as the Einstein Telescope (ET) and Cosmic Explorer (CE), may detect more than tens thousands of BNS mergers, which will enable an accurate determination of the evolution of BNS merger rate density (i.e., $R_0$, $\zeta$, $z_{\star}$, and $\xi$), and thus help to distinguish different BSE models and constrain the cosmic SFR history.

In Table~\ref{tab:t4} and Tables~\ref{tab:t9}-\ref{tab:t12}, we also provide an estimate of the Galactic BNS merger rate densities in the same approach by utilizing the selected MW-like galaxies from Millennium-II. In most previous work, the Galactic merger rates were calculated by assuming a continuous SFR and constant metallicity with the age of $10$\,Gyr \citep[e.g.,][]{Dominik2012, Vigna2018}, and the Galactic merger rate is translated to the local merger rate via the following conversion as
\begin{equation}
R_{\rm local}=10 {\rm Gpc}^{-3} {\rm yr}^{-1} \left(\frac{\rho}{0.01{\rm Mpc}^{-3}} \right) \left(\frac{R_{\rm MW}}{{\rm Myr}^{-1}} \right),
\end{equation}
where $\rho$ stands for the local density of Milky Way equivalent galaxy \citep{demink2015}. 
According to observations, there are $10^{7}$ galaxies of comparable mass to the Milky Way in the volume of $1 {\rm Gpc}^{3}$ and $\rho=0.01{\rm Mpc}^{-3}$ \citep[e.g.,][]{Abadie2010, Gupta2017, Abbott et al. 2017g}. This conversion is once again confirmed as seen from those tables (see data $R_{\rm MD,0}$ and $R_{\rm GW}$).

\subsection{Properties of Host galaxies in local Universe}
\label{subsec:hostgal}

The model $\alpha10.{\rm kb}\beta0.9$ is preferred as it is more compatible with both the Galactic BNS observations and the local BNS merger rate density inferred from GW detections, as shown in Section~\ref{subsec:GalBNSs} and \ref{subsec:Galmrgrate}. Therefore, we take model $\alpha10.{\rm kb}\beta0.9$ as our reference model for the following analysis of the host galaxy properties of BNS mergers. For comparison, we also show the results of the models $\alpha1.0{\rm kb}\beta0.9$ (i.e., a short delay time model) and $\gamma1.3{\rm kl}\beta0.9$ (i.e., a long delay time model) (see Table~\ref{tab:t4}.)

\subsubsection{Stellar mass distributions}

\begin{figure*}
	\begin{center}
		\includegraphics[width=16cm,height=6cm]{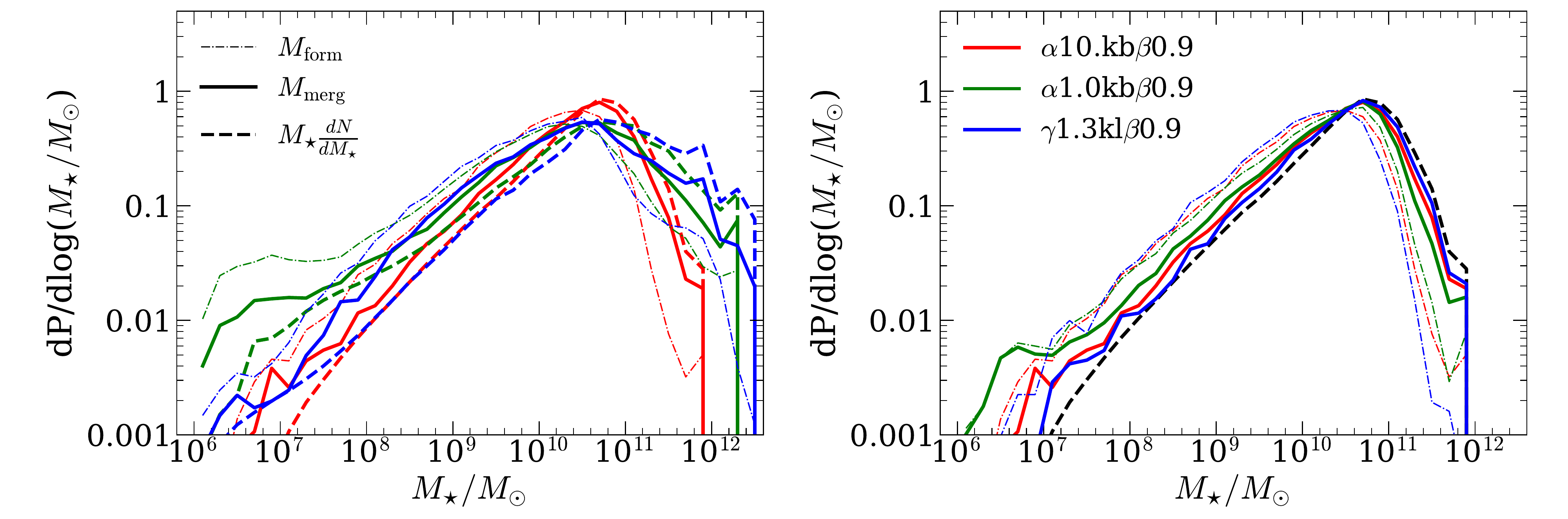}
		\caption{Stellar mass distribution functions (SMDFs) for the host galaxies of those BNSs merged at $z<0.01$ (i.e., the local GW events) at their merger time (solid lines) and formation time (dotted-dash lines), respectively. Dash lines show the SMDFs for all galaxies at redshift $z\sim0$ from the simulations. Left: SMDFs predicted by the model $\alpha10.{\rm kb}\beta0.9$ resulting from Millennium-II (red), EAGLE (green), and Illustris-TNG (blue), respectively. Right: SMDFs predicted by the Millennium-II resulting from different BSE models, $\alpha10.{\rm kb}\beta0.9$ (red), $\alpha1.0{\rm kb}\beta0.9$ (green), and $\gamma1.3{\rm kl}\beta0.9$ (blue), respectively.
		}
		\label{fig:f6}
	\end{center}
\end{figure*}

We extract the statistical information on BNS host galaxies from the mock catalogs for BNSs and their mergers. Figure~\ref{fig:f6} shows the stellar mass probability distribution functions (SMDFs) for the host galaxies of BNSs at their merger time (i.e., the GW detection time $z<0.01$; solid lines) and formation time (dotted-dash lines) for the model $\alpha10.{\rm kb}\beta0.9$. We also show the SMDFs for all galaxies at $z\sim0$, resulting from Millennium-II (red dash line), EAGLE (green dash line), and Illustris-TNG (blue dash line), respectively. 

The SMDFs of BNS host galaxies at the merger time ($z\sim 0$) resulting from Millennium-II, EAGLE, and Illustris-TNG peak at $5.0\times 10^{10}M_\odot$, $3.2\times 10^{10}M_\odot$, and $3.2\times 10^{10}M_\odot$, respectively. The five-to-ninety-five percentiles of these SMDFs are $\sim1.2\times 10^{9}$-$1.8\times10^{11} M_\odot$, $3.0\times 10^{8}$-$5.4\times 10^{11} M_\odot$, and $6.2\times 10^{8}$-$7.1\times 10^{11} M_\odot$, respectively.
Our results show that BNSs prefer merging in galaxies with $M_*\sim 10^{9.5}-10^{11}M_\odot$. The SMDF of the host galaxies of BNS mergers (solid lines) is similar to the SMDF of all galaxies (dash lines) at redshift $z\sim0$ (see Fig.~\ref{fig:f6}), but not exactly the same. The later one peaks at $5.0\times 10^{10}M_\odot$ and the five-to-ninety-five percentiles of these SMDFs resulting by using  Millennium-II, EAGLE, and Illustris-TNG  are from $\sim2.0\times 10^{9}$-$2.3\times10^{11} M_\odot$, $7.4\times 10^{8}$-$8.9\times 10^{11} M_\odot$, and $2.0\times 10^{9}$-$1.3\times 10^{12} M_\odot$, respectively. For galaxies with large stellar mass ($M_\star\ga 5\times 10^{10}M_\odot$), the BNS merger efficiency per stellar mass is lower than that for smaller galaxies. 
The main reason is that massive galaxies, especially elliptical galaxies, have relatively larger average stellar ages than that of smaller galaxies. When considering the delay time distributions of BNSs (e.g., $p(t_{\rm d})\propto t^{-1}$), we find that per stellar mass formed recently can result in more BNS merger events than per stellar mass formed at high redshift.

The results obtained from those different galaxy formation and evolution simulations/models are roughly consistent with each other, but do show some differences, probably due to the disparate SMDFs. As seen from Figure~\ref{fig:f6}, EAGLE and Illustris-TNG produce more massive galaxies at $z\sim 0$ than Millennium-II does, especially at $M_{\star} \geq 10^{11.5} M_\odot$, which leads to a shift of the SMDFs to high mass range. At the low-mass end, EAGLE produces more small galaxies than the other two models, leading to the higher probability of the low stellar mass host galaxies. The SMDFs of the host galaxies of BNS mergers shown in Figure~\ref{fig:f6} clearly reflect the differences in the resulting SMDFs of all galaxies from different galaxy formation and evolution models. The right panel of Figure~\ref{fig:f6} shows the SMDFs predicted by different BSE models, $\alpha10.{\rm kb}\beta0.9$, $\alpha1.0{\rm kb}\beta0.9$, and $\gamma1.3{\rm kl}\beta0.9$ by using Millennium-II, respectively. It is obvious that the host galaxy distribution of the BNS systems at the merger time is shifted to higher masses compared with that at their formation time and the shift increases with the increasing delay times. This is simply caused by the growth of those host galaxies after the BNS formation, especially for those BNSs with long time delay.

\begin{figure}
\begin{center}
\includegraphics[width=7cm]{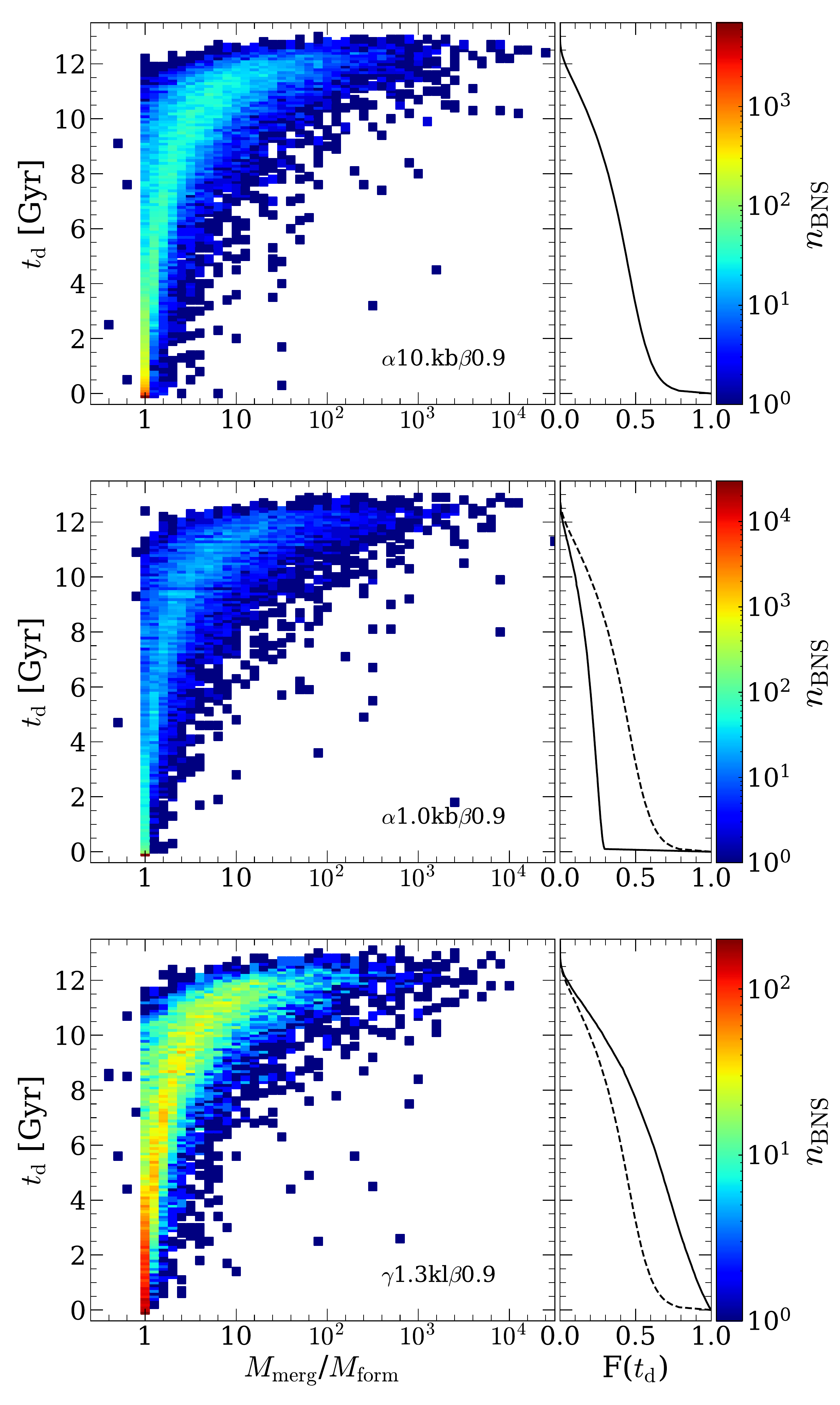}
\caption{Time delay ($t_{\rm d}$) versus mass ratio of host galaxies ($M_{\rm merg}/M_{\rm form}$) at redshift $z<0.01$ for the model $\alpha10.{\rm kb}\beta0.9$ (top),  $\alpha1.0{\rm kb}\beta0.9$ (middle), and $\gamma1.3{\rm kl}\beta0.9$ (bottom) using Millennium-II. The right panels show the cumulative distribution of delay time. The dashed line in the middle- or bottom-right panels denotes the delay time distribution shown in the top-right panel, for comparison.
}
\label{fig:f7}
\end{center}
\end{figure}

Figure~\ref{fig:f7} shows the time delay ($t_{\rm d}$) versus mass ratio ($M_{\rm merg}$/$M_{\rm form}$) (left panel) and the cumulative distribution of delay time (right panel) for the host galaxies of BNS mergers at redshift $z<0.01$ for the model $\alpha10.{\rm kb}\beta0.9$, $\alpha1.0{\rm kb}\beta0.9$, and $\gamma1.3{\rm kl}\beta0.9$ in Millennium-II as an example. 
For the model $\alpha10.{\rm kb}\beta0.9$, $\alpha1.0{\rm kb}\beta0.9$, or $\gamma1.3{\rm kl}\beta0.9$, about $\sim38.7\%$, $\sim72.9\%$, or $\sim9.0\%$ of BNS merger events are formed within $1 \rm Gyr$, while about $\sim31.9\%$, $\sim16.0\%$, or $\sim47.6\%$ of BNS merger events are formed more than $8 \rm Gyr$ ago. For the former one, these fraction of BNS mergers occur in the same galaxies when they formed, while for the latter ones, the BNS merger host galaxies may have been grown up by a factor of tens or hundreds of times in mass since the formation of those BNSs.

\subsubsection{Morphologies and types of BNS host galaxies}

The stellar mass is one important physical parameter of the host galaxy but not the only one that encode information about the formation of BNSs. The morphology of the host galaxy may be also an important parameter. Some BNSs may merge in the elliptical galaxies while some in spiral ones. For GW 170817, the confirmed host galaxy, NGC 4993, is an elliptical galaxy \citep{Im2017}. 

According to the Millennium-II galaxy formation and evolution model, we can extract the detailed information for each mock galaxy, e.g., bulge mass, stellar mass, and merger tree \citep{Guo2011}. We label a mock galaxy as elliptical or  early-type if the ratio of bulge to total stellar mass $\geq 0.8$, but a spiral or a late-type one otherwise.
In Illustris-TNG, a nonparametric measure (Gini's coefficient) of morphology \citep{Snyder2015} is provided for galaxies with stellar mass ranging from $10^{9.7}M_\odot$ to $10^{12.3}M_\odot$ at redshift $z=0$. Using the value of Gini's coefficient given in the galaxy catalog, we label a galaxy as an elliptical (bulge dominated) if $Gini\geq0.14M_{20}+0.778$ and a spiral (disc dominated) if $Gini<0.14M_{20}+0.778$. According to the results of Milleniuum-II, most galaxies with stellar mass less than $10^{9.7}M_\odot$ are late-type galaxies. Therefore, we can roughly obtain the distribution of ellipticals and spirals for the models using Illustris-TNG at low redshift as Illustris-TNG only offers the morphology of massive galaxies at a few snapshots. 

For EAGLE, no morphological information is provided.
This hinders a quantitative investigation of the host galaxy morphology of BNS mergers by using EAGLE.  

\begin{figure*}
\begin{center}
\includegraphics[width=16cm]{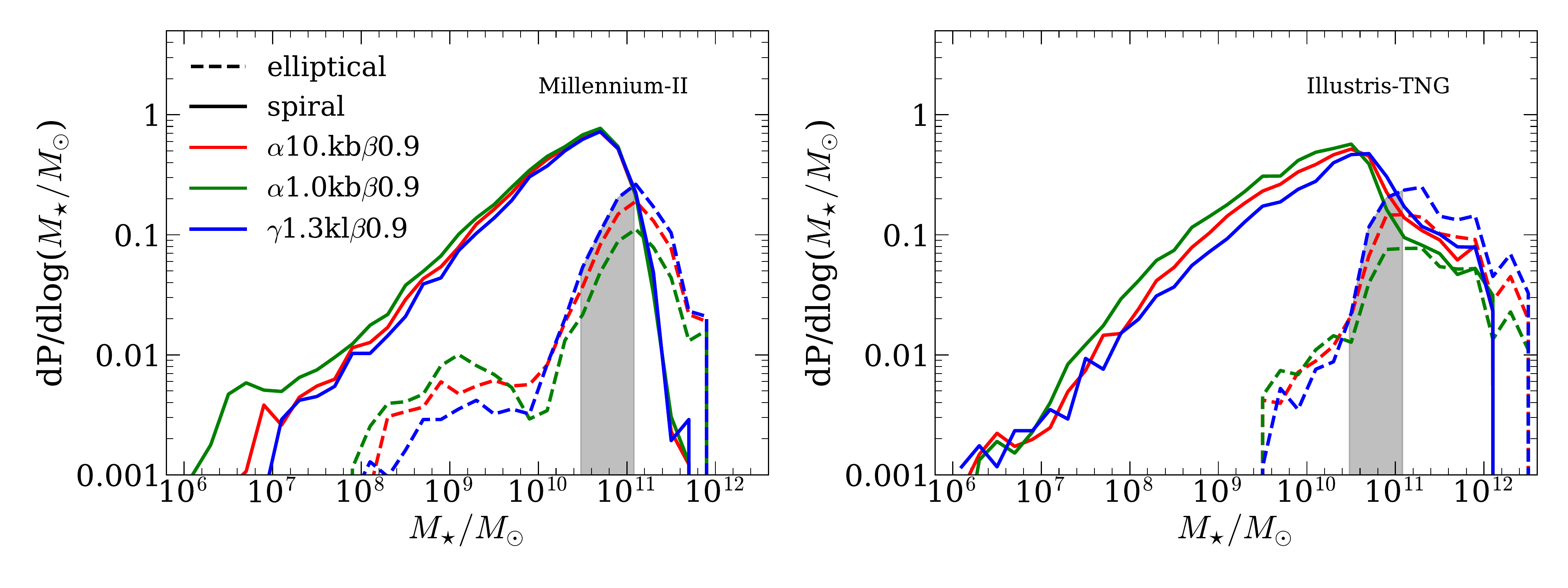}
\caption{
Probability distributions of stellar mass of host galaxies with different morphologies for BNS mergers at redshift $z<0.1$ resulting from the model  $\alpha10.{\rm kb}\beta0.9$ (red lines), $\alpha1.0{\rm kb}\beta0.9$ (green lines) and $\gamma1.3{\rm kl}\beta0.9$ (blue lines) using Millennium-II (left panel) and Illustris-TNG (right panel), respectively. Solid and dash lines in each panel represent the distributions for spiral and elliptical host galaxies, respectively. The gray region indicates the stellar mass range of NGC 4993 ($\sim 3\times 10^{10}$-$1.2\times 10^{11} M_\odot$), the host of GW\,170817  \citep{Im2017}. 
}
\label{fig:f8}
\end{center}
\end{figure*}

Figure~\ref{fig:f8} shows the probability distributions of stellar masses of host galaxies with different morphologies at redshift $z<0.01$ for the model $\alpha10.{\rm kb}\beta0.9$, $\alpha1.0{\rm kb}\beta0.9$ and $\gamma1.3{\rm kl}\beta0.9$ obtained by using Millennium-II (left panel) and Illustris-TNG (right panel), respectively.
The results show that BNSs prefer merging in spiral galaxies, especially those with low stellar masses ($\lesssim 5\times 10^{10}M_\odot$), while the fraction of elliptical hosts increases with increasing delay times. For the model $\alpha10.{\rm kb}\beta0.9$, about $84.4\%/81.2\%$ (from Millennium-II/Illustris-TNG; hereafter M/I) of BNS merger host galaxies are spirals. The spiral host stellar mass distribution at the BNS merger time peaks at $5.0\times 10^{10}M_\odot$/$3.2\times 10^{10}M_\odot$ (M/I) and the five-to-ninety-five percentiles are $1.1\times 10^{9}-1.1\times 10^{11} M_\odot$/$4.7\times 10^{8}-3.2\times 10^{11} M_\odot$ (M/I). About $15.6\%/18.8\%$ (M/I) BNS merger host galaxies are ellipticals. The elliptical host stellar mass distribution at the BNS merger time peaks at $1.3\times 10^{11}M_\odot$/$1.3\times 10^{11}M_\odot$ (M/I) and the five-to-ninety-five percentiles are  $6.5\times 10^{9}-4.1\times 10^{11} M_\odot$/$3.2\times 10^{10}-1.9\times 10^{12} M_\odot$ (M/I). For the model $\alpha1.0{\rm kb}\beta0.9$, about $90.1\%/89.3\%$ (M/I) of host galaxies are spirals and the rest $\sim9.9\%/10.7\%$ (M/I) are ellipticals. For the model $\gamma1.3{\rm kl}\beta0.9$, about $79.9\%/71.6\%$ (M/I) of BNS merger host galaxies are spirals and the rest $\sim20.1\%/28.4\%$ (M/I) are ellipticals.
The slight differences of the results on host morphologies obtained from Millennium-II and Illustris-TNG are mainly caused by that Illustris-TNG produces more massive galaxies at the high-mass end, and thus a higher fraction of elliptical galaxies ($36.5\%$) at redshift $z\sim0$ than Millennium-II does ($25.9\%$).
Despite of this discrepancy, both simulations generate enough elliptical host galaxies, even at the mass of a few times $10^{11}M_\odot$. There should be no significant bias on the results of the host mass distribution directly caused the limited simulation volumes of Millennium-II and Illustris-TNG.
For a BNS merger, the probability for its host galaxy to be similar as NGC 4993 is only about $6.6\%/5.7\%$, $3.9\%/3.1\%$ and $9.0\%/8.6\%$ (I/M; gray region in the left/right panel of Fig.~\ref{fig:f8}) for the model $\alpha10.{\rm kb}\beta0.9$, $\alpha1.0{\rm kb}\beta0.9$, and $\gamma1.3{\rm kl}\beta0.9$, respectively.

\begin{figure}
\begin{center}
\includegraphics[width=7cm]{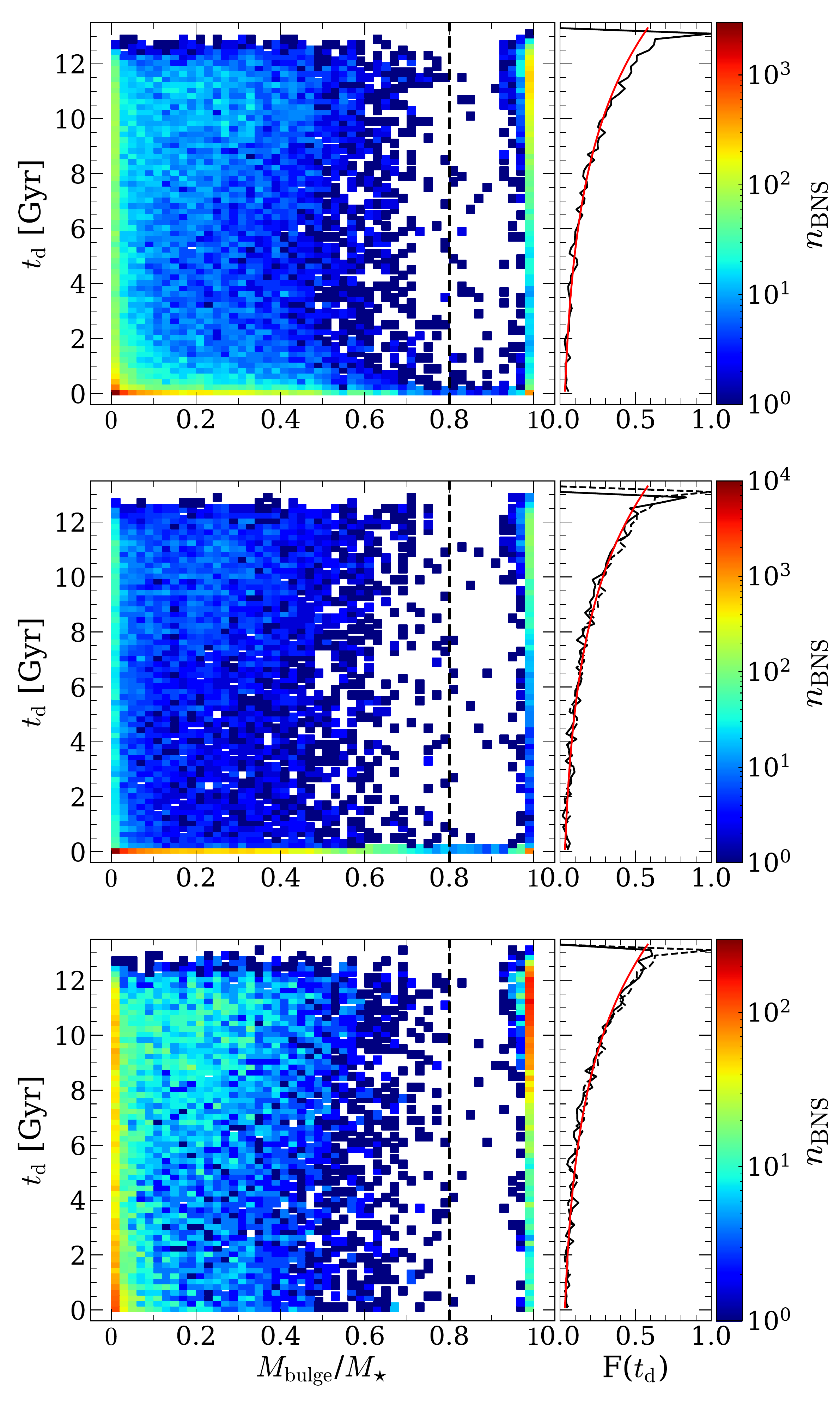}
\caption{
Time delay ($t_{\rm d}$) versus ratio of bulge to total stellar mass ($M_{\rm bulge}/M_{\star}$) for host galaxies of BNS mergers at $z<0.01$ resulting by implementation of the model $\alpha10.{\rm kb}\beta0.9$ (top), $\alpha1.0{\rm kb}\beta0.9$ (middle) and $\gamma1.3{\rm kl}\beta0.9$ (bottom) into Millennium-II. Black dash line in the bottom-left panel indicates $M_{\rm bulge}/M_{\star}=0.8$. The small right panels show the fraction of elliptical host galaxies ($f_{\rm ell}$) as a function of $t_{\rm d}$ at $z<0.01$. The dashed line in the middle- or bottom-right panels denotes the $f_{\rm ell}$ distribution versus $t_{\rm d}$ shown in the top-right panel, for comparison, and the red lines show the fitting. 
}
\label{fig:f9}
\end{center}
\end{figure}

Figure~\ref{fig:f9} shows the time delay ($t_{\rm d}$) versus ratio of bulge to total stellar mass ($M_{\rm bulge}/M_{\star}$) for host galaxies of BNS mergers at redshift $z<0.01$ resulting by implementing the model $\alpha10.{\rm kb}\beta0.9$, $\alpha1.0{\rm kb}\beta0.9$ and $\gamma1.3{\rm kl}\beta0.9$ into Millennium-II. We also plot the fraction of elliptical host galaxies as a function of delay time in this Figure. Despite the host galaxies preferring to be spirals, the probability of an elliptical host increases with increasing $t_{\rm d}$. The reason is as follows. BNSs are almost all initially formed in spiral galaxies simply due to their high SFR ($\alpha10.{\rm kb}\beta0.9$: $\sim97.1\%$; $\alpha1.0{\rm kb}\beta0.9$: $\sim95.8\%$; $\gamma1.3{\rm kl}\beta0.9$: $\sim98.1\%$ in Millennium-II). During the later evolution of BNSs, the initial spiral galaxies may merge to form or transform to ellipticals. The larger $t_{\rm d}$ and thus the longer evolution time, the higher probability for morphology transformation. We also note that for BNS mergers in local Universe, the correlation between the fraction of elliptical host galaxies ($f_{\rm ell}$) and $t_{\rm d}$ is strong and can be approximately expressed as $\log f_{\rm ell}=0.094(t_{\rm d}/1{\rm Gyr})-1.48$ (the red lines in the right panels of Fig.~\ref{fig:f9}).

In addition, some BNSs may merge in the central or satellite galaxies of big dark matter halos, while others in isolated galaxies of small halos. In our following analysis, we label a mock galaxy as the ``central galaxy'' if it is in the main subhalo of a dark matter halo, a ``satellite galaxy'' if it is in other subhalos, and an ``isolated galaxy'' if it is the only one in that halo. (We ignore the possibility that some BNSs may be ejected out of galaxies and merger at a place not being hosted by any galaxy.)

\begin{figure*}
\begin{center}
\includegraphics[width=16cm]{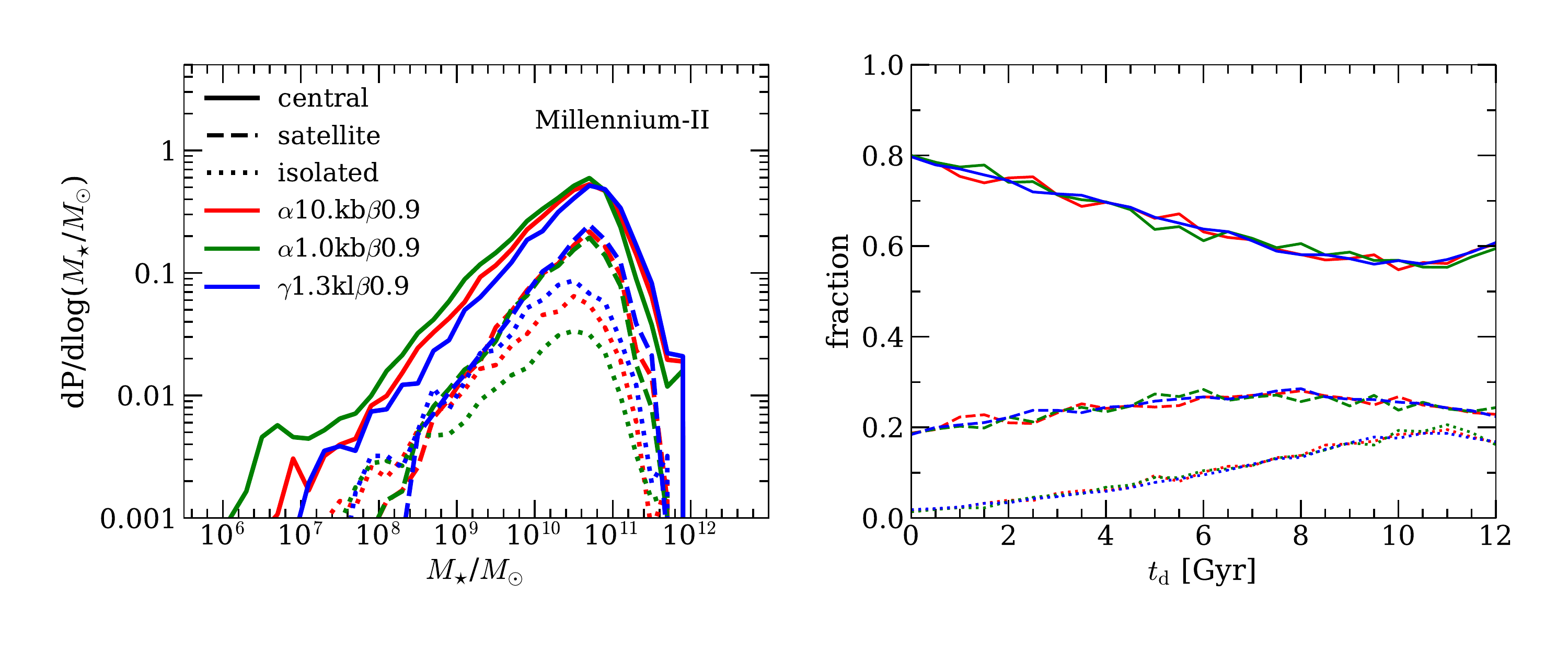}
\caption{Left: distributions of the host galaxies as a function of galaxy stellar mass for host galaxies with different types, including central galaxies, satellites and isolated galaxies, at the redshift $z<0.01$ for $\alpha10.{\rm kb}\beta0.9$, $\alpha1.0{\rm kb}\beta0.9$ and $\gamma1.3{\rm kl}\beta0.9$ model, using Millennium-II. Right: time delay ($t_{\rm d}$) versus the fraction of different types of host galaxies at redshift $z<0.01$ obtained from these models. Solid lines: central host galaxies. Dash lines: satellite host galaxies. Dotted lines: isolated host galaxies. 
}
\label{fig:f10}
\end{center}
\end{figure*}

Left panel of Figure~\ref{fig:f10} shows the distributions of host galaxies (categorized in different types, i.e., central, satellite, and isolated galaxies) of BNS mergers as a function of stellar mass at the redshift $z<0.01$ obtained by implementing the model $\alpha10.{\rm kb}\beta0.9$, $\alpha1.0{\rm kb}\beta0.9$ and $\gamma1.3{\rm kl}\beta0.9$ into Millennium-II. For the model $\alpha10.{\rm kb}\beta0.9$, most BNSs ($\sim81.0\%$) are formed in central galaxies, and only $\sim17.0\%$ BNSs are formed in satellites. A small fraction of BNSs, $\sim2.0\%$, are formed in isolated galaxies. 
However, the probability of the BNS merger host galaxy to be a central, satellite, or isolated galaxy is $69.4\%$, $22.4\%$, or $8.2\%$. For the model $\alpha1.0{\rm kb}\beta0.9$, about $80.6\%$, $17.6\%$, or $1.8\%$ BNSs are formed in central, satellite, or isolated galaxy, while the probability of the BNS merger host galaxy to be a central, satellite, or isolated galaxy is $74.8\%$, $20.3\%$, or $5.9\%$. For the model $\gamma1.3{\rm kl}\beta0.9$, about $81.8\%$, $15.9\%$, or $2.3\%$ BNSs are formed in central, satellite, or isolated galaxy, while the probability of the BNS merger host galaxy to be a central, satellite, or isolated galaxy is $63.8\%$, $24.7\%$, or $11.5\%$.

Right panel of Figure~\ref{fig:f10} also shows the fraction of different type host galaxies at redshift $z<0.01$ as a function of $t_{\rm d}$ obtained from the same models. As seen from this panel, most BNSs merge in the central galaxies. With increasing $t_{\rm d}$, the probability of BNS mergers being hosted in the central galaxies decreases but the probability in satellites or isolated galaxies increases. This tendency becomes flat or slightly reversed when $t_{\rm d} \gtrsim 11$\,Gyr, because most star forming galaxies formed at high redshift ($z\gtrsim 3$), the hosts for those BNSs with such long time delays to form, are at high density peaks and eventually evolved to be central galaxies rather than satellites or isolated ones at $z\sim 0.01$, according to Millennium-II.

\subsubsection{SFR and metallicity distributions of BNS host galaxies}

\begin{figure}
\begin{center}
\includegraphics[width=7cm]{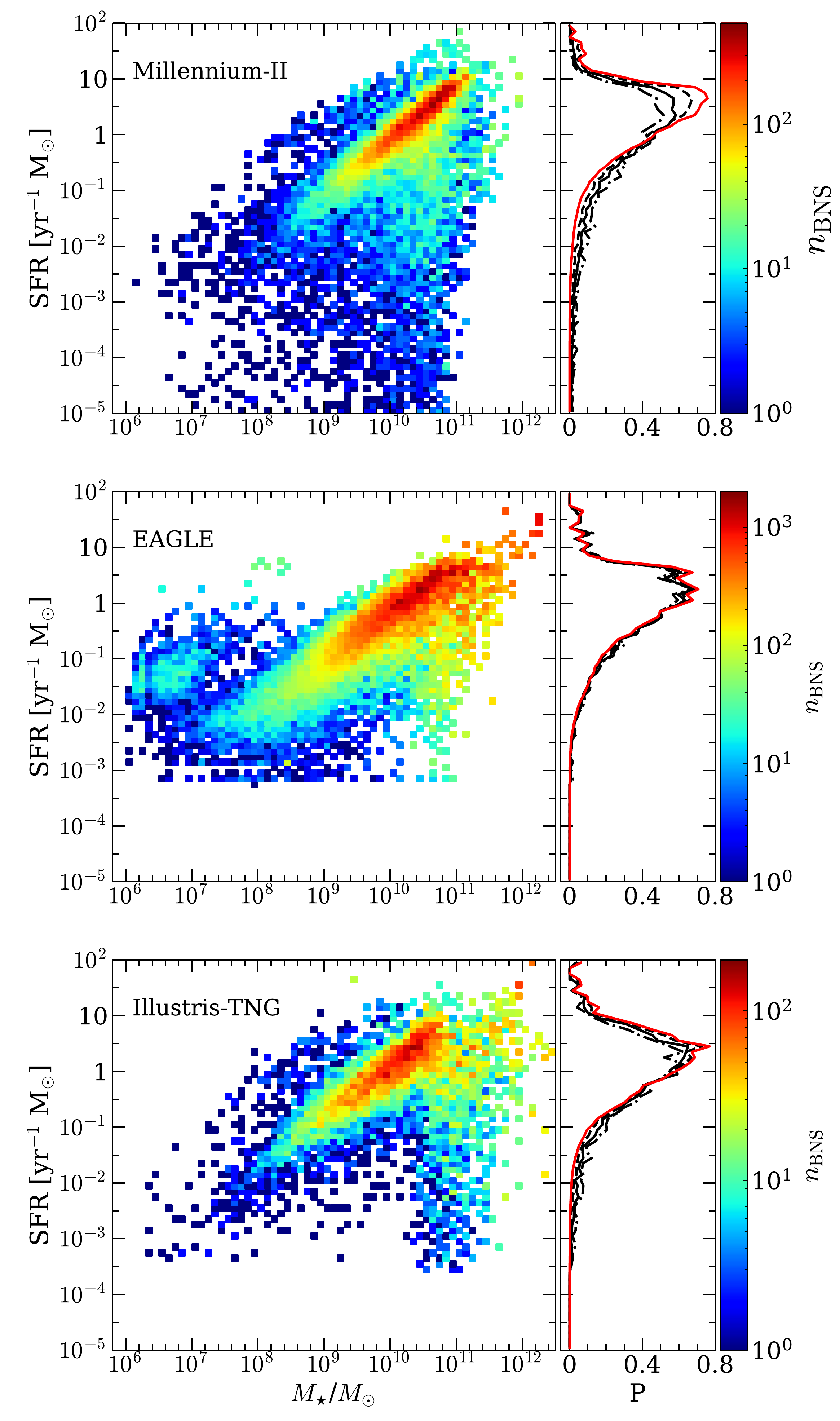}
\caption{
Distributions of BNS mergers on the SFR versus stellar mass of host galaxies (${\rm SFR}-M_\star$) plane at redshift $z<0.01$, obtained by implementing the model $\alpha10.{\rm kb}\beta0.9$ into Millennium-II (top), EAGLE (middle), and Illustris-TNG (bottom), respectively. The right small panels show the probability distributions of BNS mergers against the SFR of host galaxies, where black solid, dashed, and dashed-dotted lines show the results of model $\alpha10.{\rm kb}\beta0.9$, $\alpha1.0{\rm kb}\beta0.9$, and $\gamma1.3{\rm kl}\beta0.9$, respectively. Red lines show the SFR distribution of all star forming galaxies in Millennium-II, EAGLE, and Illustris-TNG, respectively.
}
\label{fig:f11}
\end{center}
\end{figure}

\begin{figure}
\begin{center}
\includegraphics[width=7cm]{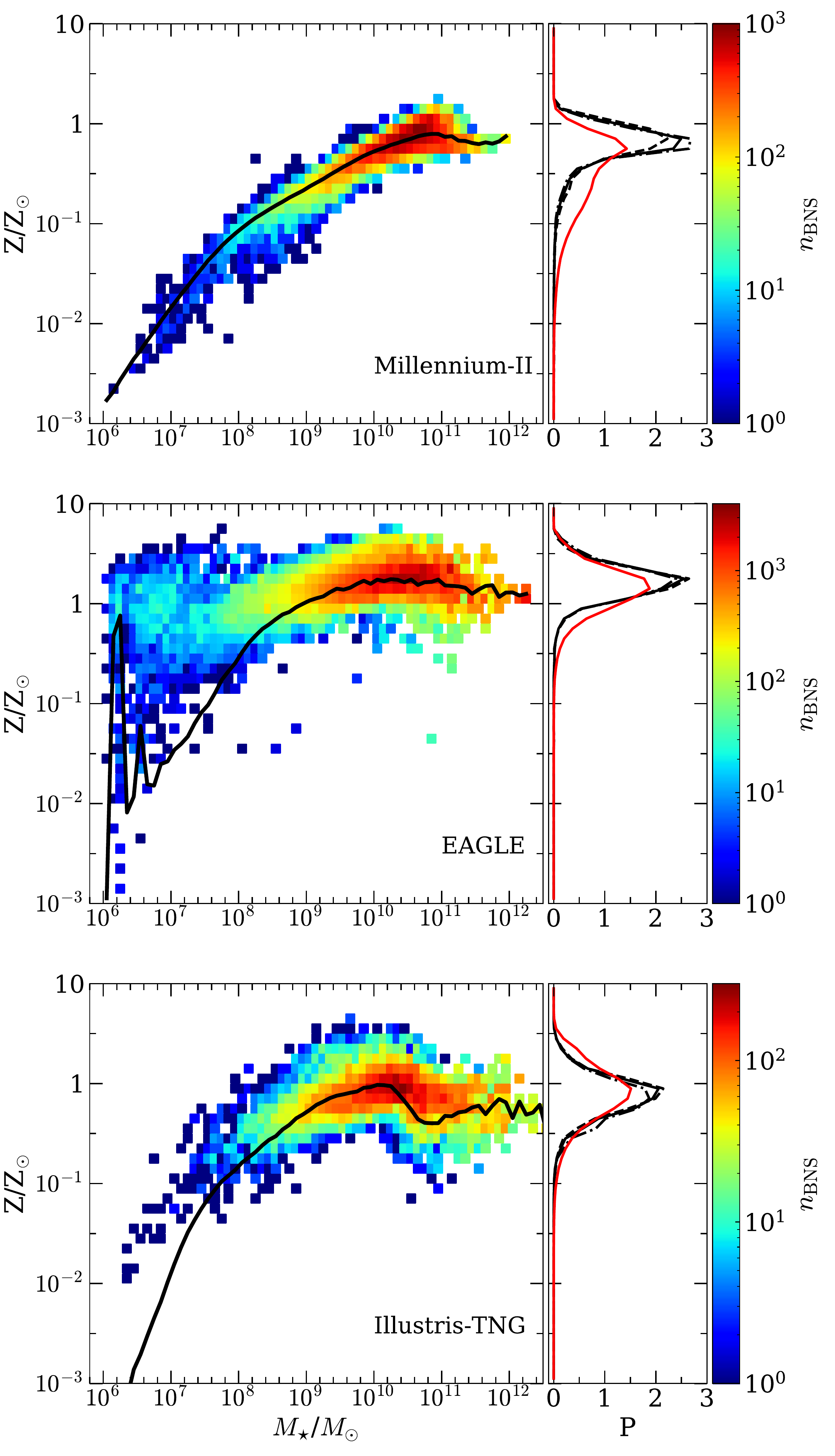}
\caption{
Legend similar to Fig.~\ref{fig:f11}, but for the distributions of BNS mergers on the metallicity versus stellar mass of host galaxies ($Z-M_\star$) plane. The black lines in left panels show the mean mass-metallicity relation at z=0 from the different simulations.
}
\label{fig:f12}
\end{center}
\end{figure}

Figure~\ref{fig:f11} shows the distributions of BNS mergers on the SFR versus stellar mass plane of their host galaxies at redshift $z<0.01$, which are obtained by implementing the reference model $\alpha10.{\rm kb}\beta0.9$ into Millennium-II (top), EAGLE (middle), and Illustris-TNG (bottom), respectively. As seen from this Figure, BNS mergers in the local Universe ($z<0.01$) are preferentially hosted in massive galaxies with SFR in the range (five-to-ninety-five percentiles) of 
$0.004-9.0 M_\odot\,\textrm{yr}^{-1}$, $0.03-7.5 M_\odot\,\textrm{yr}^{-1}$, and $0.02-9.4 M_\odot\,\textrm{yr}^{-1}$, peaking at $3.5 M_\odot\,\textrm{yr}^{-1}$, $1.8 M_\odot\,\textrm{yr}^{-1}$, and $2.8 M_\odot\,\textrm{yr}^{-1}$, for Millennium-II (top), EAGLE (middle), and Illustris-TNG (bottom), respectively. In the right marginal panels, we also show the host SFR distribution for other models and the SFR distribution of all star forming galaxies in the simulations for comparison. We find that for short time delay model (i.e., $\alpha1.0{\rm kb}\beta0.9$), the host SFR distribution is much closer to the SFR distribution of all star forming galaxies as the BNSs are almost recently formed ($\sim72.9\%$ with $t_{\rm d}<1\rm Gyr$). For long time delay model (i.e., $\gamma1.0{\rm kl}\beta0.9$), the host SFR distribution is highly related to the evolution histories of star forming galaxies at high redshift in the simulations. For intance, in Millennium-II, BNSs formed at high redshift prefer merging in galaxies with low SFR and there is an obvious offset between the host SFR distribution and the SFR distribution of all star forming galaxies, while in EAGLE, this offset is negligible.

Figure~\ref{fig:f12} shows the distributions of BNS mergers on the metallicity versus stellar mass plane of their host galaxies. As seen from this Figure, BNS mergers  at $z<0.01$ from model $\alpha10.{\rm kb}\beta0.9$ are preferentially hosted in galaxies with metallicity in the range (five-to-ninety-five percentiles) of $0.2$-$1.1 Z_\odot$, $0.8$-$3.1 Z_\odot$, and $0.3$-$1.6 Z_\odot$, for Millennium-II (top), EAGLE (middle), and Illustris-TNG (bottom), respectively. The metallicity distributions peak at $0.7 Z_\odot$, $1.8 Z_\odot$, and $0.9 Z_\odot$, respectively. Compared with the metallicity distribution of all star forming galaxies with the host metallicity distribution, there is an offset arising from the metal enrichment histories of host galaxies. 
The differences in metallicity distributions probably arise from the different models for sub-grid physics in these different galaxy formation and evolution models.
The metallicity distribution from EAGLE is obviously higher than Millennium-II and Illustris-TNG, mainly because of the high efficiency of the feed back from the star formation.

\subsubsection{BNS merger rate per galaxy versus host stellar mass, SFR, and metallicity}
 
It has been pointed out that the stellar mass $M_{\star}$ of host galaxy is an excellent tracer of BNS merger events in the local Universe, and the probability of BNS mergers is also highly correlated with the SFR and metallicity of their host galaxies \citep[e.g.,][]{Mapelli2018host1, Artale2019a, Artale2019b}. Particularly, \citet{Artale2019a} showed that the BNS merger rate per galaxy ($n_{\rm GW}$) can be well determined by the host stellar mass, SFR, and metallicity by using EAGLE simulation.  
\citet{Artale2019b} obtained such relationships by considering the relation between $n_{\rm GW}$ and $M_\star$ (one-dimensional fit; Fit 1D), between $n_{\rm GW}$ and $M_\star$-${\rm SFR}$ (two-dimensional fit; Fit 2D), or $n_{\rm GW}$ and $M_\star$-SFR-$Z$ (three dimensional fit: Fit 3D), respectively.
Following their method, here we derive the relationship between BNS merger rate per galaxy and the properties of host galaxies (Fit 1D, 2D, and 3D) for different BSE models and cosmological simulations. We also compare the linear relationship obtained from different dimensions of fitting model with each other. 

We perform a series of fits with the standard linear regression approach by using the following prescriptions.
\begin{itemize}
\item {\bf Fit 1D:} we fit the merger rate per galaxy ($n_{\rm GW}$) as a function of stellar mass ($M_{\star}$) by
\begin{equation}
\log \left( \frac{n_{\rm GW}}{{\rm Gyr}} \right) = a_{1}\log \left( \frac{M_{\star}}{M_\odot} \right) + a_{2}.
\label{eq:fit1d}
\end{equation}
\item {\bf Fit 2D:} we fit the merger rate per galaxy ($n_{\rm GW}$) as a function of both stellar mass ($M_{\star}$) and specific star formation rate (${\rm sSFR}={\rm SFR}/M_{\star}$), as
\begin{equation}
\log \left( \frac{n_{\rm GW}}{{\rm Gyr}} \right) = b_{1}\log \left( \frac{M_{\star}}{M_\odot} \right) + b_{2}\log \left( \frac{\rm sSFR}{{\rm yr}^{-1}} \right) + b_{3}.
\label{eq:fit2d}
\end{equation}
\item {\bf Fit 3D:} we fit the merger rate per galaxy ($n_{\rm GW}$) as a function of stellar mass ($M_{\star}$), sSFR and metallicity $Z$, as
\begin{equation}
\begin{split}
\log \left( \frac{n_{\rm GW}}{{\rm Gyr}} \right) = c_{1}\log \left( \frac{M_{\star}}{M_\odot} \right) + c_{2}\log \left( \frac{\rm sSFR}{{\rm yr}^{-1}} \right)\\
 + c_{3}\log \left( \frac{Z}{Z_\odot} \right) + c_{4}.
\end{split}
\label{eq:fit3d}
\end{equation}

\end{itemize}

\begin{table*}
\begin{center}
\caption{Best-fits of the BNS merger rate per galaxy as functions of host galaxy properties obtained from Millennium-II, EAGLE, and Illustris-TNG at redshift $z=0$, 1 and 2, respectively.}
\label{tab:t5}
\setlength{\tabcolsep}{1.2mm}{
\begin{tabular}{c|c|c|c|c|c} \hline
    					&		  &  Millennium-II  & EAGLE & Illustris-TNG &\\  \hline
\hline
\multicolumn{6}{c}{z=0}\\
\hline
\multirow{2}{*}{Fit 1D} & $a_{1}$ & 0.801(0.757/0.876) & 0.691(0.565/0.663) & 0.719(0.707/0.891) & 0.756(0.697/0.832) \\		
			        	& $a_{2}$ & -4.28(-4.53/-5.23) & -2.92(-2.83/-3.85) & -3.19(-3.76/-5.31) & -3.69(-3.90/-4.94) \\  \hline
\multirow{3}{*}{Fit 2D} & $b_{1}$ & 0.883(0.856/0.908) & 0.716(0.611/0.679) & 0.774(0.749/0.902) & 0.809(0.759/0.850) \\		
						& $b_{2}$ & 0.428(0.361/0.165) & 0.377(0.323/0.193) & 0.373(0.286/0.075) & 0.401(0.323/0.128) \\  
						& $b_{3}$ & 0.017(-1.19/-3.57) & 1.03(0.318/-1.86) & 0.416(-0.987/-4.58) & 0.395(-0.786/-3.64) \\  \hline	
\multirow{4}{*}{Fit 3D} & $c_{1}$ & 1.050(1.066/1.108) & 0.916(0.863/0.876) & 0.948(0.922/0.962) & 0.900(0.888/0.967)  \\		
						& $c_{2}$ & 0.410(0.334/0.143) & 0.572(0.567/0.383) & 0.522(0.434/0.126) & 0.439(0.379/0.178) \\  
						& $c_{3}$ & -0.336(-0.445/-0.406) & -0.469(-0.585/-0.459) & -0.436(-0.432/-0.149) & -0.199(-0.287/-0.257) \\ 
						& $c_{4}$ & -1.95(-3.73/-5.94) & 1.21(0.548/-1.69) & 0.205(-1.20/-4.65) & -0.118(-1.51/-4.30) \\  \hline	
\hline
\multicolumn{6}{c}{z=1}\\
\hline
\multirow{2}{*}{Fit 1D} & $a_{1}$ & 0.852(0.797/0.909) & 0.682(0.610/0.649) & 0.788(0.764/0.879) & 0.769(0.726/0.840) \\		
			        	& $a_{2}$ & -4.45(-4.67/-5.20) & -2.28(-2.65/-3.40) & -3.34(-3.89/-4.86) & -3.38(-3.79/-4.68) \\  \hline
\multirow{3}{*}{Fit 2D} & $b_{1}$ & 0.907(0.857/0.927) & 0.712(0.665/0.670) & 0.787(0.764/0.878) & 0.808(0.772/0.851) \\		
						& $b_{2}$ & 0.387(0.321/0.121) & 0.388(0.411/0.216) & 0.368(0.311/0.060) & 0.394(0.347/0.108) \\  
						& $b_{3}$ & -0.470(-1.50/-3.96) & 1.44(1.06/-1.37) & 0.516(-0.623/-4.23) & 0.573(-0.419/-3.60) \\  \hline	
\multirow{4}{*}{Fit 3D} & $c_{1}$ & 1.10(1.05/1.08) & 0.921(0.899/0.868) & 0.994(0.974/0.964) & 0.896(0.857/0.928)  \\		
						& $c_{2}$ & 0.356(0.287/0.097) & 0.601(0.649/0.416) & 0.529(0.476/0.127) & 0.435(0.387/0.143) \\  
						& $c_{3}$ & -0.446(-0.454/-0.342) & -0.582(-0.646/-0.543) & -0.540(-0.550/-0.225) & -0.220(-0.213/-0.190) \\ 
						& $c_{4}$ & -2.93(-3.98/-5.84) & 1.57(1.20/-1.27) & -0.017(-1.17/-4.45) & -0.092(-0.882/-4.02) \\  \hline	
\hline
\multicolumn{6}{c}{z=2}\\
\hline
\multirow{2}{*}{Fit 1D} & $a_{1}$ & 0.850(0.799/0.888) & 0.771(0.707/0.678) & 0.896(0.865/0.889) & 0.812(0.776/0.847) \\		
			        	& $a_{2}$ & -4.13(-4.44/-4.77) & -2.77(-3.21/-3.44) & -3.99(-4.51/-4.75) & -3.45(-3.96/-4.52) \\  \hline
\multirow{3}{*}{Fit 2D} & $b_{1}$ & 0.908(0.854/0.902) & 0.743(0.713/0.668) & 0.828(0.806/0.879) & 0.817(0.791/0.850) \\		
						& $b_{2}$ & 0.391(0.339/0.094) & 0.389(0.427/0.224) & 0.343(0.301/0.052) & 0.386(0.360/0.103) \\  
						& $b_{3}$ & -0.306(-1.18/-3.86) & 1.33(0.961/-1.14) & 0.075(-0.934/-4.12) & 0.560(-0.322/-3.46) \\  \hline	
\multirow{4}{*}{Fit 3D} & $c_{1}$ & 1.20(1.13/1.05) & 0.961(0.944/0.896) & 1.04(1.02/0.968) & 0.948(0.910/0.925)  \\		
						& $c_{2}$ & 0.337(0.284/0.066) & 0.588(0.638/0.430) & 0.514(0.476/0.126) & 0.443(0.411/0.135) \\  
						& $c_{3}$ & -0.680(-0.663/-0.347) & -0.624(-0.659/-0.647) & -0.601(-0.615/-0.257) & -0.330(-0.301/-0.189) \\ 
						& $c_{4}$ & -3.96(-4.73/-5.72) & 1.07(0.680/-1.43) & -0.519(-1.54/-4.38) & -0.231(-1.04/-3.91) \\  \hline	
\end{tabular}}
\end{center}
\begin{flushleft}
	\footnotesize{Column 3-5: fitting parameters of model $\alpha10.{\rm kb}\beta0.9$ resulting from Millennium-II, EAGLE, and Illustis-TNG, respectively; column 6: fitting parameters model $\alpha10.{\rm kb}\beta0.9$ by combining three cosmological models together. The values in brackets are the results of model $\alpha1.0{\rm kb}\beta0.9$/$\gamma1.3{\rm kl}\beta0.9$, respectively.
	}
\end{flushleft}
\end{table*}

\begin{figure*}
\begin{center}
\includegraphics[width=16cm]{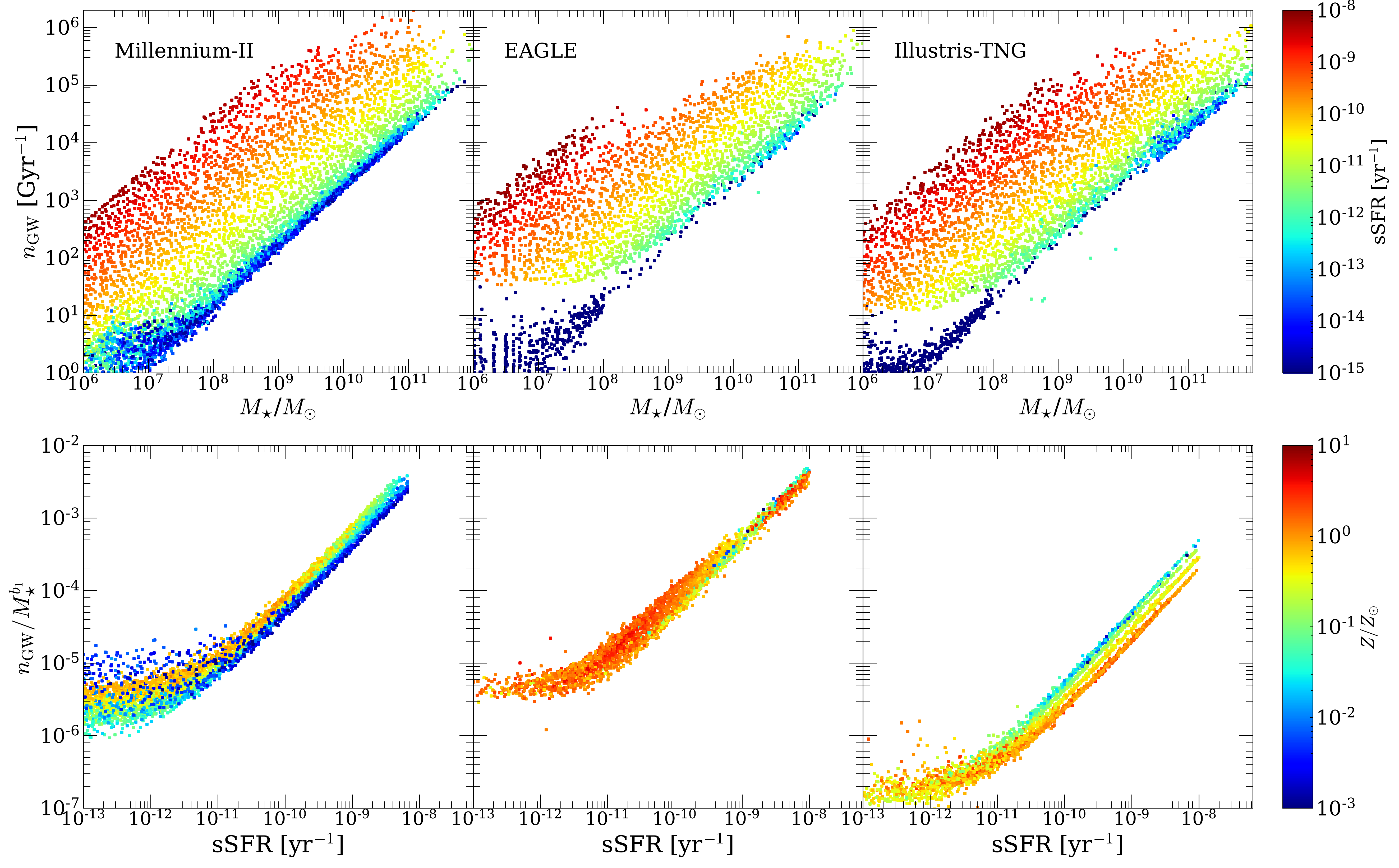}
\caption{BNS merger rate per galaxy versus stellar mass (top panels) and sSFR (bottom panels) of host galaxies at redshift $z=0$ resulting from the preferred BSE model $\alpha10.{\rm kb}\beta0.9$. Left, middle, and right panels show the results obtained by using Millennium-II, EAGLE, and Illustris-TNG, respectively.The color bars at the right side represent the sSFR and metallicity of host galaxies.
}
\label{fig:f13}
\end{center}
\end{figure*}

The best-fit values for the coefficients shown in the above Equations~\eqref{eq:fit1d}-\eqref{eq:fit3d} are listed in Table~\ref{tab:t5}.
Apparently, the slope of the 1D fit (Eq.~\ref{eq:fit1d}) becomes larger if a BSE model results in relatively longer time delays for BNS mergers than another BSE model does. This is simply because more BNSs can merge in massive quiescent galaxies while less BNSs can merge in small star-forming galaxies at $z=0$ with this BSE model.
In addition, we make a simple comparison between different models for the fit (i.e., Fit 1D, Fit 2D, and Fit 3D) by calculating the Bayesian Information Criterion \citep[BIC;][]{Schwarz1978}. We find that ${\rm BIC}_{\rm 1D}>{\rm BIC}_{\rm 2D}>{\rm BIC}_{\rm 3D}$, which indicates that multi-dimensional fitting is preferred comparing with Fit 2D and Fit 1D. 
We also find that ${\rm BIC}_{\rm 1D}>{\rm BIC}_{\rm 2D}$ (${\rm BIC}_{\rm 1D}\sim 3{\rm BIC}_{\rm 2D}$) while ${\rm BIC}_{\rm 2D}\simeq{\rm BIC}_{\rm 3D}$ (${\rm BIC}_{\rm 2D}\sim \ 0.91-0.99\ {\rm BIC}_{\rm 3D}$), which shows that BNS merger rate per galaxy is not so sensitive to metallicity.
According to these fittings, the stellar mass and sSFR are supposed to be given top priority in the search of host galaxy candidates for BNS mergers.

Figure~\ref{fig:f13} illustrates the BNS merger rate per galaxy as a function of host galaxy stellar mass and sSFR at the redshift $z=0$ resulting by implementing the reference model $\alpha10.{\rm kb}\beta0.9$ into Millennium-II (left), EAGLE (middle), and Illustris-TNG (right), respectively. As seen from the top panels, $n_{\rm GW}$ is strongly correlated with the host galaxy mass $M_{\star}$, i.e., $n_{\rm GW} \propto M_\star^{a_1}$ with $a_1\sim 0.7-0.8$. After removing the dependence on $M_{\star}$, i.e., as clearly seen from the bottom panels of this Figure, the BNS merger rate per galaxy also strongly correlated with the sSFRs of host galaxies, i.e., $n_{\rm GW} \propto {\rm sSFR}^{b_2}$ with $b_2\sim 0.37-0.43$, though the sSFR plays a secondary role comparing with $M_\star$. 
We can also see that $n_{\rm GW}/M^{b_1}_*$ correlates with sSFR tightly when the sSFR $\gtrsim 10^{-11}-10^{-10}$\,yr$^{-1}$, while such a correlation becomes weak when sSFR $\leq 10^{-11}$. The main reason is that the BNS mergers in those galaxies with high sSFRs are dominated by those with short time delays thus strongly correlate with the contemporary sSFR, while the BNS mergers in galaxies with low sSFRs are dominated by those with long time delays and thus almost independent of the contemporary sSFR. The information on both $M_\star$ and sSFR may be helpful to identify the most likely host galaxies of a given BNS merger in the sky area constrained by the GW detection. Furthermore, statistical information obtained by future multi-messenger observations of BNS mergers may provide significant information on such a relationship, which may be used to constrain BNS formation model parameters.  
If combining different cosmological models together and taking them with equal weight, then we may also obtain the best fits of model $\alpha10.{\rm kb}\beta0.9$ with $(a_1, a_2)=(0.756, -3.690)$ for Fit 1D, $(b_1, b_2, b_3)=(0.809, 0.401, 0.395)$ for Fit 2D, and $(c_1, c_2, c_3, c_4)=(0.900, 0.439, -0.199, -0.118)$ for Fit 3D, respectively. 

We also fit the BNS merger rate per galaxy at high redshifts (e.g., $z=1$ and $z=2$). For Fit 1D, we find that $a_{1}$ slightly increases with  increasing redshift. For Fit 3D, we find that $c_{3}$ increases with increasing redshift, which means  the metallicity may have relatively larger effect at high redshift than that at low redshift.

\subsection{BNS host galaxies across cosmic time}

In the above analysis, we focus on the host galaxies of BNSs mergers in the local Universe ($z<0.01$), and we extend the analysis to BNSs mergers across cosmic time in this Section.

\begin{figure*}
\begin{center}
\includegraphics[width=17cm]{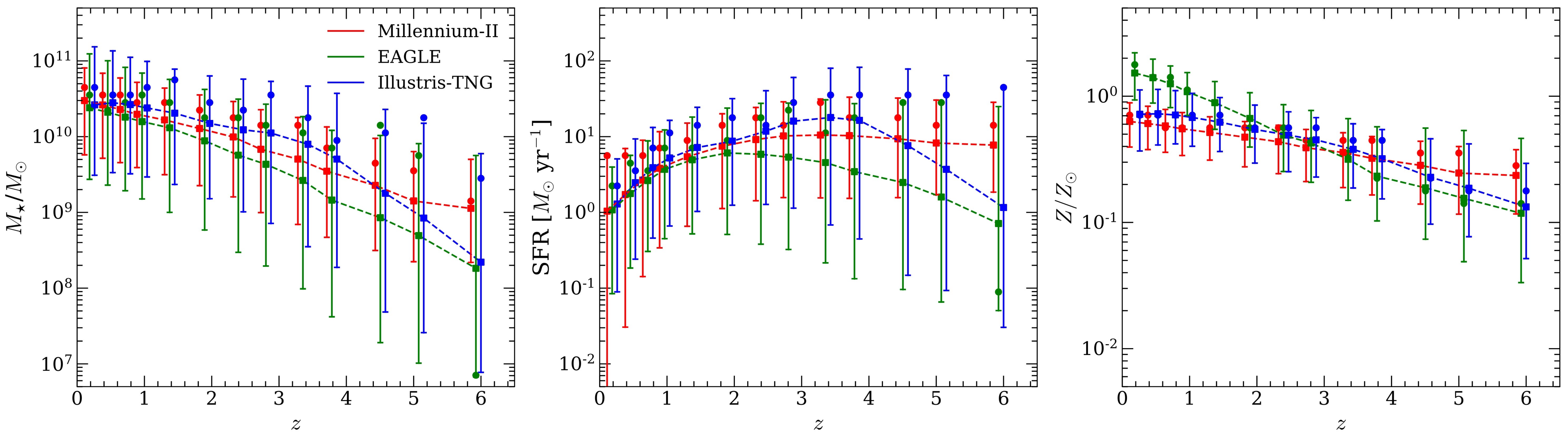}
\caption{
Peaks (circles) and medians (squares) of the stellar mass (left), SFR (middle), and metallicity (right) distributions of the host galaxies of BNS mergers as a function of redshift obtained by implementing the model $\alpha10.{\rm kb}\beta0.9$ into into Millennium, EAGLE, and Illustris-TNG, respectively. 
Vertical bars indicate the probability range of $16$-$84$ percentiles of host galaxy distributions. Different colors represent the results obtained from different galaxy formation and evolution models. Note that a small horizontal offset is added to each point obtained from EAGLE and Illustris-TNG, for clarity.
}
\label{fig:f14}
\end{center}
\end{figure*}

Figure~\ref{fig:f14} shows the peaks (circle) and medians (squares) of the host galaxy stellar mass (left), SFR (middle) and metallicity (right) distributions of BNS mergers at different GW detection time obtained from the model $\alpha10.{\rm kb}\beta0.9$. Apparently, the range of host stellar mass drops with increasing redshift because after the formation of BNSs, the host galaxies keep growing up and transforming into more massive systems in their subsequent evolution. Several peak points are out of the $16$-$84$ percentile ranges because a large fraction of BNS mergers occur in a small number of extremely massive galaxies in certain snapshots. The SFR of host galaxies increases first and then decreases while the metallicity has a decreasing tendency with increasing redshift, which well reflect the evolution of galaxies in adopted cosmological simulations. 

\begin{figure*}
\begin{center}
\includegraphics[width=13cm]{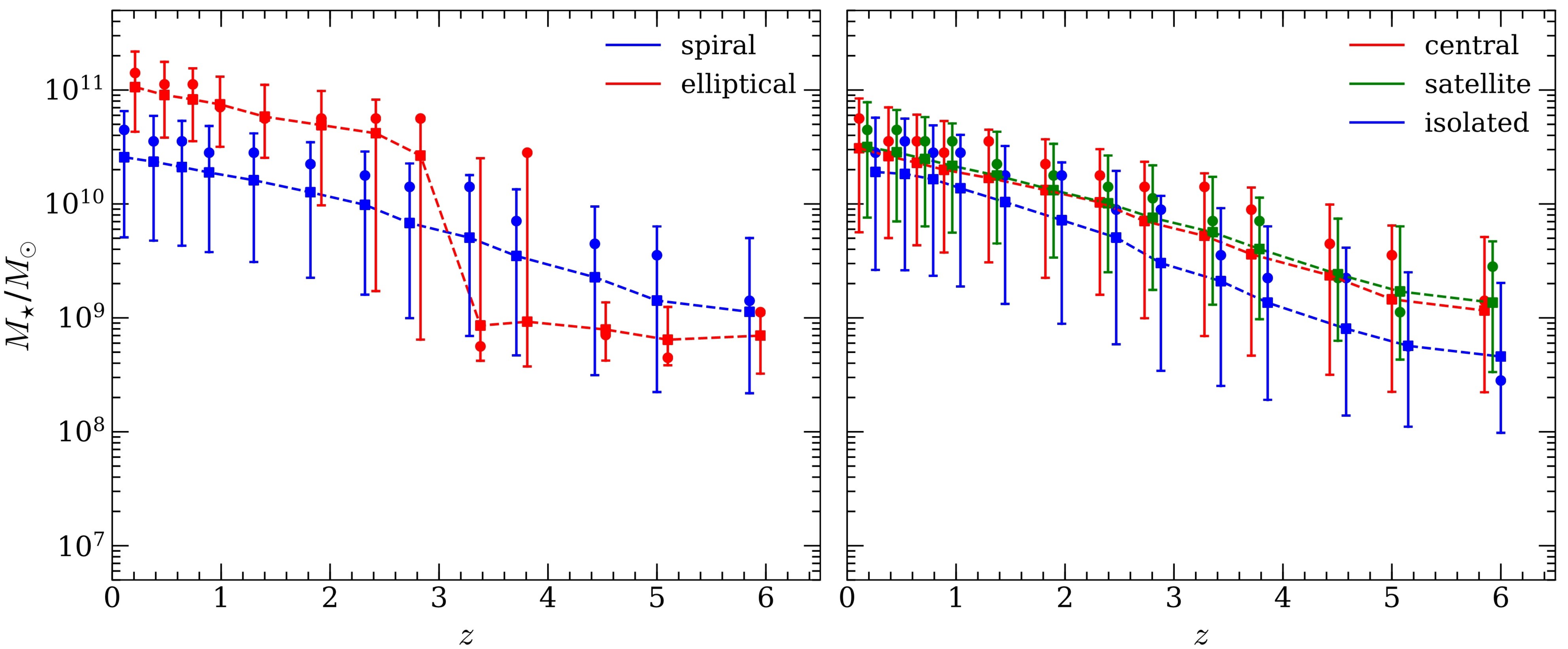}
\caption{
Peaks (circles) and medians (squares) of the stellar mass distributions for the host galaxies of BNS mergers with different morphologies and types as a function of redshift obtained by implementing the reference model $\alpha10.{\rm kb}\beta0.9$ into Millennium-II. 
Vertical bar associated with each point indicates the probability range of $16$-$84$ percentiles of host galaxies. Note that a small horizontal offset is added to each point for elliptical, satellite, and isolated host galaxies, for clarity.
}
\label{fig:f15}
\end{center}
\end{figure*}

Figure~\ref{fig:f15} shows the peaks and medians of the host galaxy stellar mass distributions of BNS mergers with different morphologies and types at different GW occurrence time obtained from the model $\alpha10.{\rm kb}\beta0.9$ by using Millennium-II. Apparently, the elliptical host galaxies are small with low stellar masses at high redshift (e.g., $z\gtrsim 4$), the upper limit of the stellar mass distribution of elliptical host galaxies increases rapidly at  redshift $2\lesssim z \lesssim 3$. During this period, elliptical galaxies grow up quickly by merging with other galaxies and some spiral galaxies may merge with each other and be transformed into large elliptical galaxies. Thus, the stellar mass distribution of elliptical host galaxy covers a considerable wide range of masses from low to high at $z\sim 2-3$. At low redshift (e.g., $z<1.5$), the elliptical host galaxies are normally more massive than others and the probability for detection of BNS mergers in small elliptical galaxies is substantially smaller. The BNS merger rate in a large elliptical host galaxy can be higher by three orders of magnitudes than that in a small elliptical host galaxy,
and the small elliptical galaxies are out of $68\%$ credible limit of elliptical host stellar mass distribution.

\begin{table*}
\begin{center}
\caption{The probability of different grouped host galaxies across the cosmic time, obtained from the model $\alpha10.{\rm kb}\beta0.9$+Millennium-II.}
\label{tab:t6}
\setlength{\tabcolsep}{3.5mm}{
\begin{tabular}{cccccccc}
\hline
Redshift & Elliptical & Spiral & Central & Satellite & Isolated & Star-forming & Quiescent \\ \hline
0        & 15.6\%  & 84.4\% &  69.4\% & 22.4\%    &  8.2\%   &   20.0\%     &80.0\% \\
0.11     & 12.8\%  & 87.2\% &  70.1\% & 22.2\%    &  7.7\%   &   32.5\%     &67.5\% \\
0.38     & 9.4\%   & 90.6\% &  70.7\% & 21.6\%    &  7.7\%   &   55.4\%     &44.6\% \\
0.64     & 6.3\%   & 93.7\% &  71.1\% & 21.2\%    &  7.7\%   &   68.3\%     &31.7\% \\
0.89     & 4.1\%   & 95.9\% &  72.6\% & 20.2\%    &  7.2\%   &   77.3\%     &22.7\% \\
1.30     & 2.5\%   & 97.5\% &  75.1\% & 18.0\%    &  6.9\%   &   85.4\%     &14.6\% \\
1.82     & 1.2\%   & 98.8\% &  77.5\% & 15.7\%    &  6.8\%   &   92.4\%     &7.6\% \\
2.32     & 0.6\%   & 99.4\% &  79.9\% & 13.5\%    &  6.6\%   &   96.0\%     &4.0\%  \\
2.73     & 0.3\%   & 99.7\% &  82.8\% & 11.1\%    &  6.1\%   &   98.4\%     &1.6\%  \\
3.28     & 0.2\%   & 99.8\% &  84.5\% &  9.8\%    &  5.7\%   &   99.3\%     &0.7\%  \\
3.71     & 0.1\%   & 99.9\% &  87.2\% &  7.9\%    &  4.9\%   &   99.7\%     &0.3\%  \\
4.43     & 0.1\%   & 99.9\% &  90.2\% &  5.8\%    &  4.0\%   &   99.8\%     &0.2\%  \\
5.00     & 0.1\%   & 99.9\% &  91.8\% &  4.8\%    &  3.4\%   &   99.9\%     &0.1\%  \\
5.85     & 0.1\%   & 99.9\% &  92.7\% &  4.0\%    &  3.3\%   &   99.9\%     &0.1\%  \\ \hline
\end{tabular}}
\end{center}
\begin{flushleft}
{\small Column 1: redshift; columns 2 and 3: the probability of host galaxies with different morphologies, including elliptical and spirals; columns 4, 5, and 6: the probability of host galaxies with different types, including central, satellite, and isolated galaxies; columns 7 and 8: the probability of host galaxies with different sSFR, including star-forming and quiescent galaxies.}
\end{flushleft}
\end{table*}

Table~\ref{tab:t6} lists the probability of different grouped host galaxies of BNS mergers across cosmic time as supplements, obtained from the reference model in Millennium-II. At high redshift, BNSs mostly merger in galaxies where they were formed in due to short time delay, such as spiral or central galaxies with high SFR. At low redshift, BNS mergers are comprised of those with both short and long delay times. The former ones prefer to be hosted in galaxies with high SFR, which means that spiral and central galaxies take important parts of the hosts at low redshift. 
The host galaxies of BNS mergers with large $t_{\rm d}$ evolve significantly along with their assembly histories. Some host galaxies that are initially spirals when BNSs were formed may transform into ellipticals, some hosts that are initially central galaxies when the BNSs were formed may change into satellites or isolated ones. This is the reason that the fractions of elliptical and satellite/isolated hosts increase with decreasing redshift. 

\citet{Artale2019a, Artale2019b} simply assumed galaxies with ${\rm sSFR}<10^{-10} {\rm yr}^{-1}$ and ${\rm sSFR}\geq10^{-10} {\rm yr}^{-1}$ as early- and late-type galaxies in their calculation. While from the morphology classification in Millennium-II and Illustris-TNG, the early-type galaxies have a sSFR about $5.75\times10^{-12} {\rm yr}^{-1}$ and $4.78\times10^{-12} {\rm yr}^{-1}$ on average. The mean sSFR of late-type galaxies is about $3.78\times10^{-11} {\rm yr}^{-1}$ and $5.68\times10^{-11} {\rm yr}^{-1}$. To compare the results with each other, we here label galaxies with ${\rm sSFR}\geq10^{-10} {\rm yr}^{-1}$ and ${\rm sSFR}<10^{-10} {\rm yr}^{-1}$ as star-forming  and quiescent galaxies, respectively. Our results are consistent with those in \citet{Artale2019b} obtained from EAGLE, in which quiescent galaxies contribute about $80\%$ of BNS mergers and star-forming galaxies only contribute $20\%$ at $z<0.1$. The fraction of quiescent galaxies has a decreasing tendency while star-forming galaxies increases with increasing redshift and the trend reverts at redshift $z\sim0.38$.

\subsection{GW170817 and its host galaxy NGC 4993}
\label{subsec:NGC4993}

Since GW170817 is the only detected BNS coalescence with both GW signal and EM counterpart, it is highly significant to study the environment of this BNS merger event, which may provide a deep insight into the formation of BNSs. The inferred host galaxy of GW170817, NGC4993, is an elliptical with stellar mass $\sim 0.3-1.2\times 10^{11} M_\odot$ and low SFR ($0.004\ M_\odot{\rm yr}^{-1} $), and its stellar population is old ($\geq3$\,Gyr) \citep{Im2017}. These values are compatible with that predicted from our reference model with either Millennium-II or Illustris-TNG (see Fig.~\ref{fig:f8}). NGC4993 is known as an early-type galaxy in the ESO508 cluster \citep{Garcia1993}, which agrees with the predicted small fraction of isolated host galaxies for BNS mergers at  $z<0.01$. Hence, NGC4993 can be marked as a typical elliptical host galaxy of BNS mergers.

\begin{figure}
\begin{center}
\includegraphics[width=7cm]{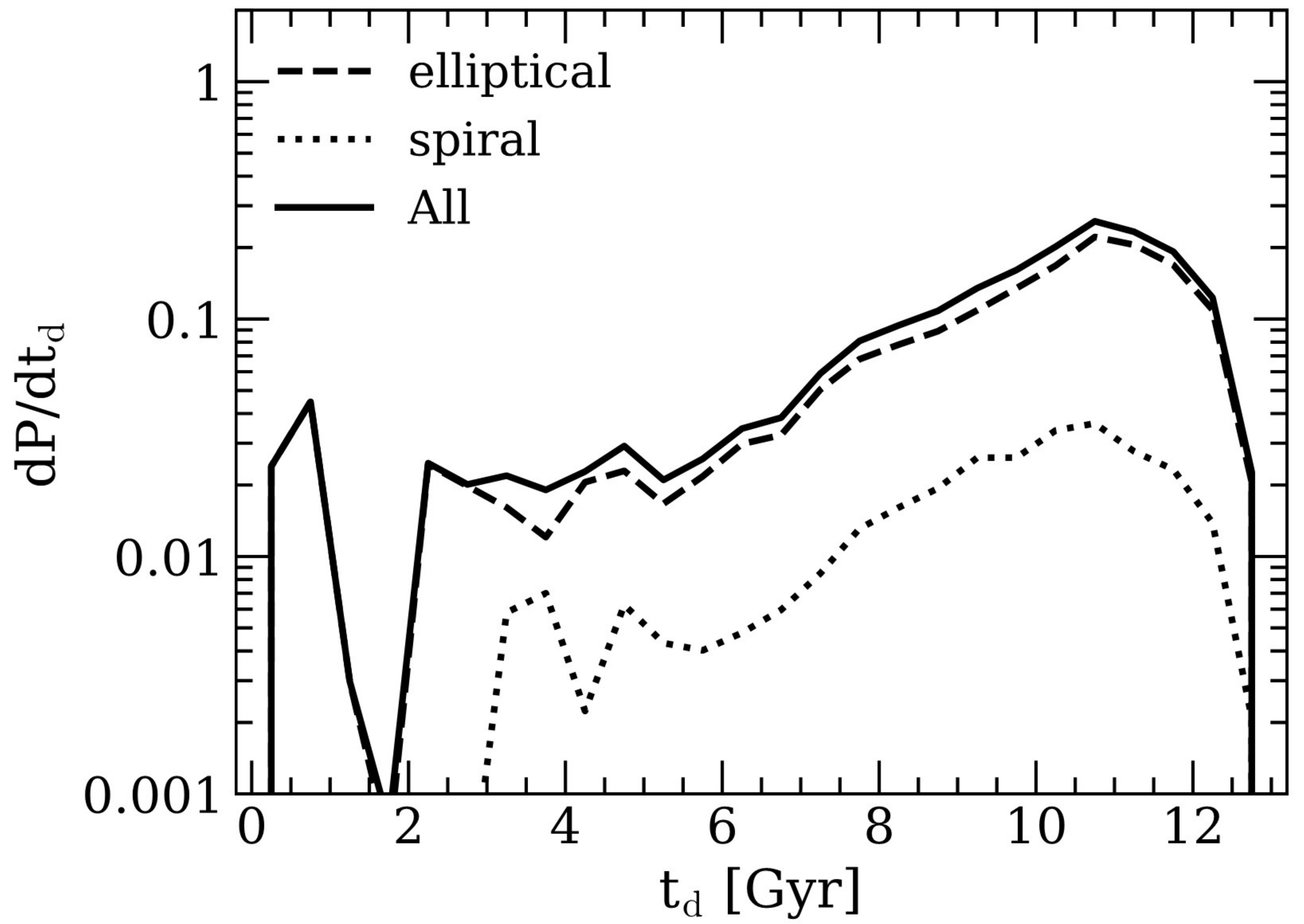}
\caption{
Distribution of delay times $t_{\rm d}$ for BNS mergers in NGC 4993-like galaxies at redshift $z<0.01$. Solid, dash, and dot lines represent that for all BNS mergers in such galaxies, those for BNS mergers from the primary ellipticals and from the recently swallowed secondary spirals, respectively.}
\label{fig:f16}
\end{center}
\end{figure}

Interestingly, NGC 4993 is also known as a shell galaxy MC 1307-231, indicating that a small late-type galaxy was swallowed in recently \citep{Malin1983}. It is possible that this small galaxy could be the original host of GW170817. According to \citet{Ebrova2018}, the event occurred at least $200$\,Myr ago with a probable time roughly around $400$\,Myr, and the estimate closer to $1$\,Gyr and higher being improbable. Using Millennium-II, we search for mock early type galaxies similar to NGC 4993, which have (1) stellar mass in the range of $0.3-1.2\times 10^{11} M_\odot$ (1226 candidates); (2) undergone a merger at the look back time from $263$\,Myr to $827$\,Myr (197 candidates remained); (3) the mass ratio of secondary to primary progenitor galaxy in the range of $0.1$-$0.25$ (13 candidates remained); (4) an elliptical primary galaxy and a spiral secondary (only 11 eligible galaxies remained). Figure~\ref{fig:f16} shows the distribution of delay times $t_{\rm d}$ for BNS mergers in NGC 4993-like galaxies at redshift $z<0.01$. We find that the probability of BNS mergers at $z<0.01$ coming from the swallowed late-type galaxy is only $\sim 14.6\%$, and in most cases ($\sim 85.4\%$) BNS mergers should come from the primary early-type progenitor due to its much larger stellar mass. For BNS merger events similar to GW170817 hosted in NGC4993-like galaxies, the probability distribution of $t_{\rm d}$ peaks at $10.5-11$\,Gyr and its five-to-ninety-five percentile range is $2.5$-$12$\,Gyr.
There is a sharp feature at $t_d\sim1-2 \rm Gyr$ caused by the special star formation histories of the selected NGC4993-like galaxies. Because of the minor merger happened recently, the star formation is triggered and there is a small probability for BNSs formed recently and merged at $z\sim0$. In contrast, if an elliptical galaxy does not undergo a merger process at low redshift, BNSs are all formed at higher redshift and there would be no such a feature shown in Figure~\ref{fig:f16}. Nevertheless, for NGC4993-like galaxies, BNS mergers are still mostly having a large delay time despite of some recent star formation triggered by the minor merger.

\subsection{Host galaxies of BNS mergers, sGRBs, and kilonovae}
\label{subsec:sGRBs}

\begin{table*}
	\begin{center}
		\caption{Host galaxies of sGRBs and kilonovae.}
		\label{tab:t8}
		\setlength{\tabcolsep}{1mm}{
			\begin{tabular}{c|c|c|c|c|c|c}
				\hline
				Event & Redshift & Morphology  & Stellar mass  & Stellar population age & SFR & Reference\\
				&          &             & $\log(M_{\star}/M_\odot)$ & $\tau _{\star}$ [Gyr]  & [$M_\odot\ yr^{-1}$] &\\
				\hline
				\multicolumn{7}{c}{sGRB host galaxies}\\
				\hline
				GRB 050509B & 0.225 & elliptical & 10.88& 3.18 & <0.15& \citet{Berger2014}, \citet{Nugent2020}  \\
				GRB 050709  & 0.161 & spiral     & 8.8  & 0.26 & 0.15 & \citet{Berger2014} \\
				GRB 050724A & 0.257 & elliptical & 11.09& 0.94 & <0.1 & \citet{Berger2014}, \citet{Nugent2020} \\
				GRB 050813  & 0.716 & elliptical & 10.22& 2.3  & -    & \citet{Berger2014}, \citet{Nugent2020} \\
				GRB 051210  & 1.34  & -          & 9.07 & 0.67 & -    & \citet{Nugent2020} \\
				GRB 051221A & 0.546 & spiral     & 9.4  & 0.17 & 0.95 & \citet{Berger2014} \\
				GRB 060502B & 0.287 & elliptical & 11.8 & 1.3  & 0.8  & \citet{Berger2014} \\
				GRB 060801  & 1.130 & spiral     & 9.1  & 0.03 & 6.1  & \citet{Berger2014} \\
				GRB 061006  & 0.438 & spiral     & 9.0  & 0.24 & 0.24 & \citet{Berger2014} \\
				GRB 061201  & 0.111 & host-less or spiral & - & - & - & \citet{Berger2014} \\
				GRB 061210  & 0.409 & spiral     & 9.6  & 0.38 & 1.2  & \citet{Berger2014} \\
				GRB 061217  & 0.827 & spiral     & 9.1  & 0.03 & 2.5  & \citet{Berger2014} \\
				GRB 070429B & 0.902 & spiral     & 10.4 & 0.46 & 1.1  & \citet{Berger2014} \\
				GRB 070714B & 0.922 & spiral     & 9.4  & 0.22 & 0.44 & \citet{Berger2014} \\
				GRB 070724A & 0.457 & spiral     & 10.1 & 0.30 & 2.5  & \citet{Berger2014} \\
				GRB 070729  & 0.8   & elliptical & 10.6 & 0.98 & <1.5 & \citet{Berger2014} \\				
				GRB 071227  & 0.381 & spiral     & 10.4 & 0.49 & 0.6  & \citet{Berger2014} \\
				GRB 080123  & 0.495 & spiral     & 10.1 & 0.31 & -    & \citet{Berger2014} \\
				GRB 080905A & 0.122 & spiral     & -    & -    & -    & \citet{Berger2014} \\
				GRB 090426A & 2.609 & spiral     & -    & -    & -    & \citet{Berger2014} \\
				GRB 090510  & 0.903 & spiral     & 9.7  & 0.14 & 0.3  & \citet{Berger2014} \\
				GRB 090515  & 0.403 & elliptical & 10.87& 4.35 & 0.1  & \citet{Berger2014}, \citet{Nugent2020} \\
				GRB 100117A & 0.915 & elliptical & 10.3 & 0.79 & <0.2 & \citet{Berger2014} \\
				GRB 100206A & 0.407 & spiral     & 10.94& 0.10 & 14$\pm$2  & \citet{Berger2014}, \citet{Chrimes2018} \\
				GRB 100625A & 0.452 & elliptical & 10.3 & 0.79 & 0.3  & \citet{Berger2014} \\
				GRB 101219A & 0.718 & spiral     & 9.2  & 0.03 & 16   & \citet{Berger2014} \\
				GRB 111117A & 1.2   & spiral     & 9.6  & 0.09 & 6.0  & \citet{Berger2014} \\
				GRB 120804A & 1.3   & spiral     & 10.8 & 0.13 & 8.0  & \citet{Berger2014} \\	
				GRB 141212A & 0.596 & elliptical & $10.15^{+0.20}_{-0.19}$ & - & 0.65$\pm$0.4  & \citet{Chrimes2018}, \citet{Pandey2019} \\
				GRB 150120A & 0.46  & -          & $10.75^{+0.40}_{-0.20}$ & - & $0.71^{+2.11}_{-0.08}$  & \citet{Chrimes2018} \\
				GRB 161104A & 0.793 & elliptical & 10.21 & 2.12 & 0.099  & \citet{Nugent2020} \\
				GRB 181123B & 1.754 & spiral     & $10.24^{+0.14}_{-0.16}$ &  0.9 & $32.82^{+16.34}_{-7.24}$  & \citet{Paterson2020} \\
				\hline
				\multicolumn{7}{c}{sGRB and kilonova host galaxies}\\
				\hline
				GRB 130603B &0.3568  &spiral     & 9.7         & -           & 1.84         & \citet{Cucchiara2013}  \\
				GRB 150101B & 0.1343 &elliptical & 10.68-10.92 & 2-2.5       &$\lesssim0.4$ & \citet{Fong2016} \\
				GRB 160821B & 0.1613 & spiral    & 8.5         & -           &$\gtrsim1.5$  & \citet{Troja2019}   \\
				GRB 170817A &0.02    &elliptical & 10.48-11.08 & $\gtrsim3$  & 0.004        & \citet{Im2017} \\
				GRB 200522A &0.5536  &spiral     & 9.656       & 0.531       & 2.1-4.8      & \citet{Fong2020} \\
				\multirow{2}{*}{GRB 070809} & 0.473 &  elliptical & 11.4 & 3.10 & <0.1 & \citet{Berger2014}, \citet{Jin2020} \\
				& 0.219 &  spiral     & 10.3 & -    &  -   & \citet{Perley2008}, \citet{Jin2020} \\
				\hline
		\end{tabular}}
	\end{center}
\end{table*}

\begin{figure*}
	\begin{center}
		\includegraphics[width=17cm]{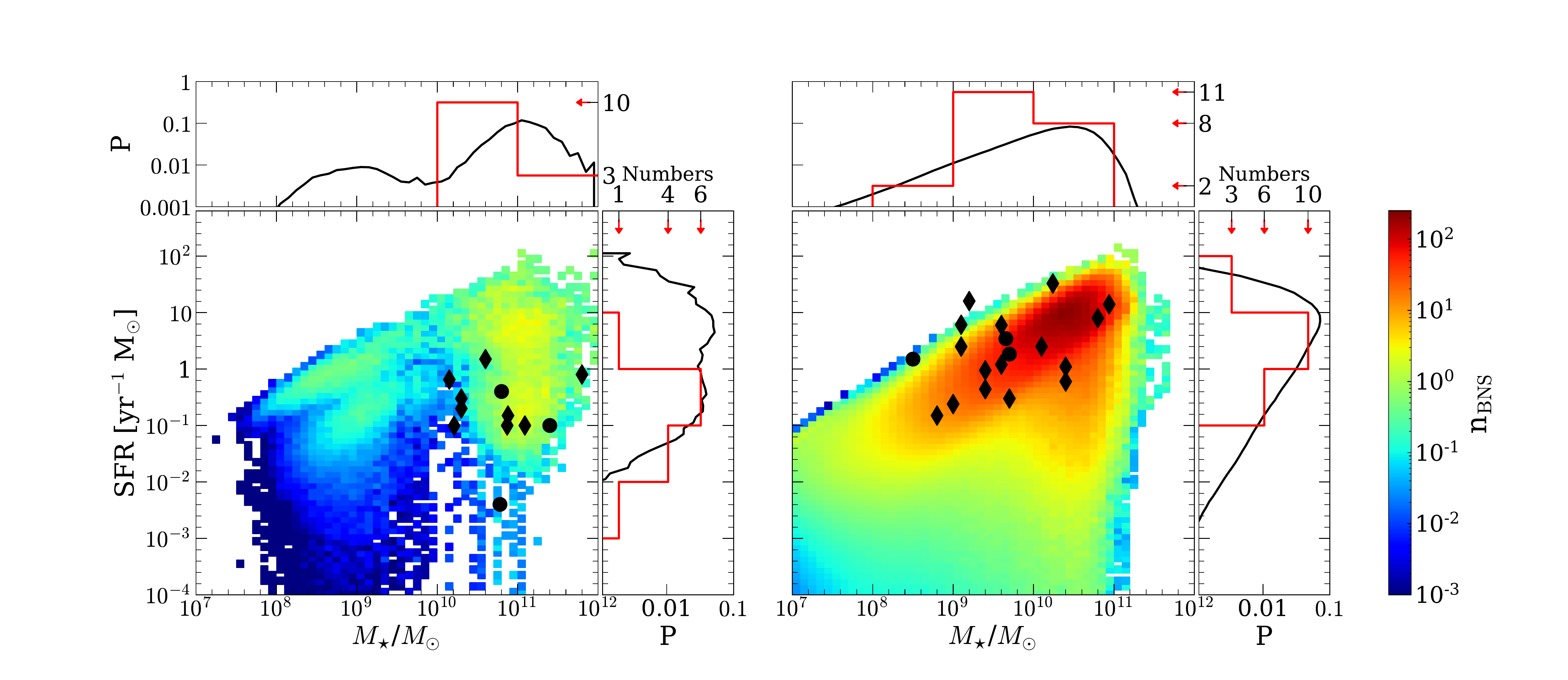}
		\caption{
			Distributions of BNS mergers (from model $\alpha10.\rm kb\beta0.9$ + Millennium-II), sGRBs and kilonovae on the SFR versus stellar mass of host galaxies (SFR-$M_{\star}$) plane, with different morphologies, i.e., elliptical (left) and spiral (right) hosts. Diamonds and circles denote the host galaxies of observed sGRBs and kilonovae, respectively. The top and left small panels show the probability distributions of BNS mergers (black lines) and sGRBs and kilonovae (red lines) against the stellar mass or SFR of host galaxies.}
		\label{fig:f17}
	\end{center}
\end{figure*}

The coalescence of some BNS systems with particular viewing angles can be detected as both sGRBs and kilonovae. Thus, the host galaxies of sGRBs and kilonovae represent those of BNS mergers. \citet{Church2011} provided the properties of host galaxies for $16$ sGRBs and among them $25\%$ are hosted in ellipticals and $75\%$ in spirals. According to \citet{Fong2013}, the hosts of sGRBs can be further classified into four types of environments, i.e., late-type ($\sim 50\%$), early-type ($\sim 15\%$), inconclusive ($\sim 20\%$), and host-less (lacking a coincident host galaxy to limits of $\geq26$\,mag; $\sim 15\%$). In addition, they also estimated the most likely ranges for the late-type and early-type fractions as $60\% $-$ 80\%$ and $20\% $-$ 40\%$, respectively.  
The spiral dominated host galaxies of sGRBs are consistent with the host distributions of BNS mergers presented in this paper.  

Figure~\ref{fig:f17} shows the comparison between the host galaxies of some detected sGRBs and kilonovae (see Tab.~\ref{tab:t8}) and the host galaxy distribution of BNS mergers obtained from the reference model ($\alpha10.{\rm kb}\beta0.9$\,$+$\,Millennium-II). 
Note that the most distant secure sGRB in elliptical host galaxy is at $z=0.915$ and thus we display the distribution of elliptical host galaxies at redshift $z<1$ in the left panel. Similarly, we consider the spiral host galaxies at redshift $z<1.8$ in the right panel as the most distant secure sGRB in spiral host galaxy is at $z=1.754$. From this Figure, we see that the host galaxies of sGRBs (diamonds) and kilonovae (circles) are typical to those of BNS merger hosts.
We find that the mass distribution of sGRBs and kilonovae hosts are consistent with the corresponding model results for BNS mergers according to the $\chi^2$-statistics ($\chi^{2}(4)=9.40$ for elliptical hosts and $\chi^{2}(4)=8.41$ for spiral hosts; both with $p>0.05$), as well as the SFR distribution of the spiral hosts ($\chi^{2}(4)=3.82$, $p>0.25$). However, the SFR distribution of the elliptical hosts appears conflict with the host distribution of BNS mergers ($\chi^{2}(4)=32.5$, $p<0.01$). 

The fraction of elliptical hosts is $\sim31-34\%$ as shown in Table~\ref{tab:t8}, substantially larger than that for local BNS mergers ($\sim14\% - 18\%$). The reason may be as follows. Normally, it is extremely hard to localize a sGRB only according to its detection in Gamma-ray. Follow-up observations of the afterglows in UV/optical bands are essential in the identification of sGRB host galaxies. For some sGRBs hosted in spiral galaxies, dust extinction may be  serious, and thus it is harder to detect their UV/optical afterglows as well as host galaxies, which may lead to an underestimate of the fraction of spiral hosts. In contrast, it may be relative easier to detect the UV/optical afterglows of those sGRBs hosted in ellipticals due to weaker contamination. As a result, the real fraction of sGRBs with spiral (or elliptical) hosts may be substantially higher (lower) than that given by current observations. 
Compared with sGRB (viewing within a small angle relative to jets), the kilonovae associated with BNS mergers are quite faint and it can be detected only for a small number of sGRBs. There are six convincing kilonova observations up to now. The most famous one is GRB170817A/GW170817, and the other five are GRB130603B \citep{Berger2013, Cucchiara2013, Postigo2014}, GRB150101B \citep{Fong2016}, GRB160821B \citep{Lamb2019, Troja2019}, GRB200522A \citep{Fong2020}, and GRB070809 \citep{Jin2020}. The host galaxy of GRB130603B is a relatively luminous, star-forming spiral, with stellar mass $\sim 5\times 10^{9} M_\odot$, SFR $\sim 1.84 M_\odot{\rm yr}^{-1}$, and metallicity $Z\sim Z_\odot$ \citep{Cucchiara2013}. Although GRB130603B is in a spiral galaxy, its explosion site has an older stellar population ($>1 \rm Gyr$) and there is no evidence for recent star formation, indicating that this BNS was formed with a long delay time. The host galaxy of GRB150101B is confirmed as an early-type one (2MASX J12320498-1056010), with stellar mass $7.1^{+1.2}_{-2.3}\times 10^{10}\ M_\odot$, stellar population age $\sim 2-2.5$\,Gyr, and SFR $\lesssim 0.4 M_\odot{\rm yr}^{-1}$ \citep{Fong2016}. The host galaxy of GRB160821B is probably a star-forming spiral with stellar mass $10^{8.5}M_\odot$, SFR $\gtrsim1.5M_\odot{\rm yr}^{-1}$, and metallicity $Z\sim Z_\odot$ \citep{Troja2019} but notably this event has a large offset ($\sim16{\rm kpc}$) from the galaxy's center. Recently, \citet{Fong2020} reported that GRB200522A locates near the center of a young (offset $\sim1{\rm kpc}$), star-forming galaxy with stellar mass $4.5\times 10^{9}M_\odot$, stellar population age $\sim 0.531$\,Gyr, SFR $2.1-4.8 M_\odot {\rm yr}^{-1}$, and metallicity $\log (Z_{\star}/Z_\odot) \approx0.02$. \citet{Jin2020} reanalyzed the data and identified the optical radiation component of GRB070809 as a kilonova, but there are two host galaxy candidates with similar possibility: one is a spiral galaxy at $z=0.219$; the other is an elliptical galaxy at $z=0.473$. 

In conclusion, GRB130603B, 150101B, 160821B, and 170817A are probably all formed with large delay time,  the hosts of GRB150101B and 170817A evolve to early-type during the longtime galaxy evolution but the hosts of GRB130603B and 160821B remain to be late-type. GRB200522A is found near the center of a late type galaxy and it can be originated from the merger of a newly formed BNS or a BNS with large delay time. The host galaxies of these kilonovae are well consistent with the distributions of BNS mergers resulting from our models as shown in Figure~\ref{fig:f17}. 

\section{Discussions}
\label{sec:discussion}

In this paper, we mainly investigate how different BSE models affect the properties of BNS systems at their formation/merger time and manage to find the preferred models to reasonably explain the Galactic BNSs observations and the local BNS merger rate density inferred from GW detections. We report the estimated BNS merger rate densities and the properties of host galaxies at the BNS merger time by using three different galaxy formation and evolution models/simulations, i.e., Millennium-II, EAGLE, and Illustris-TNG.

We investigate totally $84$ different cases in this paper and mainly focus on the CE phase, natal kick velocity, and mass ejection, which are different from those considered in previous works on BNS mergers \citep[e.g.,][]{Dominik2013, Chruslinska2018, Kruckow2018, Giacobbo2018} in some aspects. For the CE phase, we consider the $\alpha$-formalism, which has been discussed frequently in previous works. 
Our results are consistent with those obtained by \cite{Mapelli2018rate}, both supporting the large value of $\alpha$ ($=10$). Increasing $\alpha$ means that less binding energy is needed to eject the envelope for a certain binary if assuming that the left hand side of Equation~\eqref{eq:alpha-f} remains a fixed value, and  consequently the survival probability of a binary after the CE ejection increases. This is probably the main reason for the existence of short orbital period BNSs. Physical reasons for a large value of $\alpha$ may involve the CE internal energy, supernova explosion, and so on.
On the other hand, we also consider the $\gamma$-formalism, which is not often considered in the BNS studies because of two main shortcomings. One is that the $\gamma$-formalism may be understood as an intrinsic property of the formalism itself and it is hard to show clear physical picture except for a mysterious period of canonically-unstable RLOF; the other is that the definition of $\gamma$ implies a certain restriction to its value, 
\begin{equation}
\left(\frac{m_{1}+m_{2}}{m_{1}-m_{1\rm c}} \right)>\gamma>\left(\frac{m_{1}+m_{2}}{m_{1}-m_{1\rm c}} \right)\left[1-\frac{m_{1\rm c}(m_{1}+m_{2})}{m_{1}(m_{1\rm c}+m_{2})}  \right],
\end{equation}
where $m_{1}$ and $m_{2}$ are the masses before CE phase, and $m_{1\rm c}$ is the mass of the companion star after CE phase, which needs some caution \citep{Webbink2008, Ivanova2013}. 
For instance, if $m_{1\rm c}=m_{2}=m_{1}/3$, the value of $\gamma$ is constrained in the range of $2/3<\gamma<2$.

One of the goals of the current paper is to investigate the differences between the populations model with $\alpha$ and $\gamma$ formalisms. The main differences would come from the different ways to deal with the CE ejection. For both formalisms, the first CE phase can lead to the production of a low-mass binary with mass ratio close to 1. However, the $\alpha$-formalism requires a significant spiral-in stage in order to eject the CE while the $\gamma$-formalism can lead to a larger orbital separation after CE ejection. We use the radius-mass exponents and adopt the Eq. (57) in \cite{Hurley2002} as the mass transfer stability criteria. For naked helium giants, the critical mass-ratio $q_c=0.784$. We treat all cases with $q>q_c$ in which the primary star is a giant as common-envelope evolution. For NSs, we treat cases with $q>0.628$ as dynamical mass transfer.
Despite these shortcomings of the $\gamma$-formalism, in this way, our results show that adopting a large value of $\gamma$ (e.g., $1.5$) can also result in mock BNSs for MW-like galaxies agreeing well with the Galactic BNSs, which means the angular momentum evolution of a binary may play an important role in the CE evolution phase, and the $\gamma$-formalism can be an alternative way for the CE ejection models.
However, it may be still premature to conclude whether the $\alpha$- or $\gamma$-formalism dominates the CE phase, or both significantly contribute to the CE phase. Nevertheless, our investigations of the $\gamma$-formalism provides testable predictions, which may be falsified by future observations.

Furthermore, we set $\beta=\frac{M_{\rm NS,1}+M_{\rm NS,2}}{M_{\rm NS,1}+M_{\rm He}}$ as a free parameter to investigate the effects of mass ejection during the secondary SNe on the BNS formation. Our results show that if the secondary star has a small mass before the SN explosion (e.g., $\beta=0.9$), the Galactic BNSs and the local merger rate density inferred from GW detections can be well reproduced. One possible physical mechanism for such mass ejection is that the binaries undergo ultra-stripped SNe \citep[][and references therein]{Tauris2015}. Due to the helium-star expansion after core helium depletion, the compact companion heavily strips the envelope of the star by RLOF mass transfer, i.e., the so-called Case BB mass transfer. The highly stripped core eventually collapses to produce an ultra-stripped SN with a small amount of mass ejection. One of the examples might be iPTF 14gqr (SN 2014ft), which is reported as the directly observation of such ultra-stripped SNe, strongly supporting this scenario \citep{De2018}.

In Section~\ref{subsec:sGRBs}, we show the correlations of host galaxies between BNS mergers and sGRBs or kilonovae. The BNS merger rate density and its evolution may also be constrained by observations of sGRBs or kilonovae, as suggested by the detections of both a sGRB and a kilonova as the EM counterpart of GW 170817. We assume a one-to-one correspondence between sGRB/kilonova event and BNS merger event. \cite{Petrillo2013} estimated the sGRB rate by using SWIFT observations of sGRBs and found the rate is in the range of $500$-$1500\,\Gpcpyr$, with the default value of the beaming angle ($20^{\circ}$), while \citet{Coward2012} obtained a lower sGRB rate density of $8^{+5}_{-3}\,\Gpcpyr$ with an upper limit of $1100^{+700}_{-470}\,\Gpcpyr$. \citet{Fong2015} analysed the observation of $103$ sGRBs and obtained a rate of $270^{+1580}_{-180}\,\Gpcpyr$. The estimated rate of sGRB conforms to local BNS merger rates inferred from our BSE models.

We try to assess the robustness or uncertainty of the predicted BNS merger population properties associated with different models of galaxy formation and evolution, as these models both agree with a diverse set of low-redshift observations of galaxies (stellar mass, luminosity, size, star formation rate, etc.). We find that the host stellar mass distribution peaks at $10^{10.5}-10^{10.7}M_{\odot}$ and star formation rate distribution peaks $2-4 M_{\odot}\rm yr^{-1}$ for all the simulations. However, there are some discrepancies that can not be omitted. For Millennium-II, it focuses on providing well-resolved assembly histories for low-mass systems rather than good statistics for rare, high-mass galaxies, and under-estimates the abundance of galaxies at the high-mass ($\geq10^{11.5}M_{\odot}$) end. In addition, Millennium-II predicts a larger passive fraction among low mass galaxies than observations in order to limit the total production of stars, resulting in fewer BNS mergers in low mass galaxies. For EAGLE, as shown in Fig~\ref{fig:f12}, the stellar metallicities of dwarf galaxies are too high, possibly arising from limited resolution, which may be resolved simply by increasing the resolution further. For Illustris-TNG, the stellar mass function at the high-mass ($\geq10^{11.5}M_{\odot}$) end is too high, resulting in the host stellar mass distribution moving towards high-mass galaxies.

The metallicity distributions of hosts resulting from different galaxy simulations are quite different. The discrepancies in the metal enrichment procedures used in the Millennium-II, EAGLE and Illustris-TNG are responsible for the differences in the properties of BNS hosts. In Millennium-II, a semi-analytical method is used to calculate the evolution of metals. In EAGLE and Illustris-TNG, a more realistic treatment is taken into account, including winds from asymptotic giant branch (AGB) stars and massive stars, core collapse supernova (SNII) and type Ia supernova (SNIa), but the critical details differ. In EAGLE, the treatment of AGB winds follows \citet{Bergeat2005}, SNII follows \citet{Marigo2001} and \citet{Portinari1998}, and SNIa follows \citet{Thielemann2003}. In Illustris-TNG, the treatment of AGB winds follows \citet{Karakas2010}, \citet{Doherty2014}, and \citet{Fishlock2014}, SNII follows \citet{Kobayashi2006} and \citet{Portinari1998}, and SNIa follows \citet{Nomoto1997}. The metallicity distribution from EAGLE is higher than Illustris-TNG, mainly because of the high efficiency of the feed back from the star formation they adopted.

We also note that the cosmological galaxy formation and evolution models adopted above have been demonstrated good in agreement with the present-day galaxy statistics in many aspects, e.g., the galaxy stellar mass function, the galaxy luminosity function, the stellar mass-halo mass relation, the specific star formation rates and passive fractions \citep{Guo2011,Schaye2015,Pillepich2018}. However, the results of these models at high redshift may be not as reliable as that at low redshift. For instance, in Millennium-II, the abundance of low mass galaxies ($M_{\star}<10^{10}M_{\odot}$) is over-predicted at high redshift. Hence, one should be cautious about the results on the host galaxy distributions at high redshifts.

We emphasize that most host galaxies of BNS mergers are late-type galaxies (81.2\%-84.4\%) rather than early-type ones ($15.6\%$-$18.8\%$), different from that obtained by \citet{Artale2019a, Artale2019b}, in which the galaxy morphology was simply divided by using a criteria sSFR value. \cite{Mapelli2018host1} combined the {\bf MOBSE} code with Illustris-1 cosmological simulation to investigate the host galaxies (stellar mass and metallicity) of compact objects merging in the local Universe. In their research, the host galaxies of BNS mergers are predominantly massive galaxies from $\sim10^9M_{\odot}$ to $\sim10^{12}M_{\odot}$, agreeing with the Illustris-TNG ($90\%$ confidence interval: $6.2\times10^8-7.1\times10^{11}M_{\odot}$). 

To conclude this section, we note here that the models considered in the paper (both in terms of the metallicity dependent star formation history and binary evolution) may only cover a fraction of the available parameter space, which can lead to some uncertainties in our results. 

\section{Conclusions}
\label{sec:conclusion}

In this paper, we investigate the formation and evolution of BNSs and their mergers both in individual galaxies and throughout the cosmic history, by implementing parameterized population synthesis models for binary stellar evolutions (BSEs) into cosmological galaxy formation and evolution models. We specifically investigate the impacts of the common-envelope (CE) phase, natal kick, mass ejection during the secondary SN explosion, and metallicity, on the properties of resulting BNSs. By performing $84$ sets of BSE simulations, we find that the assumptions for the CE phase, kick velocities, and mass ejection during the secondary SN explosion all strongly influence the formation efficiency and  properties of resulting BNSs, while that the metallicity may only have a secondary effect on the resulting BNS properties. 

We find that the Galactic BNSs have already imposed more stringent constraints on the formation of BNSs than the GW detections do as follows: 
\begin{itemize}
\item an efficient CE depletion, i.e., a large value of $\alpha$ (i.e., $\alpha=10$) or $\gamma$ (i.e., $\gamma=1.5$), is required. In contrast, a small value of $\alpha$ (i.e., $\alpha=0.1$) implies that the orbits of progenitor binary stars shrink considerably before the BNS formation because of the difficulty in CE ejection, and a small value of $\gamma$ (i.e., $\gamma=1.1$) implies that the semimajor axes of the binaries are still substantially large after the CE phase, which both obstruct the formation of BNSs; 
\item the bimodal distribution of natal kick velocities is preferred as it is more compatible with both the eccentricity-orbital period ($e-P_{\rm orb}$) distribution of the Galactic BNSs, especially those short-period, high eccentricity ones, and  the local BNS merger rate density;
\item a low mass ejection during the secondary SN explosion is preferred, because most Galactic BNSs are detected with low orbital eccentricities, which indicates that the donor may be heavily stripped by the compact companion before the secondary SN explosion. 
\end{itemize}
By utilizing the Bayes methods, we find that model $\alpha10.\rm kb\beta0.9$ and $\gamma1.5\rm kb\beta0.9$ are preferable BSE models, of which the model results on BNSs can be compatible with both the Galactic BNS observations and the local BNS merger rate inferred from GW detections (model $\alpha10.\rm kb\beta0.9$: $316-784\,\Gpcpyr$; model $\gamma1.5\rm kb\beta0.9$: $137-429\,\Gpcpyr$). If only using the current estimate on the local BNS merger rate via GW observations ($320_{-240}^{+490}\Gpcpyr$), the obtained constraints on the BSE model parameters may have large uncertainties, which are much less stringent that those obtained by using both the Galactic BNS observations and the estimate of the local BNS merger rate density. This is different from those results for the requirement of a large $\alpha$ in previous works mainly because the current local BNS merger rate inferred from LIGO/Virgo O1, O2, and O3 observations is narrowed to a lower value compared with that from O1 and O2 observations, but still with large uncertainties.

The evolution of BNS merger rate density resulting from different models can be all well described by a function of redshift $z$, i.e., $R(z)=R_{0}\left[1+\frac{1}{(1+z_*)^\xi}\right]\frac{(1+z)^\zeta}{1+\left[(1+z)/(1+z_*)\right]^\xi}$, where $R_{0}$ represents the local merger rate density. The fits for different BSE models show that the shape parameters $(z_*, \zeta, \xi)$ can be used to distinguish different models and adequately understand the cosmic SFR history with future BNS merger detections. At low redshift, the merger rate can be described by $R(z)\simeq R_{0}(1+z)^{\zeta}$ because of the large value of $z_*$ and $\xi$, and the index $\zeta$ varies from $\sim0.7$ for models resulting BNSs with large delay time to $\sim3.0$ for models resulting BNSs with short delay time, especially $\zeta\sim1.34-2.03$ for model $\alpha10.\rm kb\beta0.9$ and $\zeta\sim1.43-2.08$ for model $\gamma1.5\rm kb\beta0.9$. 

We find that the BNS merger rate in individual galaxies are mainly determined by their mass, specific star formation rate (sSFR), and metallicity. The stellar mass has the decisive influence on the rate, which can be used to give the top priority for possible host galaxy candidate. For host galaxy candidates with similar mass, BNSs prefer merging in star-forming galaxies rather than quiescent galaxies. Metallicity has little effect on the merger rate as the evolution of BNSs is not as sensitive as the evolution of binary black holes (BBHs).

We find that most BNS mergers at low redshift ($z\lesssim 1$) happen in galaxies with large stellar mass ($10^{9}M_\odot<M_{\star}<10^{11}M_\odot$), high SFR ($0.1{\rm yr}^{-1} M_\odot<\rm SFR<10{\rm yr}^{-1} M_\odot$) and near-solar metallicity ($Z\sim Z_\odot$). For our reference model $\alpha10.\rm kb\beta0.9$, we emphasize that the host galaxies of most BNS mergers are late-type galaxies ($81.2\%-84.4\%$) rather than early-type galaxies ($15.6\%-18.8\%$), which is also consistent with the observational distribution of sGRB and kilonova host galaxies. The host galaxies of most BNS mergers are central ones ($69.4\%$) rather than satellites ($22.4\%$) or isolated galaxies ($8.2\%$). The fraction of late-type or central host galaxies increases along with increasing redshift. 
For a short time delay model (e.g., $\alpha1.0\rm kb\beta0.9$), about $90\%$ of host galaxies are late-type galaxies and only $\sim10\%$ are early-type galaxies. For a long time delay model (e.g.,  $\gamma1.3\rm kl\beta0.9$), about $72-80\%$ of host galaxies are late-type galaxies and $20-28\%$ are early-type galaxies.

By combining the special merger history of NGC 4993, we find that there is a high probability of an original, elliptical host galaxy (85.4\%) rather than a swallowed, spiral host galaxy (14.6\%) for GW170817 and the delay time for GW170817 like events peaks at 10.5-11.0 $\rm Gyr$, ranging from 2.5 $\rm Gyr$ to 12.0 $\rm Gyr$ with a $5$ to $95$ percentile.
 
\section*{Acknowledgements}
We thank the referee for helpful comments and suggestions.
This work is partly supported by the National Natural Science Foundation of China (Grant No. 11690024, 11673031, and 11873056), the National Key Program for Science and Technology Research and Development (Grant No. 2020YFC2201400), and
the Strategic Priority Program of the Chinese Academy of Sciences (Grant No. XDB 23040100).

\noindent{\bf DATA AVAILABILITY}

\noindent The data underlying this article will be shared on reasonable request to the corresponding author.





%




%
\appendix

\section{Likelihood calculation}
\label{sec:appA}
\setcounter{table}{0}
\renewcommand{\thetable}{A\arabic{table}}

Combining the outcomes from the BSE models with the star formation and metallicity enrichment histories of each individual galaxy given by the cosmological galaxy formation and evolution models/simulations, we obtain the properties of BNSs at their formation time in each galaxy. The subsequent (mean) evolution of the semimajor axes ($a$) and eccentricities ($e$) of these BNSs is assumed to be controlled by the GW radiation as \citep{Peters1964}
\begin{eqnarray}
\left< \frac{da}{dt} \right>& = & -\frac{64}{5}\frac{G^{3}m_{1}m_{2}(m_{1}+m_{2})}{c^{5}a^{3}(1-e^{2})^{7/2}}
\left(1+\frac{73}{24}e^{2}+\frac{37}{96}e^{4}\right), \nonumber \\
\left< \frac{de}{dt} \right>& = & -\frac{304}{15}e\frac{G^{3}m_{1}m_{2}(m_{1}+m_{2})}{c^{5}a^{4}(1-e^{2})^{5/2}}
\left(1+\frac{121}{304}e^{2}\right). \nonumber
\label{eq:aeevol}
\end{eqnarray}
Here $a$, $e$, $m_1$, and $m_2$ denote the semimajor axis, eccentricity, primary and secondary component masses of the binary, $G$ the gravitational constant, $c$ the speed of light. At any given cosmic time $t$, the properties of surviving BNSs in each mock galaxy can be obtained by convolving the formation of BNSs at earlier time with the orbital evolution of each BNS.

We evolve the eccentricity and period of each BNS in mock MW-like galaxy to redshift $z=0$ and calculate the likelihood $L_{i}$ of the model $\rm M_{i}$ as:
\begin{eqnarray}
& & \log L_i   =  \sum_{j=1}^{N_{\rm BNS}}\left.\log p(\log P_{{\rm orb},j},e_j) \right|_{{\rm M}_i} \nonumber \\
& = &   \sum_{j=1}^{N_{\rm obs}}  \sum_{l=1}^{N_{P}} \sum_{k=1}^{N_{e}}  N(\log P_{{\rm orb},j},e_{j}|\log P_{l}, e_k, \sigma_{\log P_{l}}^{2},\sigma_{e_k}^{2}) \left.\right|_{{\rm M}_i}. \nonumber \\
\end{eqnarray}
Here, the summation is over all $(\log P_{l}, e_k)$ pixels and all observed Galactic BNSs $N_{\rm BNS}$. For each pixel $(\log P_{l}, e_k)$, we further adopt a two-dimension normal distribution $N(\log P_{\rm orb}, e| P_{l}, e_{k}, \sigma_{\log P_{l}}^{2},\sigma_{e_k}^{2}) \left. \right |_{{\rm M}_i} $ to avoid zero probability at some pixels resulting from the $i$-th model and get a good behaved $e-P_{\rm orb}$ probability density distribution. We assume that the standard deviations of $\log P_{l}$ and $e_k$ are constant and assign them as $\sigma_{\log P_{l}} = 0.3$ and $\sigma_{e_k}=0.03$, respectively. 

Note here that the theoretical distributions obtained in \citet{Vigna2018} were calculated by weighing the evolution timescale of BNS at each $(\log P_{l}, e_k)$ pixel to the probability density map. In this way, they did not consider the uncertainties in the galactic star formation and metal enrichment histories. Moreover, the probability density distribution takes no account of radio lifetimes of BNSs. At certain critical points in BNS evolutionary history, the evolution timescales exceed the radio lifetimes and these parts make few contributions to $e-P_{\rm orb}$ distribution of Galactic BNSs due to the limitations in observation. We also enlarge the number of mock BNSs to better represent the real distribution.

As mentioned in Section~\ref{subsec:observe_BNS}, the short-period, high eccentricity BNSs (J0509$+$3801, J1757$-$1854, and B1913$+$16) may be better explained via the dynamical formation channel \citep[see][]{AndrewsMandel2019}. Here, we adopt the same method in Section~\ref{subsec:GalBNSs} to quantitatively compare the models by removing these BNSs with possible dynamical origin.

\begin{table*}
	\begin{center}
		\caption{Similar to Table~\ref{tab:t2} but calculated with 14 detected Galactic BNSs, except for three candidates through dynamical formation channel. Here, $\log L_{\alpha10.\rm kb}=-57.15$.}
		\label{tab:a1}
		\setlength{\tabcolsep}{1.8mm}{
			\begin{tabular}{l|c|c|c|c|c|c|c|c|c}
				\hline
				\multirow{2}{*}{Model Name} & \multicolumn{4}{c}{$\log K_i$ for survived pulsar-BNSs}  & & \multicolumn{4}{c}{$\log K_i$ for all survived BNSs} \\ \cline{2-5} \cline{7-10}
				&   & $\beta=0.6$  & $\beta=0.8$  & $\beta=0.9$ & &     & $\beta=0.6$  & $\beta=0.8$  & $\beta=0.9$ \\
				\hline
				$\alpha0.1{\textrm{kl}}$& -57.82 & -53.23 & -27.15 & -13.47 &  & -60.65 & -28.26 & -10.28 & -20.03 \\
				$\alpha0.1{\textrm{kh}}$& -71.55 & -75.78 & -55.81 & -75.51 &  & -41.00 & -45.45 & -46.00 & -43.84 \\
				$\alpha0.1{\textrm{kb}}$& -61.59 & -48.35 & -23.17 & -15.09 &  & -53.25 & -32.13 & -14.79 & -13.36 \\
				$\alpha1.0{\textrm{kl}}$& -58.88 & -29.08 & 5.13   & 8.84   &  & -53.27 & -10.86 & -6.75  & -11.74 \\
				$\alpha1.0{\textrm{kh}}$& -58.61 & -54.69 & -76.34 & -58.15 &  & -42.95 & -41.67 & -44.53 & -49.39 \\
				$\alpha1.0{\textrm{kb}}$& -23.56 & -11.57 & 1.76   & 7.23   &  & -27.79 & -3.54  & -7.44  & -6.05  \\
				$\alpha10.{\textrm{kl}}$& -23.51 & -22.14 & 10.53  & 12.22  &  & -5.35  & -3.78  & 0.62   & -5.84  \\
				$\alpha10.{\textrm{kh}}$& -21.77 & -11.96 & -5.23  & -7.68  &  & -13.89 & -26.07 & -13.01 & -12.65 \\
				$\alpha10.{\textrm{kb}}$& 0      & -0.50  & 9.49   & 11.52  &  & -10.93 & -2.79  & -4.57  & 5.94   \\
				$\gamma1.1{\textrm{kl}}$& -38.15 & -39.02 & -31.39 & -31.05 &  & -38.22 & -32.96 & -31.31 & -30.60 \\
				$\gamma1.1{\textrm{kh}}$& -37.38 & -44.08 & -33.03 & -36.72 &  & -10.71 & -15.62 & -8.04  & -13.81 \\
				$\gamma1.1{\textrm{kb}}$& -37.63 & -37.80 & -30.40 & -30.29 &  & -27.59 & -27.27 & -19.62 & -24.39 \\
				$\gamma1.3{\textrm{kl}}$& -22.60 & -22.12 & 6.34   & 8.56   &  & -15.75 & -15.67 & -5.77  & -7.43  \\
				$\gamma1.3{\textrm{kh}}$& -3.83  & -3.27  & -4.60  & -6.90  &  & -7.39  & -1.68  & -2.52  & -7.86  \\
				$\gamma1.3{\textrm{kb}}$& -6.33  & -7.37  & 7.27   & 9.16   &  & -4.83  & -3.73  & -5.49  & 4.58   \\
				$\gamma1.5{\textrm{kl}}$& -25.64 & -23.44 & 9.60   & 12.11  &  & -18.07 & -0.37  & -6.81  & -12.37 \\
				$\gamma1.5{\textrm{kh}}$& -37.38 & -37.59 & -40.53 & -18.34 &  & -18.86 & -24.15 & -8.14  & -19.00 \\
				$\gamma1.5{\textrm{kb}}$& -21.24 & -5.52  & 8.02   & 9.78   &  & -22.37 & -2.59  & 0.12   & -5.53  \\
				$\gamma1.7{\textrm{kl}}$& -21.46 & -30.84 & 4.59   & 6.56   &  & -10.20 & -9.50  & -1.80  & -11.87 \\
				$\gamma1.7{\textrm{kh}}$& -22.64 & -3.21  & -17.19 & -4.11  &  & -14.54 & -2.77  & -20.59 & -18.25 \\
				$\gamma1.7{\textrm{kb}}$& -4.39  & -0.14  & 5.91   & 8.19   &  & -4.47  & 2.13   & -0.85  & -0.72  \\
				\hline
		\end{tabular}}
	\end{center}
\end{table*}

Table~\ref{tab:a1} lists the values of the logarithmic Bayes factors, $\log K_{i}$, calculated with 14 detected Galactic BNSs, excluded those three BNSs that may be formed through the dynamical channel. Compared to Table~\ref{tab:t2}, for $\alpha$-formalism, the best fitting model becomes model $\alpha10.\rm kl\beta0.9$ and for $\gamma$-formalism, the best fitting model becomes model $\gamma1.5\rm kl\beta0.9$. These models provide additional possibilities to reproduce Galactic BNSs.

\section{BNS merger rate in the local Universe and the fitting parameters}
\label{sec:appB}
\setcounter{table}{0}
\renewcommand{\thetable}{B\arabic{table}}

\begin{table*}
\begin{center}
\caption{BNS merger rate densities in unit of ${\rm Gpc}^{-3}\,{\rm yr}^{-1}$ in the local Universe, ${\rm Myr}^{-1}$ in the MW-like galaxy at redshift $z=0$ and the fitting parameters of the evolution of the merger rate density as a function of redshift resulting from the various models.}
\label{tab:t9}
\setlength{\tabcolsep}{1mm}{
\begin{tabular}{c|c|c|c|c|c|c|c|c|c|c|c|c|c|c|c|c|c}
\hline
\multirow{2}{*}{Model Name}& \multicolumn{4}{c}{Millennium-II}  & \multicolumn{4}{c}{EAGLE}  & \multicolumn{4}{c}{Illustris-TNG}  & \multicolumn{4}{c}{Madau \& Dickinson (2014)}  & $R_{\rm MW}$\\ \cline{2-17}
&$R_{\rm Mil,0}$&$\zeta$&$z_{\star}$&$\xi$  &$R_{\rm EAG,0}$&$\zeta$&$z_{\star}$&$\xi$  &$R_{\rm Ill,0}$&$\zeta$&$z_{\star}$&$\xi$ & $R_{\rm MD,0}$&$\zeta$&$z_{\star}$&$\xi$ &$[{\rm Myr}^{-1}]$\\
\hline
$\alpha0.1{\textrm{kl}}$ & 23.7  & 2.09 & 4.33 & 6.33 & 19.7  & 2.87 & 2.62 & 4.61 & 35.6  & 1.79 & 4.07 & 5.55  & 41.5  & 2.62 & 2.86 & 5.99 & 5.0   \\
$\alpha0.1{\textrm{kh}}$ & 1.1   & 1.82 & 4.26 & 6.22 & 0.9   & 2.44 & 2.68 & 4.51 & 1.5   & 1.55 & 4.05 & 5.67  & 1.8   & 2.14 & 2.85 & 5.53 & 0.2  \\
$\alpha0.1{\textrm{kb}}$ & 13.9  & 2.00 & 4.36 & 6.30 & 11.5  & 2.73 & 2.65 & 4.49 & 20.1  & 1.72 & 4.10 & 5.52  & 25.2  & 2.47 & 2.91 & 5.95 & 2.9  \\
$\alpha1.0{\textrm{kl}}$ & 214   & 2.19 & 4.33 & 6.41 & 160   & 3.02 & 2.52 & 4.27 & 302   & 1.91 & 4.01 & 4.99  & 437   & 2.72 & 3.10 & 4.91 & 40.1 \\
$\alpha1.0{\textrm{kh}}$ & 27.8  & 2.17 & 4.39 & 6.46 & 19.3  & 2.80 & 2.79 & 4.22 & 37.3  & 1.85 & 4.22 & 5.31  & 61.6  & 2.68 & 2.99 & 5.27 & 4.9  \\
$\alpha1.0{\textrm{kb}}$ & 125   & 2.11 & 4.37 & 6.39 & 92.6  & 2.88 & 2.59 & 4.18 & 171   & 1.84 & 4.07 & 5.01  & 258   & 2.64 & 3.07 & 4.88 & 23.3 \\
$\alpha10.{\textrm{kl}}$ & 441   & 1.25 & 3.95 & 5.47 & 358   & 1.69 & 2.42 & 3.37 & 518   & 1.00 & 3.72 & 4.38  & 713   & 1.62 & 3.13 & 5.10 & 90.0   \\
$\alpha10.{\textrm{kh}}$ & 82.4  & 1.74 & 4.16 & 5.96 & 63.9  & 2.24 & 2.67 & 3.85 & 102   & 1.48 & 3.95 & 4.90  & 164   & 2.17 & 3.00 & 5.13 & 15.8 \\
$\alpha10.{\textrm{kb}}$ & 255   & 1.44 & 4.06 & 5.69 & 201   & 1.90 & 2.57 & 3.61 & 303   & 1.20 & 3.84 & 4.65  & 446   & 1.84 & 3.05 & 5.02 & 50.5 \\
$\gamma1.1{\textrm{kl}}$ & 0.4   & 0.84 & 3.67 & 6.95 & 0.4   & 0.50 & 3.69 & 4.77 & 0.5   & 0.47 & 3.81 & 6.03  & 1.0   & 1.48 & 2.39 & 4.48 & 0.1  \\
$\gamma1.1{\textrm{kh}}$ & 21.5  & 1.21 & 4.12 & 5.53 & 14.1  & 1.68 & 2.75 & 3.31 & 21.7  & 1.08 & 3.94 & 4.33  & 57.6  & 1.48 & 3.19 & 4.62 & 3.7  \\
$\gamma1.1{\textrm{kb}}$ & 13.4  & 1.22 & 4.15 & 5.56 & 8.9   & 1.67 & 2.76 & 3.34 & 13.7  & 1.08 & 3.97 & 4.38  & 34.4  & 1.48 & 3.16 & 4.54 & 2.3  \\
$\gamma1.3{\textrm{kl}}$ & 188   & 0.60 & 3.13 & 4.44 & 126   & 1.48 & 1.81 & 3.16 & 181   & -    & -    & -     & 381   & 0.97 & 2.83 & 3.61 & 34.2 \\
$\gamma1.3{\textrm{kh}}$ & 71.2  & 1.50 & 4.11 & 5.85 & 52.7  & 1.87 & 2.81 & 3.70 & 81.5  & 1.26 & 3.97 & 4.86  & 143   & 1.92 & 2.95 & 5.05 & 13.4 \\
$\gamma1.3{\textrm{kb}}$ & 143   & 1.06 & 4.00 & 5.42 & 104   & 1.53 & 2.60 & 3.43 & 153   & 0.91 & 3.79 & 4.41  & 280   & 1.45 & 2.95 & 4.44 & 27.1 \\
$\gamma1.5{\textrm{kl}}$ & 161   & 1.31 & 4.21 & 5.90 & 118   & 2.00 & 2.63 & 4.26 & 183   & 1.21 & 3.89 & 5.22  & 332   & 1.51 & 2.66 & 4.41 & 31.2 \\
$\gamma1.5{\textrm{kh}}$ & 23.3  & 1.78 & 4.19 & 6.15 & 16.4  & 2.37 & 2.77 & 4.17 & 27.6  & 1.57 & 4.02 & 5.21  & 61.8  & 2.05 & 2.90 & 4.69 & 4.1  \\
$\gamma1.5{\textrm{kb}}$ & 96.2  & 1.51 & 4.18 & 5.98 & 67.7  & 2.10 & 2.74 & 4.17 & 109   & 1.35 & 3.96 & 5.22  & 214   & 1.78 & 2.78 & 4.50 & 17.7 \\
$\gamma1.7{\textrm{kl}}$ & 172   & 0.99 & 5.05 & 6.23 & 81.1  & 1.22 & 4.92 & 3.95 & 128   & 0.96 & 5.30 & 5.14  & 688   & 1.36 & 3.02 & 4.19 & 21.0   \\
$\gamma1.7{\textrm{kh}}$ & 76.9  & 1.58 & 4.34 & 5.99 & 51.5  & 1.79 & 3.34 & 3.64 & 83.7  & 1.33 & 4.29 & 4.90  & 167   & 2.14 & 2.98 & 5.23 & 13.5 \\
$\gamma1.7{\textrm{kb}}$ & 161   & 1.32 & 4.54 & 5.95 & 94.6  & 1.43 & 4.04 & 3.74 & 152   & 1.14 & 4.62 & 4.96  & 439   & 1.75 & 2.96 & 4.74 & 25.4 \\  \hline
\end{tabular}}
\end{center}
\begin{flushleft}
\footnotesize{Column 1: model name; columns 2-5, 6-9, 10-13, and 14-17: the local merger rate density and fitting parameters resulting from Millennium-II, EAGLE, Illustis-TNG and observational extinction-corrected SFR in \citet{Madau2014} and the metallicity redshift evolution in \citet{Belczynski2016a}, respectively; column 18: the Galactic BNS merger rate in unit of $\rm Myr^{-1}$.}
\end{flushleft}	
\end{table*}

\begin{table*}
\begin{center}
\caption{BNS merger rate densities in unit of ${\rm Gpc}^{-3}\,{\rm yr}^{-1}$ in the local Universe, ${\rm Myr}^{-1}$ in the MW-like galaxy at redshift $z=0$ and the fitting parameters of the evolution of the merger rate density as a function of redshift resulting from the various models with $\beta=0.6$.}
\label{tab:t10}
\setlength{\tabcolsep}{1mm}{
\begin{tabular}{c|c|c|c|c|c|c|c|c|c|c|c|c|c|c|c|c|c}
\hline
\multirow{2}{*}{Model Name}& \multicolumn{4}{c}{Millennium-II}  & \multicolumn{4}{c}{EAGLE}  & \multicolumn{4}{c}{Illustris-TNG}  & \multicolumn{4}{c}{Madau \& Dickinson (2014)}  & $R_{\rm MW}$\\ \cline{2-17}
&$R_{\rm Mil,0}$&$\zeta$&$z_{\star}$&$\xi$  &$R_{\rm EAG,0}$&$\zeta$&$z_{\star}$&$\xi$  &$R_{\rm Ill,0}$&$\zeta$&$z_{\star}$&$\xi$ & $R_{\rm MD,0}$&$\zeta$&$z_{\star}$&$\xi$ &$[{\rm Myr}^{-1}]$\\
\hline
$\alpha0.1{\textrm{kl}}\beta0.6$ & 46.6  & 1.97 & 4.57 & 6.44 & 34.4  & 2.63  & 2.81 & 4.14  & 61.7  & 1.71 & 4.31 & 5.35  & 111    & 2.36 & 3.11 & 5.86 & 8.7   \\
$\alpha0.1{\textrm{kh}}\beta0.6$ & 2.0   & 1.86 & 4.50 & 6.34 & 1.6   & 2.51  & 2.72 & 4.06  & 2.6   & 1.61 & 4.24 & 5.25  & 4.0    & 2.37 & 3.04 & 5.61 & 0.4   \\
$\alpha0.1{\textrm{kb}}\beta0.6$ & 24.4  & 1.98 & 4.52 & 6.40 & 18.7  & 2.62  & 2.78 & 4.14  & 32.9  & 1.71 & 4.28 & 5.32  & 56.0   & 2.42 & 3.08 & 5.87 & 4.7   \\
$\alpha1.0{\textrm{kl}}\beta0.6$ & 346   & 2.05 & 4.46 & 6.44 & 255   & 2.76  & 2.68 & 4.09  & 466   & 1.79 & 4.18 & 5.08  & 747    & 2.57 & 3.06 & 4.93 & 63.9  \\
$\alpha1.0{\textrm{kh}}\beta0.6$ & 38.0  & 2.20 & 4.41 & 6.47 & 25.7  & 2.78  & 2.86 & 4.16  & 50.4  & 1.87 & 4.27 & 5.30  & 88.2   & 2.73 & 3.01 & 5.50 & 6.4   \\
$\alpha1.0{\textrm{kb}}\beta0.6$ & 190   & 2.06 & 4.44 & 6.42 & 139   & 2.74  & 2.70 & 4.10  & 255   & 1.78 & 4.19 & 5.11  & 408    & 2.60 & 3.03 & 5.01 & 34.9  \\
$\alpha10.{\textrm{kl}}\beta0.6$ & 589   & 1.41 & 4.03 & 5.63 & 461   & 1.74  & 2.62 & 3.35  & 686   & 1.14 & 3.85 & 4.49  & 1050   & 1.74 & 3.17 & 5.32 & 115 \\
$\alpha10.{\textrm{kh}}\beta0.6$ & 103   & 1.80 & 4.21 & 6.05 & 78    & 2.27  & 2.74 & 3.86  & 127   & 1.52 & 4.03 & 4.97  & 210    & 2.24 & 3.01 & 5.24 & 19.4  \\
$\alpha10.{\textrm{kb}}\beta0.6$ & 324   & 1.55 & 4.12 & 5.81 & 252   & 1.92  & 2.70 & 3.58  & 386   & 1.28 & 3.95 & 4.73  & 602    & 1.93 & 3.09 & 5.21 & 62.8  \\
$\gamma1.1{\textrm{kl}}\beta0.6$ & 0.6   & 1.00 & 2.94 & 5.31 & 0.6   & -     & -    & -     & 0.7   & -    & -    & -     & 1.1    & 1.40 & 2.39 & 4.17 & 0.1   \\
$\gamma1.1{\textrm{kh}}\beta0.6$ & 24.4  & 1.21 & 4.27 & 5.83 & 15.1  & 1.59  & 3.05 & 3.36  & 24.0  & 1.07 & 4.16 & 4.61  & 65.0   & 1.49 & 3.21 & 4.62 & 4.1   \\
$\gamma1.1{\textrm{kb}}\beta0.6$ & 14.7  & 1.18 & 4.27 & 5.75 & 9.4   & 1.54  & 3.05 & 3.30  & 14.5  & 1.04 & 4.15 & 4.52  & 39.0   & 1.43 & 3.24 & 4.59 & 2.5   \\
$\gamma1.3{\textrm{kl}}\beta0.6$ & 279   & 0.70 & 3.11 & 4.21 & 179   & 1.51  & 1.85 & 3.04  & 263   & 0.77 & 2.60 & 3.15  & 570    & 1.09 & 2.84 & 3.73 & 49.8  \\
$\gamma1.3{\textrm{kh}}\beta0.6$ & 82.8  & 1.58 & 4.16 & 5.91 & 60    & 1.92  & 2.89 & 3.73  & 94.7  & 1.33 & 4.05 & 4.94  & 169    & 2.00 & 2.95 & 5.14 & 15.3  \\
$\gamma1.3{\textrm{kb}}\beta0.6$ & 187   & 1.08 & 4.10 & 5.46 & 130   & 1.48  & 2.75 & 3.34  & 195   & 0.92 & 3.94 & 4.43  & 367    & 1.52 & 2.95 & 4.49 & 34.7  \\
$\gamma1.5{\textrm{kl}}\beta0.6$ & 307   & 1.49 & 4.19 & 5.85 & 204   & 2.04  & 2.84 & 4.31  & 340   & 1.31 & 4.04 & 5.42  & 625    & 1.68 & 2.72 & 4.95 & 55.1  \\
$\gamma1.5{\textrm{kh}}\beta0.6$ & 30.5  & 1.85 & 4.23 & 6.18 & 21.4  & 2.36  & 2.86 & 4.15  & 36.6  & 1.60 & 4.09 & 5.26  & 78.1   & 2.18 & 2.86 & 4.82 & 5.3   \\
$\gamma1.5{\textrm{kb}}\beta0.6$ & 154   & 1.61 & 4.21 & 5.99 & 103   & 2.13  & 2.87 & 4.24  & 174   & 1.41 & 4.08 & 5.38  & 330    & 1.89 & 2.76 & 4.80 & 27.4  \\
$\gamma1.7{\textrm{kl}}\beta0.6$ & 339   & 1.33 & 4.90 & 6.43 & 176   & 1.48  & 4.73 & 4.39  & 287   & 1.20 & 5.12 & 5.69  & 1132   & 1.89 & 2.81 & 4.93 & 48.1  \\
$\gamma1.7{\textrm{kh}}\beta0.6$ & 95.4  & 1.66 & 4.37 & 6.06 & 64.7  & 1.83  & 3.43 & 3.75  & 106   & 1.38 & 4.36 & 5.06  & 205    & 2.23 & 2.95 & 5.34 & 16.8  \\
$\gamma1.7{\textrm{kb}}\beta0.6$ & 241   & 1.44 & 4.64 & 6.16 & 143   & 1.51  & 4.28 & 4.03  & 234   & 1.22 & 4.78 & 5.33  & 628    & 2.01 & 2.86 & 5.09 & 38.8  \\  \hline
\end{tabular}}
\end{center}
\begin{flushleft}
\footnotesize{Similar to Table~\ref{tab:t9} but for $\beta=0.6$ sub-models.}
\end{flushleft}	
\end{table*}

\begin{table*}
\begin{center}
\caption{BNS merger rate densities in unit of ${\rm Gpc}^{-3}\,{\rm yr}^{-1}$ in the local Universe, ${\rm Myr}^{-1}$ in the MW-like galaxy at redshift $z=0$ and the fitting parameters of the evolution of the merger rate density as a function of redshift resulting from the various models with $\beta=0.8$.}
\label{tab:t11}
\setlength{\tabcolsep}{1mm}{
\begin{tabular}{c|c|c|c|c|c|c|c|c|c|c|c|c|c|c|c|c|c}
\hline
\multirow{2}{*}{Model Name}& \multicolumn{4}{c}{Millennium-II}  & \multicolumn{4}{c}{EAGLE}  & \multicolumn{4}{c}{Illustris-TNG}  & \multicolumn{4}{c}{Madau \& Dickinson (2014)}  & $R_{\rm MW}$\\ \cline{2-17}
&$R_{\rm Mil,0}$&$\zeta$&$z_{\star}$&$\xi$  &$R_{\rm EAG,0}$&$\zeta$&$z_{\star}$&$\xi$  &$R_{\rm Ill,0}$&$\zeta$&$z_{\star}$&$\xi$ & $R_{\rm MD,0}$&$\zeta$&$z_{\star}$&$\xi$ &$[{\rm Myr}^{-1}]$\\
\hline
$\alpha0.1{\textrm{kl}}\beta0.8$ & 57.4  & 1.82  & 4.63 & 6.38  & 45.3  & 2.33  & 2.90 & 3.88  & 75.2  & 1.55  & 4.39 & 5.22  & 127    & 2.36  & 3.07 & 5.61  & 11.0 \\
$\alpha0.1{\textrm{kh}}\beta0.8$ & 2.3   & 1.93  & 4.49 & 6.30  & 1.9   & 2.40  & 2.86 & 4.01  & 3.1   & 1.62  & 4.28 & 5.24  & 4.8    & 2.47  & 2.97 & 5.57  & 0.5  \\
$\alpha0.1{\textrm{kb}}\beta0.8$ & 28.2  & 1.89  & 4.58 & 6.36  & 22.3  & 2.46  & 2.82 & 3.96  & 37.8  & 1.62  & 4.32 & 5.18  & 61.7   & 2.46  & 3.05 & 5.68  & 5.5  \\
$\alpha1.0{\textrm{kl}}\beta0.8$ & 432   & 1.88  & 4.55 & 6.41  & 323   & 2.51  & 2.73 & 3.89  & 559   & 1.63  & 4.28 & 5.02  & 899    & 2.41  & 3.10 & 4.81  & 81.3 \\
$\alpha1.0{\textrm{kh}}\beta0.8$ & 43.7  & 2.19  & 4.41 & 6.47  & 29.7  & 2.77  & 2.85 & 4.15  & 58.0  & 1.86  & 4.27 & 5.29  & 100    & 2.72  & 3.00 & 5.44  & 7.5  \\
$\alpha1.0{\textrm{kb}}\beta0.8$ & 229   & 1.97  & 4.49 & 6.40  & 169   & 2.62  & 2.72 & 3.99  & 301   & 1.70  & 4.23 & 5.06  & 480    & 2.52  & 3.05 & 4.94  & 42.7 \\
$\alpha10.{\textrm{kl}}\beta0.8$ & 733   & 1.43  & 4.23 & 5.85  & 559   & 1.85  & 2.70 & 3.58  & 849   & 1.20  & 4.01 & 4.78  & 1336   & 1.75  & 3.13 & 5.16  & 142  \\
$\alpha10.{\textrm{kh}}\beta0.8$ & 128   & 1.84  & 4.29 & 6.15  & 95.3  & 2.33  & 2.78 & 3.90  & 159   & 1.56  & 4.11 & 5.06  & 262    & 2.31  & 3.02 & 5.25  & 23.9 \\
$\alpha10.{\textrm{kb}}\beta0.8$ & 400   & 1.59  & 4.25 & 5.96  & 304   & 2.02  & 2.74 & 3.71  & 475   & 1.34  & 4.06 & 4.91  & 749    & 1.97  & 3.08 & 5.14  & 76.8 \\
$\gamma1.1{\textrm{kl}}\beta0.8$ & 3.4   & -     & -    & -     & 2.6   & -     & -    & -     & 3.5   & -     & -    & -     & 9.4    & -     & -    & -     & 0.7  \\
$\gamma1.1{\textrm{kh}}\beta0.8$ & 27.0  & 1.25  & 4.28 & 5.71  & 16.7  & 1.65  & 2.98 & 3.25  & 26.5  & 1.12  & 4.14 & 4.40  & 68.7   & 1.53  & 3.25 & 4.56  & 4.5  \\
$\gamma1.1{\textrm{kb}}\beta0.8$ & 17.1  & 1.09  & 4.39 & 5.71  & 10.8  & 1.45  & 3.07 & 3.09  & 16.7  & 0.96  & 4.26 & 4.30  & 49.3   & 1.18  & 3.46 & 4.34  & 2.8  \\
$\gamma1.3{\textrm{kl}}\beta0.8$ & 417   & 1.02  & 3.37 & 5.05  & 300   & 1.34  & 2.36 & 3.20  & 437   & 0.84  & 3.14 & 3.88  & 738    & 1.55  & 2.70 & 4.24  & 79.3 \\
$\gamma1.3{\textrm{kh}}\beta0.8$ & 104   & 1.64  & 4.22 & 5.98  & 75.7  & 1.98  & 2.92 & 3.76  & 121   & 1.37  & 4.12 & 5.01  & 208    & 2.09  & 2.96 & 5.16  & 19.3 \\
$\gamma1.3{\textrm{kb}}\beta0.8$ & 251   & 1.24  & 4.00 & 5.47  & 184   & 1.57  & 2.75 & 3.43  & 277   & 1.01  & 3.87 & 4.48  & 455    & 1.75  & 2.88 & 4.64  & 47.9 \\
$\gamma1.5{\textrm{kl}}\beta0.8$ & 380   & 1.54  & 4.15 & 5.97  & 258   & 1.92  & 3.01 & 4.26  & 423   & 1.31  & 4.10 & 5.51  & 795    & 1.78  & 2.75 & 4.86  & 67.9 \\
$\gamma1.5{\textrm{kh}}\beta0.8$ & 37.9  & 1.91  & 4.26 & 6.24  & 26.4  & 2.45  & 2.85 & 4.16  & 46.1  & 1.66  & 4.12 & 5.27  & 89.1   & 2.27  & 2.88 & 4.83  & 6.6  \\
$\gamma1.5{\textrm{kb}}\beta0.8$ & 189   & 1.68  & 4.19 & 6.08  & 129   & 2.10  & 2.97 & 4.21  & 216   & 1.44  & 4.12 & 5.44  & 411    & 1.98  & 2.77 & 4.75  & 33.4 \\
$\gamma1.7{\textrm{kl}}\beta0.8$ & 585   & 1.06  & 5.28 & 7.01  & 334   & 1.06  & 5.61 & 4.76  & 529   & 0.86  & 5.69 & 6.29  & 1205   & 1.94  & 2.88 & 5.05  & 95.1 \\
$\gamma1.7{\textrm{kh}}\beta0.8$ & 119   & 1.71  & 4.46 & 6.19  & 78.9  & 1.90  & 3.50 & 3.80  & 133   & 1.44  & 4.45 & 5.16  & 220    & 2.34  & 2.95 & 5.42  & 20.9 \\
$\gamma1.7{\textrm{kb}}\beta0.8$ & 342   & 1.35  & 4.86 & 6.43  & 208   & 1.35  & 4.70 & 4.14  & 339   & 1.11  & 5.06 & 5.55  & 655    & 2.10  & 2.89 & 5.18  & 57.5\\  \hline
\end{tabular}}
\end{center}
\begin{flushleft}
\footnotesize{Similar to Table~\ref{tab:t9} but for $\beta=0.8$ sub-models.}
\end{flushleft}	
\end{table*}

\begin{table*}
\begin{center}
\caption{BNS merger rate densities in unit of ${\rm Gpc}^{-3}\,{\rm yr}^{-1}$ in the local Universe, ${\rm Myr}^{-1}$ in the MW-like galaxy at redshift $z=0$ and the fitting parameters of the evolution of the merger rate density as a function of redshift resulting from the various models with $\beta=0.9$.}
\label{tab:t12}
\setlength{\tabcolsep}{1mm}{
\begin{tabular}{c|c|c|c|c|c|c|c|c|c|c|c|c|c|c|c|c|c}
\hline
\multirow{2}{*}{Model Name}& \multicolumn{4}{c}{Millennium-II}  & \multicolumn{4}{c}{EAGLE}  & \multicolumn{4}{c}{Illustris-TNG}  & \multicolumn{4}{c}{Madau \& Dickinson (2014)}  & $R_{\rm MW}$\\ \cline{2-17}
&$R_{\rm Mil,0}$&$\zeta$&$z_{\star}$&$\xi$  &$R_{\rm EAG,0}$&$\zeta$&$z_{\star}$&$\xi$  &$R_{\rm Ill,0}$&$\zeta$&$z_{\star}$&$\xi$ & $R_{\rm MD,0}$&$\zeta$&$z_{\star}$&$\xi$ &$[{\rm Myr}^{-1}]$\\
\hline
$\alpha0.1{\textrm{kl}}\beta0.9$ & 59.8  & 1.80 & 4.62 & 6.36 & 47.5  & 2.30  & 2.91 & 3.85  & 78.1  & 1.53  & 4.39 & 5.19  & 127    & 2.39  & 3.05 & 5.62  & 11.7  \\
$\alpha0.1{\textrm{kh}}\beta0.9$ & 2.5   & 1.82 & 4.59 & 6.35 & 2.0   & 2.35  & 2.88 & 3.95  & 3.3   & 1.56  & 4.35 & 5.22  & 4.9    & 2.45  & 2.98 & 5.54  & 0.5   \\
$\alpha0.1{\textrm{kb}}\beta0.9$ & 28.9  & 1.88 & 4.58 & 6.38 & 23.0  & 2.42  & 2.86 & 3.94  & 38.6  & 1.60  & 4.35 & 5.22  & 62.0   & 2.47  & 3.04 & 5.68  & 5.6   \\
$\alpha1.0{\textrm{kl}}\beta0.9$ & 456   & 1.84 & 4.57 & 6.39 & 342   & 2.46  & 2.74 & 3.85  & 586   & 1.60  & 4.29 & 5.00  & 928    & 2.39  & 3.10 & 4.80  & 86.4  \\
$\alpha1.0{\textrm{kh}}\beta0.9$ & 44.5  & 2.18 & 4.41 & 6.47 & 30.2  & 2.78  & 2.84 & 4.16  & 59.0  & 1.86  & 4.26 & 5.29  & 101    & 2.72  & 3.00 & 5.44  & 7.6   \\
$\alpha1.0{\textrm{kb}}\beta0.9$ & 237   & 1.95 & 4.49 & 6.39 & 177   & 2.59  & 2.72 & 3.96  & 311   & 1.69  & 4.23 & 5.05  & 491    & 2.51  & 3.05 & 4.94  & 44.4  \\
$\alpha10.{\textrm{kl}}\beta0.9$ & 782   & 1.42 & 4.26 & 5.86 & 588   & 1.86  & 2.71 & 3.61  & 901   & 1.20  & 4.04 & 4.82  & 1404   & 1.76  & 3.11 & 5.12  & 151 \\
$\alpha10.{\textrm{kh}}\beta0.9$ & 135   & 1.84 & 4.30 & 6.17 & 99.8  & 2.31  & 2.80 & 3.90  & 167   & 1.56  & 4.14 & 5.08  & 272    & 2.32  & 3.01 & 5.24  & 25.2  \\
$\alpha10.{\textrm{kb}}\beta0.9$ & 421   & 1.58 & 4.28 & 5.98 & 316   & 2.03  & 2.75 & 3.72  & 499   & 1.34  & 4.08 & 4.93  & 784    & 1.98  & 3.06 & 5.10  & 80.5  \\
$\gamma1.1{\textrm{kl}}\beta0.9$ & 5.9   & 0.23 & 3.53 & 4.93 & 4.6   & -     & -    & -     & 6.1   & -     & -    & -     & 15.8   & -     & -    & -     & 1.1   \\
$\gamma1.1{\textrm{kh}}\beta0.9$ & 26.3  & 1.26 & 4.31 & 5.79 & 15.8  & 1.69  & 2.99 & 3.27  & 25.4  & 1.14  & 4.16 & 4.44  & 69.5   & 1.52  & 3.28 & 4.59  & 4.3   \\
$\gamma1.1{\textrm{kb}}\beta0.9$ & 18.6  & 0.97 & 4.50 & 5.85 & 11.8  & 1.27  & 3.28 & 3.03  & 17.7  & 0.87  & 4.37 & 4.37  & 52.4   & 1.05  & 3.51 & 4.30  & 3.1   \\
$\gamma1.3{\textrm{kl}}\beta0.9$ & 447   & 1.07 & 3.46 & 5.21 & 332   & 1.33  & 2.45 & 3.23  & 480   & 0.85  & 3.27 & 4.03  & 784    & 1.57  & 2.74 & 4.33  & 86.2  \\
$\gamma1.3{\textrm{kh}}\beta0.9$ & 107   & 1.67 & 4.21 & 6.00 & 76.4  & 2.03  & 2.91 & 3.80  & 124   & 1.40  & 4.11 & 5.04  & 215    & 2.11  & 2.96 & 5.19  & 19.6  \\
$\gamma1.3{\textrm{kb}}\beta0.9$ & 261   & 1.28 & 3.99 & 5.52 & 194   & 1.58  & 2.77 & 3.44  & 291   & 1.04  & 3.87 & 4.52  & 477    & 1.76  & 2.90 & 4.67  & 50.1  \\
$\gamma1.5{\textrm{kl}}\beta0.9$ & 401   & 1.53 & 4.18 & 6.01 & 272   & 1.90  & 3.04 & 4.23  & 445   & 1.30  & 4.13 & 5.51  & 845    & 1.78  & 2.77 & 4.82  & 71.5  \\
$\gamma1.5{\textrm{kh}}\beta0.9$ & 39.0  & 1.94 & 4.26 & 6.26 & 27.6  & 2.45  & 2.86 & 4.16  & 48.1  & 1.67  & 4.13 & 5.28  & 91.2   & 2.31  & 2.88 & 4.85  & 6.8   \\
$\gamma1.5{\textrm{kb}}\beta0.9$ & 199   & 1.66 & 4.23 & 6.11 & 137   & 2.08  & 2.98 & 4.18  & 228   & 1.43  & 4.15 & 5.44  & 429    & 1.98  & 2.79 & 4.75  & 35.5  \\
$\gamma1.7{\textrm{kl}}\beta0.9$ & 661   & 1.01 & 5.34 & 7.01 & 385   & 0.98  & 5.78 & 4.79  & 608   & 0.79  & 5.79 & 6.33  & 1258   & 1.94  & 2.90 & 5.06  & 111 \\
$\gamma1.7{\textrm{kh}}\beta0.9$ & 123   & 1.74 & 4.45 & 6.18 & 81.3  & 1.91  & 3.51 & 3.82  & 138   & 1.45  & 4.45 & 5.17  & 229    & 2.35  & 2.95 & 5.42  & 21.3  \\
$\gamma1.7{\textrm{kb}}\beta0.9$ & 369   & 1.33 & 4.90 & 6.44 & 226   & 1.30  & 4.81 & 4.14  & 368   & 1.08  & 5.13 & 5.57  & 683    & 2.10  & 2.90 & 5.18  & 62.9  \\  \hline
\end{tabular}}
\end{center}
\begin{flushleft}
\footnotesize{Similar to Table~\ref{tab:t9} but for $\beta=0.9$ sub-models.}
\end{flushleft}	
\end{table*}


\bsp	
\label{lastpage}
\end{document}